\documentclass[graybox, envcountchap]{svmult}

\DeclareRobustCommand{\ion}[2]{\textup{#1\,\textsc{\lowercase{#2}}}}
\newcommand*\element[1][]{%
  \def\aa@element@tr{#1}%
  \aa@element
}

\usepackage{mathptmx}        
\usepackage{amsmath}
\usepackage{amssymb}
\usepackage{color}
\usepackage{helvet}          
\usepackage{courier}         
\usepackage{dirtree}

\usepackage{makeidx}        
\usepackage{graphicx}        
\usepackage{subfig}

\usepackage{multicol}        
\usepackage[bottom]{footmisc}

\usepackage{hyperref}        
\hypersetup{colorlinks=true,urlcolor=blue}

\usepackage[misc]{ifsym}

\makeindex             

\def\ch{{\it Chandra}}
\def\xmm{{\it XMM-Newton}}

\begin{document}


\title{Active Galactic Nuclei with High-Resolution X-ray Spectroscopy}
\titlerunning{AGN with high-resolution X-ray spectroscopy}
\author{Luigi C. Gallo,\thanks{corresponding author}  Jon M. Miller, and Elisa Costantini}
\authorrunning{L. C. Gallo, J. M. Miller \& E. Costantini}
\institute{Luigi C. Gallo (\Letter) \at Department of Astronomy \& Physics, Saint Mary’s University, 923 Robie Street, Halifax, NS B3H 3C3, Canada, \email{luigi.gallo@smu.ca}
\and Jon M. Miller \at Department of Astronomy, University of Michigan, 500 Church Street Ann Arbor, MI 48109, USA, \email{jonmm@umich.edu}
\and Elisa Costantini \at  SRON Netherlands Institute for Space Research, Niels Bohrweg 4, 2333CA Leiden, The Netherlands  \\
Anton Pannekoek Astronomical Institute, University of Amsterdam,  Postbus 94249, NL-1090 GE Amsterdam, The Netherlands, \email{e.costantini@sron.nl}}
%
%
\maketitle

\abstract{The imminent launch of XRISM will usher in an era of high-resolution X-ray spectroscopy.  For active galactic nuclei (AGN) this is an exciting epoch that is full of massive potential for uncovering the ins and outs of supermassive black hole accretion.  In this work, we review AGN research topics that are certain to advance in the coming years with XRISM and prognosticate the possibilities with Athena and Arcus. Specifically, our discussion focuses on: (i) the relatively slow moving ionised winds known as warm absorbers and obscurers; (ii) the iron emitting from different regions of the inner and outer disc, broad line region, and torus; and (iii) the ultrafast outflows that may be the key to understanding AGN feedback.}


\section{Introduction}
\label{sec:intro}

Active galactic nuclei (AGN) are unlike any other class of astronomical object.  They cannot be described by a single, dominating process.  Instead, AGN radiate energy over the entire electromagnetic spectrum, and are the sites of pair-production, cosmic rays, and gravitational waves.  Radiation is created through multiple processes, including blackbody, synchrotron, Comptonisation, bremsstrahlung, and line emission.   Gravity attracts material inward toward the black hole, but mass and energy can also be ejected into the host galaxy and beyond.   Gas can be viewed in absorption and emission, and it exists in various physical states that can be optically thick and thin, as well as neutral and completely ionised.  Ionisation (see Eq.~\ref{eq:xi}) can occur through collisional and radiative excitation.  Moreover, all these physical processes are subject to extreme gravity and magnetic fields, often invoking special and general relativity and relativistic magnetohydrodynamics (MHD).  Researchers must draw from many areas of physics to understand AGN.

Though fascinating in their own right, AGN have far-reaching influence in other fields of astronomy.  The AGN system is tiny in comparison to the host galaxy mass  ($M_{BH} / M_{gal} \sim 0.001$; e.g. \cite{FM00,Gebhardt00}), but its rest mass energy is comparable to the gravitational binding energy of the host galaxy.   If a small fraction of the SMBH energy is deposited into the host galaxy or intracluster medium, this AGN feedback will influence how galaxies evolve.  AGN feedback (e.g. \cite{Begelman04, King10, Fabian12}) will be important if the kinetic luminosity 
\begin{equation}\label{eq:lkin}
L_{KE} = \frac{1}{2} \dot{M}_{out}v_{r}^{2}
\end{equation}
of the wind deposits into the host galaxy approximately $0.5\%-5\%$ of the AGN bolometric luminosity ($L_{bol}$)  \cite{SR98, DiMatteo05, SH05, HE10, KP15}.  In the kinetic luminosity expression (Eq.~\ref{eq:lkin}), $v_{r}$ is the radial velocity of the wind and 
\begin{equation}\label{eq:mdot}
\dot{M}_{out}= \mu N_{H}m_{p}\Omega r f v_{r}
\end{equation}
is the mass outflow rate in terms of the total column density ($N_H$), solid angle subtended by the outflow ($\Omega$), distance from the black hole ($r$), proton mass ($m_p$), the volume filling factor ($f$), $v_r$, and the mean atomic weight correction ($\mu$), which is $\sim 1.23$ for cosmic abundances \cite{blustin05}.

\begin{figure*}
   \centering
   \advance\leftskip-0.2cm
   {\scalebox{0.43}{\includegraphics[trim= 0.cm 3cm 1cm 2cm, angle=0, clip=true]{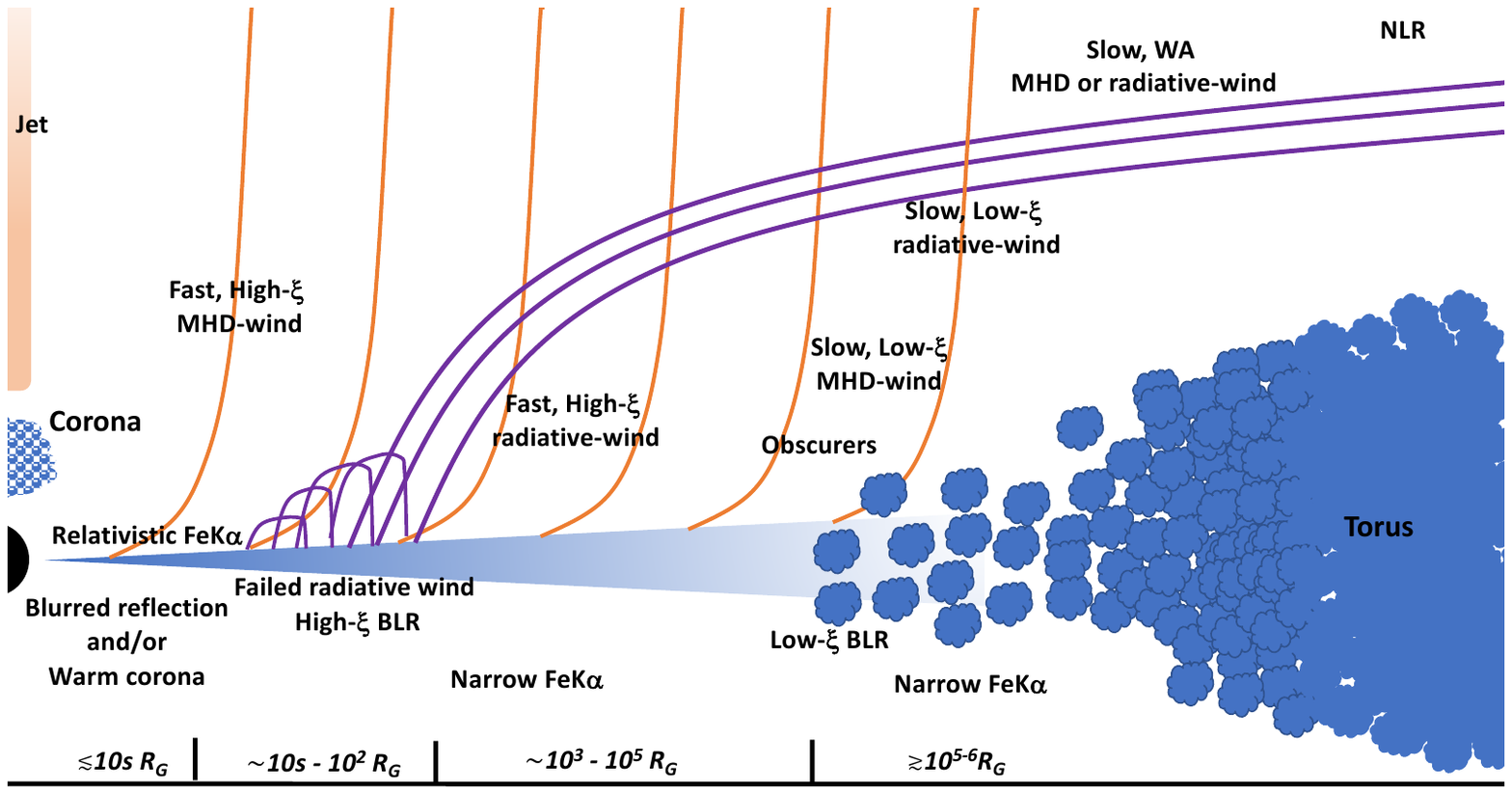}}}
   \caption{A simple illustration of the AGN region depicting the scales and main components responsible for X-ray emission.  The standard accretion disc extends down to the innermost stable circular orbit (ISCO) between $\sim 1.2-6~R_G$ depending on the black hole spin. The optically thin, hot corona is the primary X-ray source, and it is located close to the SMBH.  A collimated jet whose base may coincide with the corona, may or may not be present.  The magnetic field lines (orange curves) are the sites of the MHD-driven winds (Section~\ref{subsubsec:mhd}), and they thread the accretion disc over a large range.  The streamlines from radiation-driven winds (purple curves; Section~\ref{subsubsec:rdw}) are successfully launched at larger radii where the radiation pressure exceeds the gravitation force in the disc.  Within this launching radius, the wind falls back toward the disc, forming a failed wind, which may constitute the inner broad-line region (BLR, Section~\ref{subsec:narrowfek}) and/or shielding gas \cite{Proga04}.  The disc morphs into the obscuring torus (Section~\ref{subsec:narrowfek}) at large distances, and the traditional warm absorber (WA, Section~\ref{sec:wa}) on similar scales.  The narrow-line region (NLR) occupies galactic scales, but is photoionised by the AGN.  The observer's line-of-sight will dictate the type of system that is seen and the dominant processes at work.  High-resolution spectroscopy will probe the X-ray emission from all these regions.
}
   \label{fig:agn}
\end{figure*}

A simple illustration of the AGN region is depicted in Figure~\ref{fig:agn}.  The so-called central engine of the AGN is defined by the accretion disc that funnels material toward the supermassive ($M_{BH}\simeq 10^{6-9}~M_{\odot}$) black hole (SMBH).   The accretion rate, which is often parameterized by the ratio of the bolometric luminosity over the Eddington luminosity\footnote{The Eddington luminosity, $L_{Edd}=1.26\times10^{38} M_{BH}/M_{\odot}\rm erg~s^{-1}$, is the maximum luminosity a system can have such that the gravitational infall of ionised hydrogen gas is exactly balanced by the outward radiation pressure.} ($\lambda=L_{bol} /L_{Edd}$), will determine the structure of the accretion disc.  For moderate values of the Eddington ratio ($\sim 0.01 – 0.3$), the disc can be approximated as a standard Shakura-Sunyaev disc \cite{SS73} that is optically-thick, geometrically-thin, and radiative efficient. 

The X-ray emission from the inner-most region within 10's of gravitational radii ($R_G = GM/c^2$) is dominated by the hot corona.  This primary X-ray source can illuminate the inner disc leading to the production of the reflection spectrum that is blurred by relativistic effects close to the black hole (e.g. \cite{RossFabian93, Ballantyne01, RossFabian05}).  The dominant spectral feature here is the relativistic Fe~K$\alpha$ emission line (e.g. \cite{Fabian89, Laor}). Alternatively, the inner disc region might be blanketed by a warm corona that is optically-thick and conceals the fast inward flow (e.g. \cite{wc1, wc2, Ballantyne20}).

Winds in the accretion disc are important for transporting angular momentum outward so that material can flow inward (e.g. \cite{BH91, BP82, Murray95}).   Magnetic fields (orange curves in Figure~\ref{fig:agn}) that are capable of launching material out of the system (Section~\ref{subsubsec:mhd})  will thread the accretion disc over large distances.  The material in the MHD-driven wind closest to the black hole and moving fastest will also be the most highly ionised since it is closest to the primary X-ray source (i.e. the corona).   

Radiative winds (Section~\ref{subsubsec:rdw}) will also be launched from a distance corresponding to where the outward velocity from radiation pressure exceeds the escape velocity in the disc (purple streamlines in Figure~\ref{fig:agn}).   At distances closer than the launching radius, the radiation-driven wind will fall back onto the disc.  This failed wind region may manifest as many AGN components \cite{GP19,GP21} and may form part of the highly ionised broad-line region (BLR; Section~\ref{subsec:narrowfek}) in AGN.  

On parsec scales, the outer disc morphs into the torus, which is significantly optically thick and neutral.   Compton scattering and neutral iron emission are evident here (Section~\ref{subsec:narrowfek}). The traditional warm absorber (WA) will occupy scales similar to these outer disc regions (e.g. \cite{Laha14, Laha16}).  The WA is responsible for the ``normal’’ velocity winds (Section~\ref{sec:wa}) that may be driven by either radiative or MHD effects.  The narrow-line region (NLR) occupies galactic scales, but it is still photoionised by the AGN.  

There are many intrinsic variations in the AGN phenomenon that likely arise from different accretion rates.  However, the observer's line-of-sight also plays a significant factor  \cite{Antonucci93, Urry95}.  The observer's view through the obscuring torus will determine which disc regions are observable and if the AGN is defined as an unobscured Type I (e.g. Seyfert 1) or obscured Type II (e.g. Seyfert 2).  Whether the jet is present, its relative dominance over other AGN components, and the observer's perspective will dictate if the AGN is radio-loud (jetted) or radio-quiet (non-jetted). The line-of-sight through the winds to the primary X-ray source might also influence the types of winds that are observed.

In the past few decades, tremendous advances were made regarding the X-ray studies of AGN, particularly relating to broadband spectroscopy and variability.  The transmission gratings on \ch\ (see Chap. 3) and the reflection gratings on \xmm\ (see Chap. 2) provided glimpses into the discovery space opened by high-resolution spectroscopy.  AGN grating data were rich in spectral features, delivering information on gas temperatures, densities, dynamics, and origins.  AGN X-ray spectra are not just ``power laws’’. 

As we enter the era of calorimeter spectrometers with XRISM \cite{xrism} and Athena \cite{athena}, a new discovery space will be unveiled.  With a resolving power $E / \Delta E \approx 1400$, corresponding to $\sim5$~eV resolution at $\sim7$~keV, and a collecting area about $10$-times that of \ch\ in the Fe~K$\alpha$ band, XRISM will transform AGN science in the coming year.  We had but a brief, exciting view of this with {\it Hitomi} (e.g. \cite{hitomi, perseus,n1275}).

In this chapter, we will review and explore the areas of AGN research where high-resolution spectroscopy will make a certain impact.  In Section~\ref{sec:wa}, the ionised (warm) absorber science that has benefited greatly from grating spectrometers will be reviewed.  Later sections focus on the Fe~K$\alpha$ band, where calorimeters will resolve these data for the first time.  In Section~\ref{sec:narrow}, the emphasis will be on the nature and origin of the narrow (and broad) Fe~K$\alpha$ emission lines.  In Section~\ref{sec:ufo}, the highly-ionised iron seen in absorption and forming ultrafast outflows is examined.


\section{Warm absorbers}
\label{sec:wa}

The first detection of absorption from ionised gas in an AGN X-ray spectrum \cite{halpern84} opened a new window to study highly ionised nuclear winds. These outflows were subsequently detected and studied by all moderate-resolution CCD cameras: {\it ASCA} \cite{reynolds97,george98}, {\it BeppoSAX} (e.g., \cite{nicastro00,costantini00}), and {\it ROSAT} (e.g., \cite{komossa00}). From these early measurements, the clearest feature in the spectrum was identified as the photoelectric bound-free transition of \ion{O}{vii}  at $E=0.74$~keV. This feature, detected with different optical depths in most pointed observations of bright AGN, indicated gas with column densities $N_{\rm H}>10^{21}$\,cm$^{-2}$ \cite{george98}. A second feature at $E=0.87$~keV was attributed to the \ion{O}{viii} photoelectric edge, suggesting a more ionized component \cite{reynolds97}. 

\begin{figure}
\hbox{
\hspace{-0.3cm}
    \includegraphics[width=0.5\textwidth,angle=0]{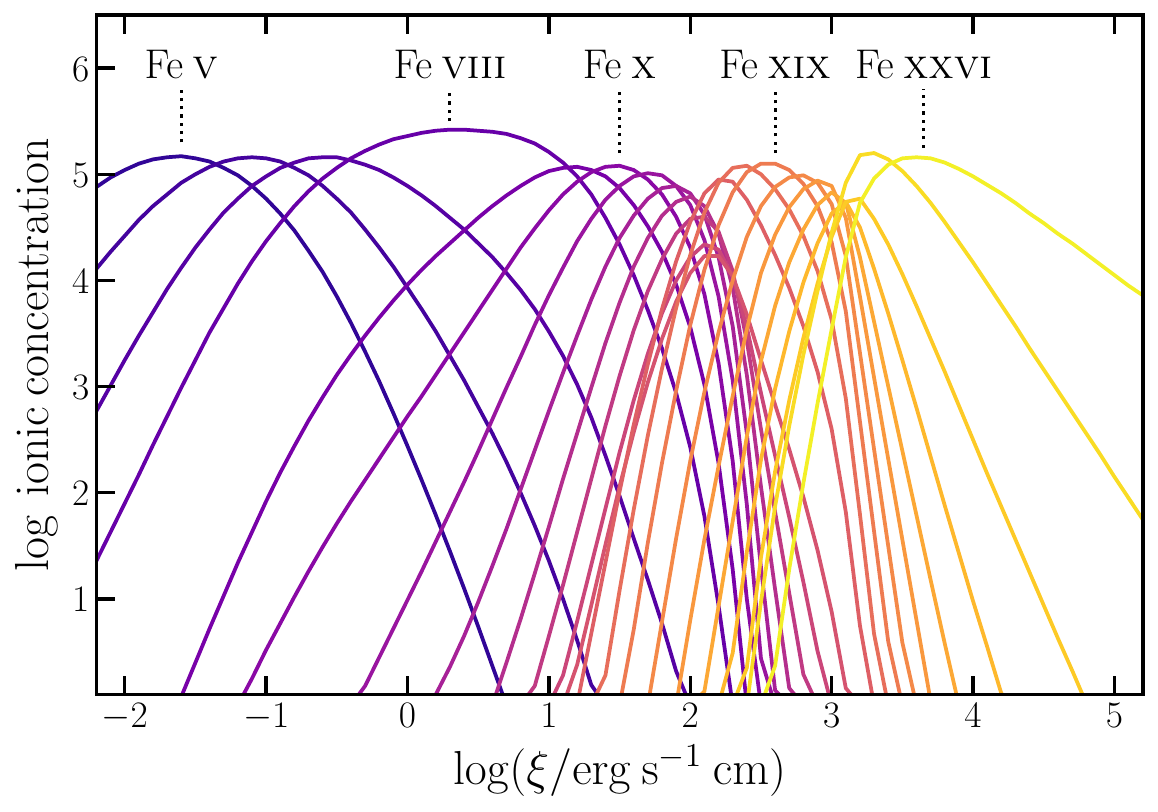}
     \includegraphics[width=0.38\textwidth,angle=90]{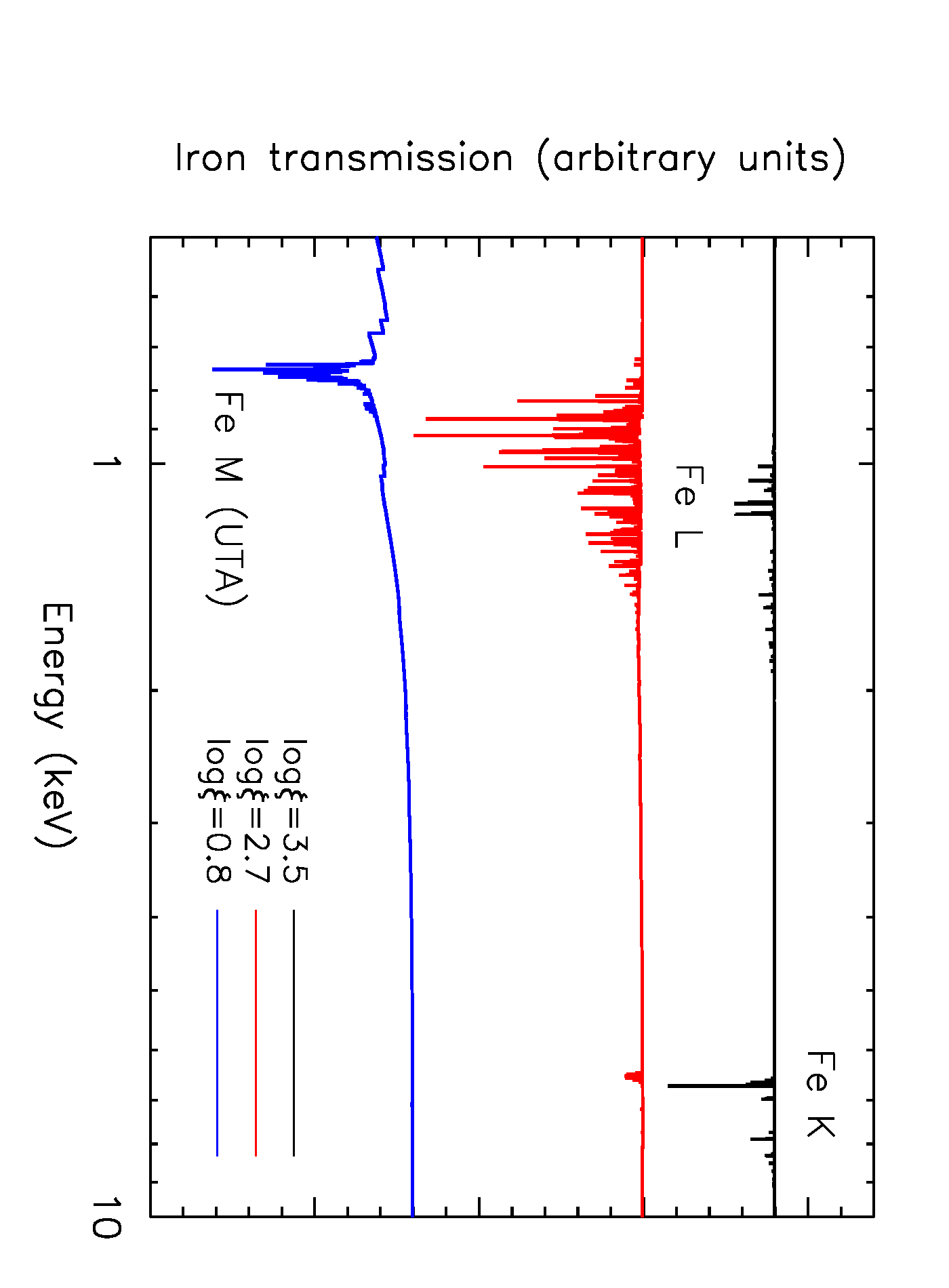}
 }
\caption{\label{fig:xi}Left panel:  The distribution in the concentration of iron ions as a function of log$\xi$. As the ionisation parameters increases, the ion concentration of more ionised Fe is enhanced (courtesy of D.~Rogantini). Right panel: The iron transmission spectra for selected values of log$\xi$ highlighting the Fe-UTA, L- and K-shell transitions.}
 
\end{figure}

A more quantitative characterisation of these WA came with the advent of the high-energy resolution grating spectrometers: \xmm-RGS,  \ch-HETG and \ch-LETG \cite{kaastra00, behar01, Kaspi01}. Dozens of transitions originating from carbon, nitrogen, oxygen, iron and neon were identified, and the deep absorption edges were no longer the only predictors of ionised absorbers. The feature identified as \ion{O}{vii} in earlier studies turned out to be heavily blended with iron transitions from the L-shell to the M-shell, called the iron UTA (unidentified transition array) \cite{sako01, behar01}. Refinements in the atomic database (e.g., \cite{gu01,gu06})  and subsequent studies determined that the iron UTA ions are very sensitive to changes in the ionisation parameter (Figure~\ref{fig:xi}), allowing a robust determination of the state of the gas \cite{behar01,steenbrugge05} and a diagnostic for gas changes over time (e.g., \cite{krongold07}).

The column density provides a measure of the quantity of gas intrinsic to the source along our line of sight. This does not provide any information on the geometry of the system since neither the thickness nor the density of the gas is directly measurable. The only constraint is that the thickness ($\Delta r$) of the gas cannot be greater than the gas distance from the central source ($r$) \cite{blustin05}:
\begin{equation}\label{eq:radius}
\Delta r<r.    
\end{equation}
A range of column densities spanning more than two orders of magnitude have been reported \cite{Laha14}, with the bulk of the gas being in the interval $\rm log( {\it N}_H / cm^{-2}) \sim 20-22$.

In this paper, the ionisation state of the gas is parameterised by \cite{tarter69}:\\
\begin{equation}\label{eq:xi}
\xi=L_{\rm ion}/nr^2\hspace{1cm}\rm{[erg\,cm\,s^{-1}]},
\end{equation}
where $\xi$ is the ionisation parameter, $L_{\rm ion}$ is the ionizing luminosity in the 1--1000\,Ryd interval, and $n$ is the gas density.
For different values of $\xi$, distinctive groups of ions will be present in the X-ray spectrum\footnote{In this section, all illustrations are produced using the SPEX package \cite{kaastra96}}. In Figure~\ref{fig:xi} (left panel), we show the distribution of ionic column densities for the iron ions as a function of log\,$\xi$. From the spectral point of view, this results in Fe absorption features, among others, distributed all over the X-ray spectrum. In Figure~\ref{fig:xi} (right panel), we show how iron significantly characterises the absorbed  spectrum. For instance, at log\,$\xi\sim 0.8$, the lower ionisation ions (marked by the iron UTA transitions) are more visible. As log\,$\xi$ increases, higher ionisation ions are present, originating from other iron L- and K-shell transitions.

The spectral energy distribution (SED) ranging from the ionisation threshold energy of hydrogen (1\,Ryd=13.6\,eV) to the end of the canonical X-ray band (1000\, Ryd) has a profound impact on the characteristics of the WA. The effects of the UV and soft X-ray spectral shape, as well as the high energy tail of the distribution, have been extensively studied (e.g. \cite{nicastro99a,susmita09,susmita12}). Alongside the distribution of the ionising photons as a function of energy, the metallicity of the WA itself also has a significant influence (e.g. \cite{komossa01,susmita09}).

\begin{figure}
\hbox{
\hspace{-0.9cm}
\includegraphics[width=0.4\textwidth,angle=90]{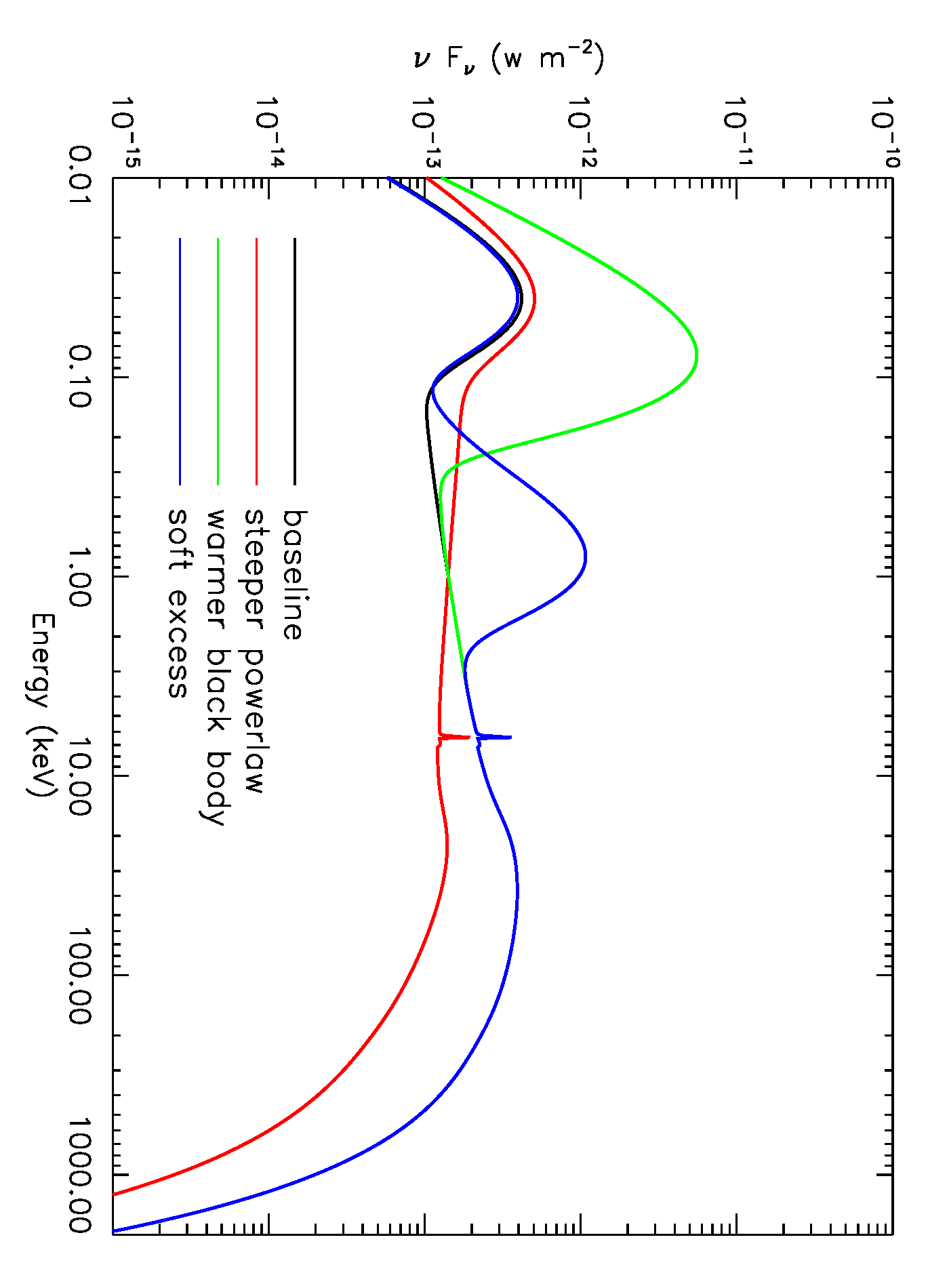}
     \hspace{+0.1cm}
     \includegraphics[width=0.4\textwidth,trim= 0.5cm 0.5cm 0.5cm 1.5cm, angle=90]{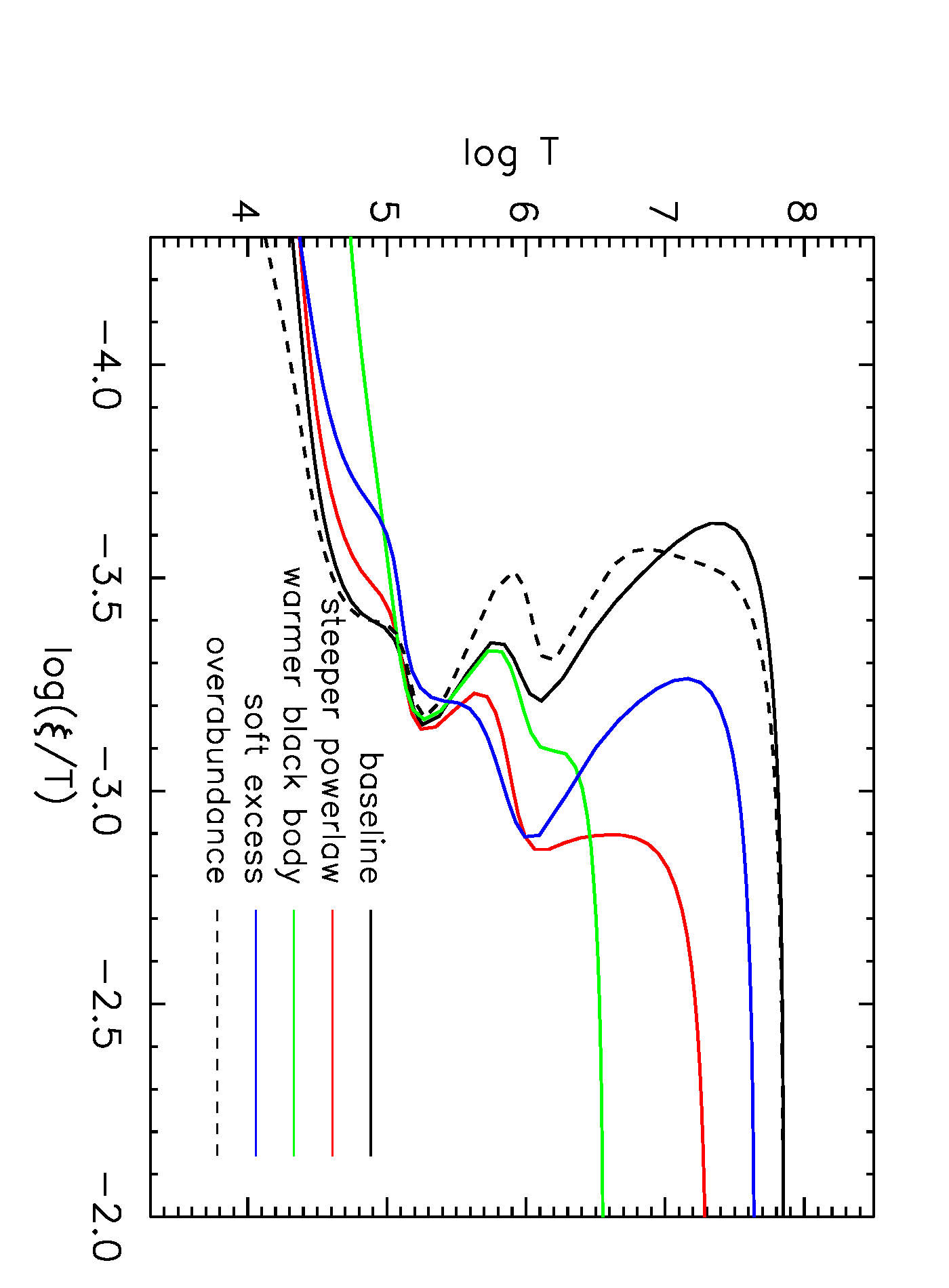}
 
}
\caption{\label{f:stability} Left panel: Examples of SEDs illuminating the ionized gas (following \cite{susmita09,susmita12}). Here, the baseline SED is produced by black body radiation with a temperature 0.01~keV, and a Comptonised component acting at soft X-ray energies with a seed temperature commensurate with the disc. The medium energy is characterized by a power law with a photon index $\Gamma=1.8$ and a reflection component, including the iron emission line, which is cutoff at $E=300$~keV (black solid curve). Right panel: The thermal stability curves for the SEDs displayed in the left panel (solid curves). The curves have been shifted along the x axis, to highlight the different SED effects \cite{susmita12}. In addition, the effect of a metal overabundance with respect to solar values is displayed with a dashed curve.}
\end{figure}

A useful visualisation of these influencing factors is given by the so-called stability curve. This describes, in a  log$\xi/T$ {\it vs} log$T$ plane, where a WA can exist in equilibrium, for a given SED and metallicity set. The term $\xi/T$ is proportional to the ionisation pressure parameter, $\Xi$, which is the ratio of the radiation pressure over the gas thermal pressure (e.g. \cite{rozanska08}):
\begin{equation}
\Xi\equiv\frac{P_{\rm rad}}{P_{\rm th}}=\frac{L}{4\pi r^2 c}\frac{1}{nk_{B}T}\propto\frac{\xi}{T}.
\end{equation}
Here, $k_B$ is the Boltzmann constant and Eq.\ref{eq:xi} is used to simplify the final formulation.  Following \cite{susmita09,susmita12}, Figure~\ref{f:stability} illustrates the behaviour of the thermal stability curve (right panel) as a function of a few key parameters of the SED (left panel). 

The presence of a stable branch in the curve, that is an almost constant pressure zone for a range of temperatures (e.g., \cite{krolik81}), is enhanced as the power law slope is relatively flat ($\Gamma=1.8$ in this example). Under these conditions, more WA components with different ionisations can co-exist in pressure equilibrium \cite{krongold05}. However, the power law slopes are often significantly steeper (up to $\Gamma\sim2.8$ \cite{bianchi09}). In these cases ($\Gamma=2.1$, red solid curve in Figure~\ref{f:stability}), the pressure equilibrium zone is disrupted.

The effect of a higher temperature seed black body, or of an enhanced soft excess (green and blue solid curve, respectively in Figure~\ref{f:stability}), impact the curve at lower values of $\xi/T$, where warm absorbers exist at higher temperatures than in the baseline case, due to heating by iron ions \cite{susmita12}.

Element abundances in the WA also have an impact on the stability curve. To emphasize this effect, in Figure~\ref{f:stability} (right panel) an overabundance by a factor of five with respect to solar values is displayed (dashed line). The effect is to create a larger zone of pressure stability.

The shape of the SED also influences the type of ions that appear in the X-ray absorption spectrum \cite{nicastro99a}. A SED typical of a Seyfert~1 galaxy shows a number of absorption lines and photoelectric edges from C, N, and O.  Instead, a steeper energy spectrum, for example, characteristic of narrow-line Seyfert~1 (e.g. \cite{boller96, brandt97, Gallo18}), stimulates iron transitions at different ionisation stages, producing the typical UTA and L- transition arrays at soft energies \cite{nicastro99a}. For higher ionisation parameters, narrow-line Seyfert~1s display more pronounced iron absorption in the 6--7\,keV range (e.g. \cite{gallo06}). 

The spectra obtained in the X-rays with high-energy-resolution clearly show that absorption lines displayed a significant blueshift in velocity with respect to the redshift of the AGN host \cite{kaastra00,Kaspi01}. The blueshift corresponds to WA with velocities in the range $10^{2-3}$\,km\,s$^{-1}$ (e.g. \cite{Laha14}). This parameter added the important information that other than an absorbing gas along the line of sight, these components were ejected from the nuclear region towards the host environment. In the same object, different WA components do not necessarily share the same outflow velocity. The distribution of outflow velocities for WA was found to be weakly correlated with the ionisation parameter \cite{Laha14}. 

Some high-quality spectra of Seyfert~1 galaxies are known to host multiple WA components (e.g. \cite{Detmers11,kaastra14}), that differ in ionisation, column density, and outflow velocity. How these components are organized in the AGN system is one of the open questions in this field.  As seen above, WA components are only sometimes found to be in pressure equilibrium, where a diffuse, more ionized absorber contains a colder, possibly clumpy, component. Pressure equilibrium would ensure a long-lived gas outflowing structure. In the same object, some WA components are found in equilibrium with each other, in the constant pressure branch of the curve, while others sit in the low ionisation branch (e.g. \cite{Detmers11}). 

An intuitive picture (Figure~\ref{fig:agn}) sees the outflows along our line of sight as part of a continuous stream, with the more ionized components located closer to the ionising source (Eq~\ref{eq:xi}). Using Eq.~\ref{eq:radius} as an upper limit for the radius and assuming that the outflow velocity $v_{out}$ is larger than or equal to the gas escape velocity, a rough range for the location of the absorbers can be found, using this simple geometry \cite{blustin05,Laha16}. The lower limit on the radius is then given by $r>2GM_{BH}/v^2_{out}$. With these estimates, the WA location encompasses a radius range of more than two orders of magnitude, with lower ionisation components located further away \cite{Laha16}. Their location seems to be roughly between the broad line region and the molecular torus \cite{blustin05}.
 
The WA components are generally modelled as discrete components, with well-defined parameters. The absorption measure distribution (AMD, \cite{holczer07,Behar09,keshet22}), describes instead the absorption spectra in terms of a continuous distribution of column densities per unit of log$\xi$ (${\rm d}N_{\rm H}/{\rm d}({\rm log}\xi)$) as a function of log$\xi$. The integral over log$\xi$ would result in the total column density of the WA components. A linear fit of the AMD distribution may be only a simple parameterization, but provides a slope $a$ that may be used as a diagnostic to be compared with theoretical models (see below and \cite{Behar09}). The value of $a$ has been reported to range between $0-0.4$ \cite{Behar09}. 

Absorption by ionized gas has been mostly associated with radio-quiet objects. The sparse detection in radio-loud, non-blazar objects (e.g. \cite{reeves09,torresi12,digesu16}), suggests that radio loudness could somehow interfere with the detection of WA, for example, if the object was observed at an unfavourable angle. The X-ray radiation of radio-loud objects being more intense (\cite{gupta18} and references therein), was thought to fully ionise the surrounding medium. A systematic study of radio-loud objects in X-rays showed a robust anti-correlation between the power of the radio emission and the column density of the warm absorbers \cite{mehdipour19}, independent of the inclination angle and X-ray luminosity.  
This anti-correlation pointed to a bi-modality between the radio activity and the disk ultimately originating the WA, providing further clues on the origin of the winds.  

\subsection{Obscurers}
A category of ionized gas that has been relatively overlooked due to their unpredictable occurrence, has been the so-called ``obscurers'' (e.g., \cite{longinotti13,kaastra14, mehdipour17, kara21}). This obscuring gas can be formed by one or more high column density ($N_{\rm H}\sim10^{22-23}$\,cm$^{-2}$) gas mass, that temporarily obliterates the X-ray soft energy spectrum. This phenomenon became well known from long-term {\it RXTE} and {\it Swift} light curves, as seemingly normal Seyfert~1 sources, would undergo periods of very hard spectra (high hardness ratios), due to the suppression of the soft X-rays \cite{markowitz14}. These sources were sometimes caught in this state, for example by  \xmm-pn (e.g. NGC~3516\,\cite{turner11}, Mrk766\,\cite{risaliti11}, 1H0419-577\,\cite{digesu14a}), but without any UV spectroscopic coverage. The limited energy coverage  raises different interpretations on whether the hardness ratio is due to absorption or to intrinsic changes in the source (e.g. \cite{jiang19}). 

The fortuitous simultaneous observations of NGC~5548 in this spectral state with \xmm, {\it HST}-COS, {\it NuSTAR} and {\it Integral} \cite{kaastra14}, allowed for the first time a comprehensive study of this phenomenon. The {\it HST}-COS spectrum showed, for every major transition (from \ion{C}{ii} up to \ion{C}{iv}, \ion{N}{v}, and Ly\,$\alpha$ \cite{kaastra14}), the presence of deep outflows, with velocities in the range 1000--5000\,km\,s$^{-1}$. The ionisation parameter of this high-column density gas component is relatively low with log$\xi\sim-(2-1)$, possibly pointing at a dense gas mass. This mass is also not completely covering our line-of-sight ($C_{V}\sim0.5-1$), changing in both covering factor and column density on a time scale of days--months \cite{digesu15}, hinting at a patchy nature for the obscurer.  

The degree of absorption correlated with obscuration in the soft X-rays, unequivocally linked the two absorbing agents as the same gas. The location of the obscurer has been estimated to be between the UV-emitting broad line region and the WA. The obscuring gas indeed absorbs the blue side of the broad emission lines in the UV (see Fig.~1 in \cite{kaastra14}). At the same time, the WA components have been found to be photoionised by the central source SED, but significantly modified in shape by the obscurer. This meant that the WA components must have been located at larger distances with respect to the obscuring gas. Instabilities and eruptions in the accretion disks have been invoked to explain the occasional rise of this high-column density gas. However, it is still uncertain why some obscuration events last for almost 10 years (e.g., NGC~5548 \cite{mehdipour22}) and others, occurring in objects of seemingly similar evolutionary state, last only weeks (e.g., NGC~3783 \cite{mehdipour17}). The decline of these disk-wind obscurers is not directly connected to the SED changes, nor to the frequency of their appearance \cite{kaastra18,mehdipour22}.

Dedicated campaigns, covering the UV and the X-ray band, revealed that many Seyfert~1 galaxies may undergo periods of obscuration during their active life. The study of obscurers also brought to light a complex interplay between the illumination and the covering factors of the UV and X-ray obscurer \cite{mao22,dehghanian19,mehdipour22}. The presence of the obscurer may also describe the X-ray shielding invoked to explain the survival of UV absorbers \cite{Proga04,dehghanian19}.
In addition to the cold patchy components, the occurrence of obscurers has also been associated with an additional component of very high-ionisation gas, with velocity consistent with the cold component,  indicating inhomogeneity in the medium, where lower density hems become more ionised \cite{mehdipour17}.

\subsection{The importance of WA and the density determination}
As seen above, WA and outflows in general are promising conveyors of feedback into the host galaxy, with important implications for galaxy evolution and formation. In any model predicting the launching mechanism and the impact of outflows on the surrounding medium, it is fundamental to know at what distance the outflow is launched. The rough estimates reported above indicate a range of distances that span orders of magnitudes. This uncertainty is then reflected in Eq.~\ref{eq:mdot}. On the other hand, the distance $r$ in Eq.~\ref{eq:xi} can be calculated from observable parameters only if the density $n$ is known.

A method that has been successfully used in the UV band is density evaluation through density-sensitive absorption lines. These absorption lines are the result of the electron population of a so-called metastable level, above the ground level. The population of this unstable level may be due to both an excess of optical photons in the SED or to the gas with a density above a given threshold, which is different depending on the ion. The column density of the metastable level line is compared to the ground transition and to theoretical curves to find the best fit for the gas density (e.g. \cite{korista08}).
Metastable levels of \ion{C}{iii}, \ion{N}{iii} have been regularly used (e.g. \cite{gabel03}) as well as a number of other UV (\ion{Si}{ii}, \ion{S}{iii}, \ion{P}{iii}, \ion{Fe}{iii} \cite{arav15}), and optical transitions (e.g. \ion{Fe}{II} \cite{arav08}). Sometimes, the metastable line happens to be part of a WA component that is visible both in the UV and the X-rays, providing a density estimate also for the X-ray absorber \cite{costantini10,edmonds11,digesu13}.  In the X-ray band, only one detection of \ion{O}{v} has been reported for an AGN so far \cite{kaastra04}, leading to a lower limit for the gas density.

Methods that may be easier to apply, make use of the response of the gas as a function of the ionizing luminosity (Eq.~\ref{eq:xi}). Every ion in a gas will have its own reaction time, depending on the density of the gas \cite{krolik_kriss95,nicastro99b,rogantini22}. In particular, the time taken for the gas to react is inversely proportional to the gas density.  The time dependence of the ionic concentration of a certain element $X$, $n_{\mathrm{X}^{i}}$ can be written as:  
\begin{equation}\label{eq:kk}
		\displaystyle \frac{d n_{\mathrm{X}^{i}}}{d t} = - n_{\mathrm{e}} n_{\mathrm{X}^{i}} \alpha_{\mathrm{rec},\mathrm{X}^{i-1}} - n_{\mathrm{X}^{i}} I_{\mathrm{X}^i} + n_{\mathrm{e}} n_{\mathrm{X}^{i+1}} \alpha_{\mathrm{rec,X}^{i}} + n_{\mathrm{X}^{i-1}} I_{\mathrm{X}^{i-1}}. 
	\end{equation}
Here, $n_{\mathrm{e}}$ is the electron density, while $I_{\mathrm{X}^{i}}$ and $\alpha_{\mathrm{rec,X}^{i}}$ are the ionisation and recombination rates 
between state $i$ and $i+1$.

\begin{figure}
   \advance\leftskip0.5cm
\includegraphics[scale=.45, trim= 0cm 0cm 0cm 0cm]{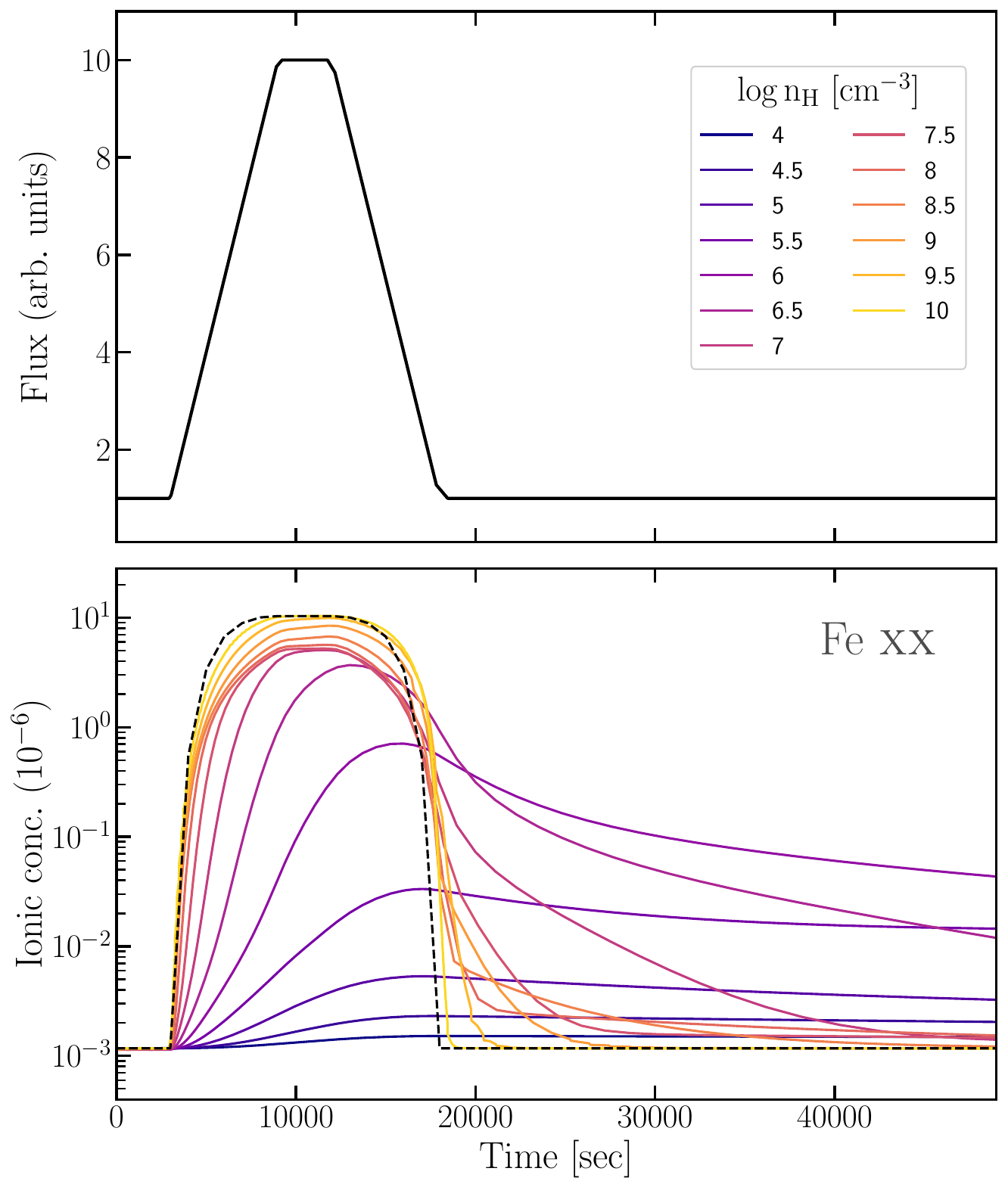}
\caption{\label{fig:tpho} Upper panel: An example of a flaring light curve. Lower panel: The ionic concentration of \ion{Fe}{xx} response to the time evolution of the flux change. The equilibrium condition (i.e. instantaneous response, following Eq.~\ref{eq:xi}) is marked with a dashed line. Depending on the density, the gas will respond with a different delay, as highlighted by the colour-coded solid lines. Adapted from \cite{rogantini22}.}
\end{figure}

In Fig.~\ref{fig:tpho}, the behaviour of \ion{Fe}{xx} as a function of time is illustrated for a range of gas densities. The higher the column density the more the gas response approaches an instantaneous change as a function of the flux variation, leaving the gas in equilibrium conditions at every time. If, on the contrary, the gas density is very low, the signal is diluted and no variation in the gas is observed. For a range of densities 
(log($N_{\rm H}/{\rm cm}^{-2})\sim5-7$), \ion{Fe}{xx} reaction time is significantly delayed.

In principle, Eq.~\ref{eq:kk} can be used for every source whose flux significantly changes in time. In practice, this method, that relies on time-resolved spectroscopy, is often limited by the signal-to-noise per time bin of the spectrum. A slow variation may lead to subtle variations in the WA that are difficult to detect \cite{kaastra12}, while strong and sudden variations require that the WA is analysed on few-kilosecond time bins, therefore reducing the quality of the spectrum \cite{krongold07}. Several estimates of the density, and therefore on the distance, of the WA have been reported. 
Only a few experiments were successful in putting limits on the distance of the WA. For example, \cite{krongold05} derived an upper limit of $r<6$\,pc for a gas component in NGC~3783, consistent with the limit ($r<25$\,pc) derived for the same gas component from the analysis of the UV data \cite{gabel05}.

In Mrk~509, the general lack of variability in the WA set the lower limits to relatively far distances for the different components (ranging from about 5\,pc up to 70\,pc, with some components at kpc scales \cite{kaastra12}). A similar range, spanning from parsecs to tens of parsecs, has been found for the WA components of NGC~985 \cite{ebrero21}. 
Smaller black hole systems, like in the narrow-line Seyfert~1 galaxies, show instead the desired large variation in flux on time scales of few kiloseconds. 
An intensively studied object of this class is NGC~4051, where the WA components were found to be closer to the black hole, around 1\,light day distance (e.g., \cite{krongold07,steenbrugge09}).

\subsection{Future outlook on warm absorbers}

\begin{figure}
   \advance\leftskip2.5cm
\includegraphics[scale=0.51,angle=0]{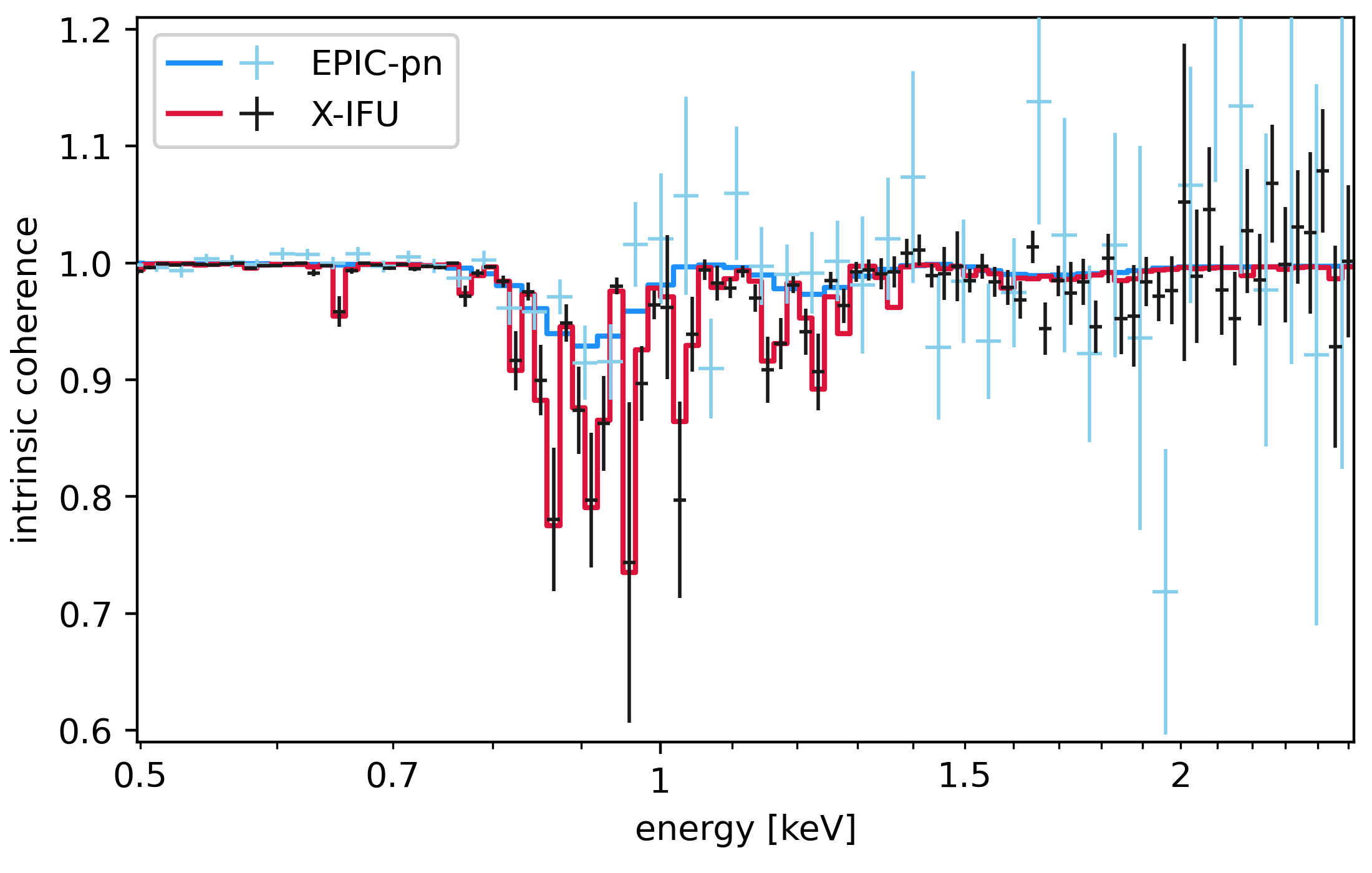}
\caption{\label{fig:coherence} Athena-XIFU simulation (red data points and solid line) of the coherence as a function of energy for a typical narrow-line Seyfert~1 source hosting a WA. This is compared to the capabilities of current high-sensitivity instruments (e.g., \xmm~EPIC-pn, blue data points and solid line). The shape of the coherence spectrum as a function of frequency can be modelled to solve for the gas density. Modified from \cite{juranova22}.}
\end{figure}

The design of future high-resolution instruments would certainly reward the study of ionised absorbers.  High-sensitivity, high-resolution ($\Delta E\leq5$\,eV) calorimeters (XRISM, \cite{xrism} and Athena-XIFU \cite{athena}) will bring significant advancement, for example, in the study of variable WA and the determination of the gas density. In the future, time-resolved spectra will allow us to follow the evolution of multi-components in WA as a function of time \cite{rogantini22}. At the same time, different applications of Eq.~\ref{eq:kk} will be possible, for instance combining timing and spectroscopy \cite{silva16}. The coherence between two signals, in this case between the continuum and the warm absorber, as a function of energy, bears the signature of the time delay of the recombining gas and therefore of the gas density \cite{juranova22}. This can be detected and studied provided high-sensitivity and resolution data (Figure~\ref{fig:coherence}). 

\begin{figure}
   \advance\leftskip1cm
\includegraphics[scale=.5,angle=0]{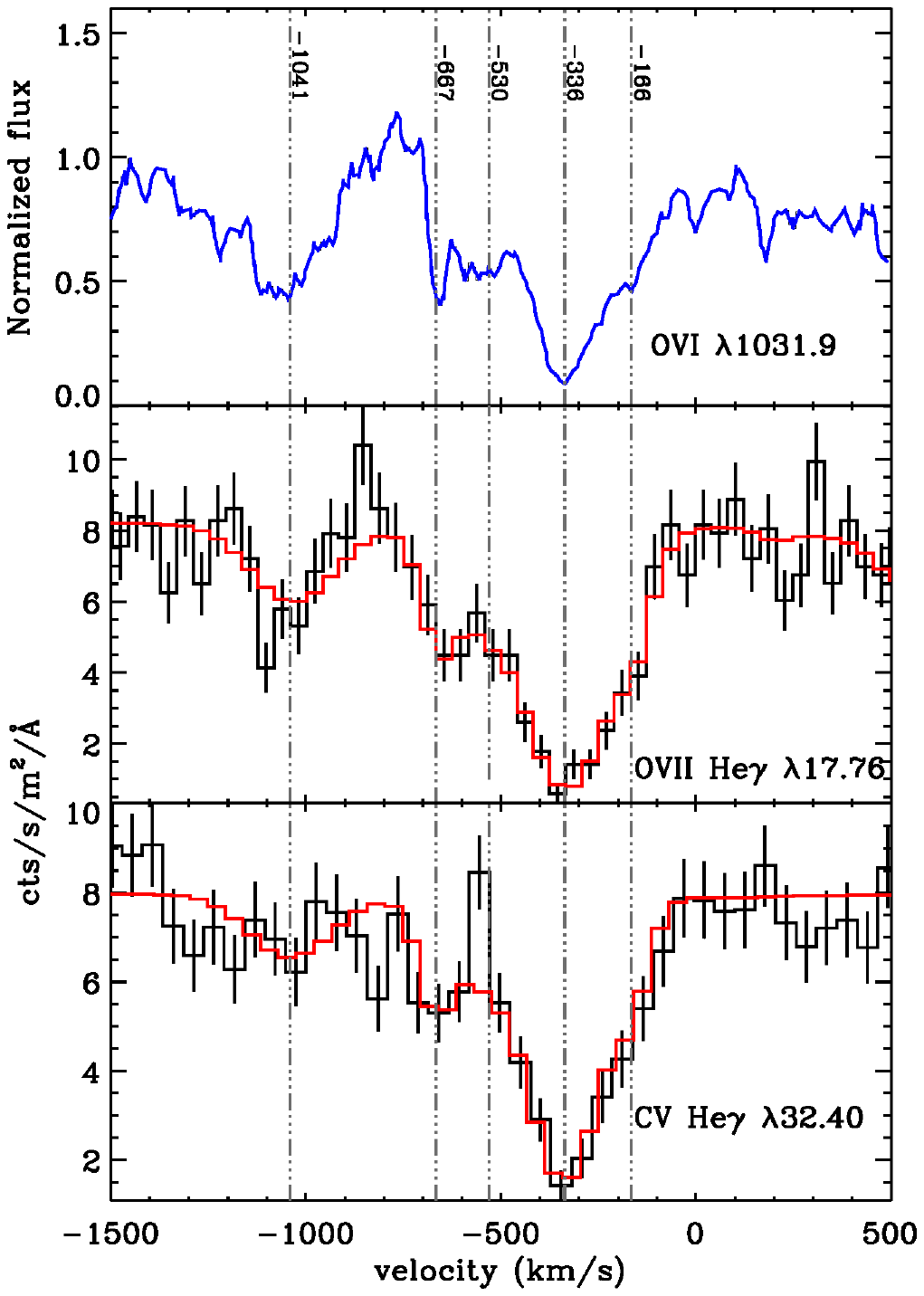}
\caption{\label{fig:arcus_vel} Arcus simulation of the non-saturated $\gamma$ lines of the He-like series for \ion{O}{vii} and \ion{C}{v}, respectively (middle and lower panels), compared to the {\it HST}-STIS velocity resolved profile of one of the doublet components of \ion{O}{vi} ($\lambda 1031.9\AA$) (upper panel \cite{crenshaw03}).}
\end{figure}
\begin{figure}
\includegraphics[scale=.45, angle=90]{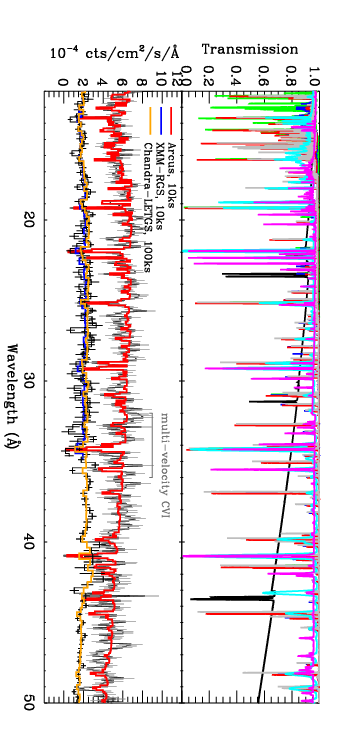}
\caption{\label{fig:arcus} A 10\,ks simulation using a next generation soft-X-ray grating ($\lambda/\Delta\lambda>2500$). The simulation includes a six-component WA \cite{kaastra14} and Galactic cold absorption (upper panel). The grating simulation is compared with \xmm-RGS and \ch-LETGS (lower panel). The data have been rebinned and shifted along the vertical axis for clarity.}
\end{figure}

New generation grating spectra (as in the mission concept Arcus \cite{smith16}), operating in the soft-X-ray energy, will permit studying the kinematics of the absorbers (Figure~\ref{fig:arcus_vel}) and line absorption profiles, as routinely performed in the UV band. The envisaged resolution ($\lambda/\Delta\lambda>2500$) and a large effective area compared to \xmm-RGS, will allow us to perform different density diagnostic tests, including the detection of metastable levels \cite{mao17} and time-resolved spectroscopy \cite{rogantini22} on a large number of sources (Figure~\ref{fig:arcus}).


\section{Fe K emission lines in AGN}
\label{sec:narrow}

\subsection{The atomic physics of Fe K emission lines}
\label{subsec:iron}

Narrow emission lines from the Fe K shell are the most prominent atomic feature in the X-ray spectra of AGN.  Typically, a single emission line is seen in addition to a local continuum that can be approximated with a power law.  These lines are produced when ionizing X-ray radiation illuminates relatively cold, dense gas.  In this sense, narrow Fe K lines trace the interaction of the central engine with the accretion flow on all scales, and can serve to test models for its (sub-)structure and evolution.  Indeed, hard X-ray emission from the central engine can even excite narrow Fe K lines on scales that are better associated with the larger host galaxy than the accretion flow. The ``X-ray reflection nebulae'' in the center of the Milky Way, for instance, likely indicate that Sgr A* was much more luminous $few\times 10^{4}$~years ago \cite{Koyama96} (also see \cite{Ponti15}).

It is important to clarify that Fe K-shell lines are only prominent within the spectra of AGN owing to a combination of three key factors: (1) a relatively high fluorescence yield, (2) a relatively high abundance, and (3) the Fe K lines fall within a part of the spectrum that is otherwise relatively simple.  The fluorescence yield of an atomic shell is simply the probability that a vacancy results in a radiative transition, rather than the ejection of an Auger electron.  The K-shell fluorescence yield is positively correlated with atomic number \cite{Bambynek72}.  By a factor of $\sim5$, Fe is the element with the highest product of fluorescence yield and abundance  \cite{GF91}.  Fluorescence lines from the K-shell of more abundant elements with lower yields are also excited when Fe K-shell lines are excited; they are just less prominent.

A narrow Fe~K emission line in the spectrum of a given AGN is often referred to as ``{\it the} neutral Fe K'' line or ``the neutral Fe~K$\alpha$ line.''  These colloquial terms are convenient, especially at modest sensitivity and/or modest spectral resolution.  However, the situation is more complex, and this may become readily apparent in the era of calorimeter spectroscopy.  It is therefore worth undertaking a quick overview of the key atomic physics, before reviewing some key recent developments with \ch, \xmm, and other telescopes.

The neutral Fe~K$\alpha$ fluorescence line arises from a 2p-1s electron transition, and the spin-orbit interaction therefore creates two lines with a small energy difference: K$\alpha_{1}$ at 6.404~keV and K$\alpha_{2}$ and 6.391~keV, with a 2:1 branching ratio \cite{Bambynek72}.  This difference, just $\Delta E = 13$~eV, exceeds the $\Delta E = 5$~eV energy resolution that was achieved with the calorimeter aboard {\it Hitomi} \cite{hitomi}; the ``Resolve’’ calorimeter that will fly aboard XRISM in 2023 is expected to have the same resolution \cite{xrism}.  The X-ray Integral Field Unit (XIFU) spectrometer expected to fly aboard Athena in the 2030s is expected to have a resolution of just $\Delta E = 2.5$~eV. 

\begin{figure*}
   \centering
   \advance\leftskip-0.cm
   {\scalebox{0.4}{\includegraphics[trim= 2cm 1cm 1cm 2cm, angle=-90, clip=true]{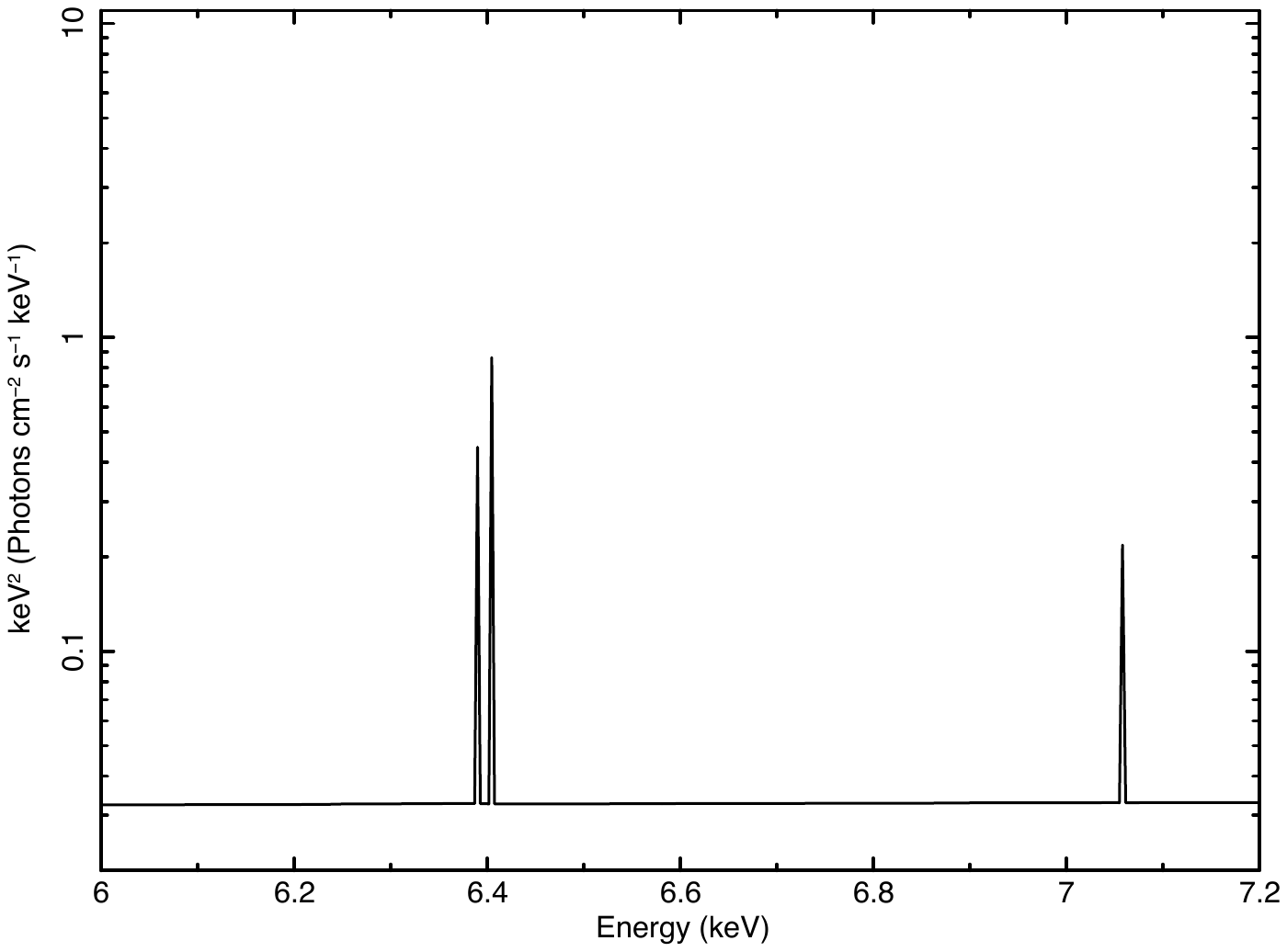}}}
   {\scalebox{0.4}{\includegraphics[trim= 2cm 1cm 1cm 2cm, angle=-90,  clip=true]{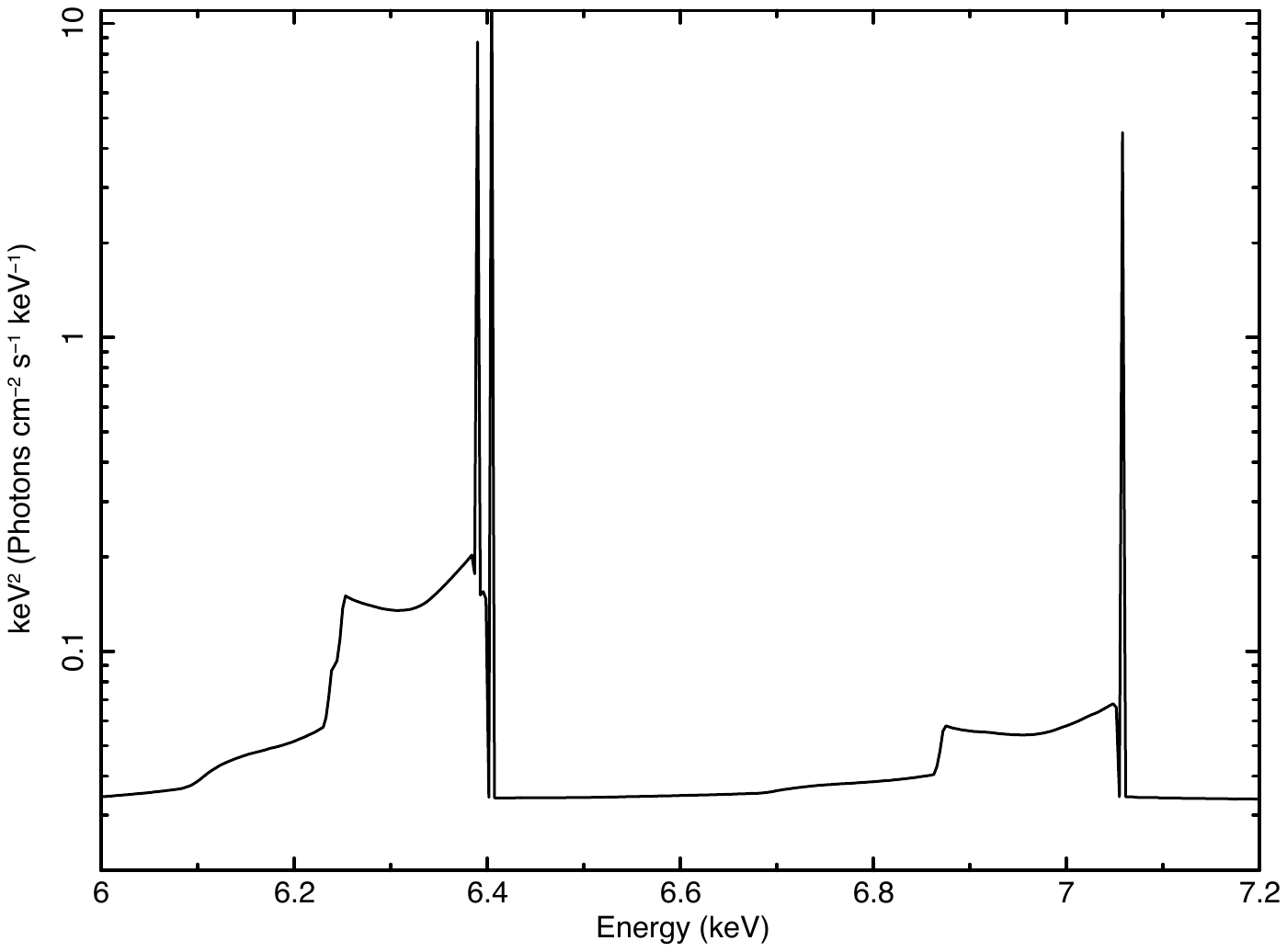}}}       
   \caption{Spectral models depicting line production in the Fe~K region of AGN spectra, generated using the {\sc mytorus} code \cite{mytorus, Yaqoob12}, also see \cite{Kammoun20}. In the upper panel, the line production region has column density of $N_{H} = 1\times 10^{22}~{\rm cm}^{-2}$.  In the lower panel, the line production region has a column density of $N_{H} = 2\times 10^{24}~{\rm cm}^{-2}$, making it Compton-thick.  These models were constructed assuming neutral gas and a viewing angle of $\theta = 30^{\circ}$.  In both cases, the two Fe~K$\alpha$ lines are apparent, as well as the Fe K$\beta$ line.  However, in the Compton-thick case, the first Compton shoulder is visible, owing to $180^{\circ}$ scattering in the gas. }
   \label{fig:jmm1}
\end{figure*}

The upper panel of Figure~\ref{fig:jmm1} shows a model spectrum constructed using {\sc mytorus}  \cite{mytorus} owing to its physical self-consistency, and because it has a native resolution of just 2~eV and is therefore suited to even calorimeter spectra \cite{mytorus, Yaqoob12}.  It depicts an Fe~K emission spectrum from neutral gas, broadly characteristic of the spectra observed in Seyfert~1 AGN.  The model was constructed according to the prescriptions in \cite{Kammoun20} assuming an obscuring column density of $N_{H} = 1\times 10^{22}~{\rm cm}^{-2}$ and an inclination of $\theta = 30^{\circ}$.  At high resolution, the Fe~K$\alpha_{1}$ and Fe~K$\alpha_{2}$ lines are easily separated, and the corresponding Fe~K$\beta$ line is also clearly represented.

The weighted average of these two lines is $E = 6.3997 \simeq 6.40$~keV, but if the two lines are fit with a single Gaussian -- common in non-calorimeter data – it is then important to account for the two lines in any determination of the line production radius based on virial or Keplerian motions.  The energy difference between the two corresponds to a broadening of $\Delta E / E \simeq 600~{\rm km}~{\rm s}^{-1}$ that must be subtracted before estimating the production radius.

Most of the strong Fe K lines that are observed in AGN are statistically consistent with being neutral; however, this does not automatically signal that the emitting gas is neutral.  The weighted mean line energy of \ion{Fe}{i} is 6.40~keV, but this only changes to 6.43~keV for Ne-like \ion{Fe}{xvii} \cite{Bautista03, Bautista04, Palmeri03a, Palmeri03b, Mendoza04} (also see \cite{NIST}).  Even for grating spectrometers and CCD spectrometers, a difference of just $\Delta E = 30$~eV is often within the measurement uncertainty in data of modest sensitivity.  

As we look ahead to the era of calorimeter spectroscopy, will realistic doubts about the ionisation of the emitting gas be resolved?  Potentially, but not necessarily.  Consider a line of sight that views the optical broad line region at an inclination of $\theta = 45^{\circ}$, potentially appropriate for a Seyfert~1.5 AGN.  Our line of sight may reveal more of the face of the BLR that is on the far side of the central engine.  If the BLR is an outflow, an Fe K line produced within it may be red-shifted relative to the host frame of reference.  A line from Ne-like \ion{Fe}{xvii} at 6.43~keV would only have to arise in an outflow with a velocity of $v \simeq 1400~{\rm km}~{\rm s}^{-1}$ to appear to be consistent with neutral \ion{Fe}{i}.  The K-shell edge for \ion{Fe}{xvii} lies at 7.8~keV, whereas the edge for \ion{Fe}{i} lies at 7.1~keV.  Emission lines are often much easier to detect than associated edges, but the detection of K-shell edges would nominally distinguish neutral gas with no velocity shift from ionized gas that is shifted to be coincident with other charge states.

It is also important to appreciate that the fluorescence yield changes with ionisation, not just atomic number.  Between \ion{Fe}{i} and \ion{Fe}{xxii}, the fluorescence yield slowly rises from $Y_{\ion{Fe}{i}} = 0.34$ to $Y_{\ion{Fe}{xxii}} = 0.49$ \cite{Bambynek72}, but  falls to $Y_{\ion{Fe}{xxiii}} = 0.11$ for \ion{Fe}{xxiii}, recovering to $Y_{\ion{Fe}{xxiv}} = 0.75$ for \ion{Fe}{xxiv}, and has values of $Y_{\ion{Fe}{xxv}} = 0.5$ and $Y_{\ion{Fe}{xxvi}} = 0.7$ for He-like \ion{Fe}{xxv} and H-like \ion{Fe}{xxvi} (see \cite{KK87}, and references therein).  This has important consequences for breaking degeneracies between charge state and velocity shifts in data with modest sensitivity.  In a given scenario, it may be more likely that an observed line represents blue- or red-shifted emission from a charge state with a high yield, rather than a line from a charge state with a low yield from gas that is largely at rest.  

When Fe~K lines are produced in optically thick gas (e.g., the accretion disk, the molecular ``torus'', or even cold clumps within the optical broad line region or disk winds), they are part of a larger reaction spectrum that is called ``X-ray reflection'' (e.g. \cite{GF91}, and many others).  As noted above, this process includes the production of lines from other abundant elements, but there are two other prominent attributes.  One is an absorption trough owing to the Fe K-shell photoelectric absorption edges (7.1-9.3 keV, depending on the ion).  The other is known as the ``Compton hump,'' generally peaking between 20-30~keV.  This is not a true flux excess; rather, it is the result of the albedo of the cold gas peaking in this range.  Higher energy X-rays penetrate deeply into the accretion disk and thermalize.  In combination with the photoelectric absorption trough at low energy, the effects combine to yield a ``hump'' that appears above the power law observed from the central engine.

The preceding discussion has focused on the narrow core of Fe~K lines.  The narrow core represents the line flux that escapes from the irradiated cold, dense gas without being scattered within the emitting region.  Some fraction of the line photons will scatter, though, giving rise to a series of Compton shoulders (e.g., \cite{GF91, Matt02}).  These shoulders take the form of a plateau that extends redward of the core, abruptly ending in a cliff at 6.24 keV, set by the maximum energy that a photon loses in a $180^{\circ}$ scattering event.  Up to the optically thick limit, it is more likely that a photon will scatter once than multiple times, and the first Compton shoulder is the only one that is anticipated in actual data.  The relative strength of the Compton shoulder and narrow core is a function of the column density of the emitting region, and the inclination angle.  Relative to the narrow core, the first shoulder is more pronounced with increasing column density, peaking in material that is Compton-thick, and at low inclinations, because the scattered photons originate deeper in the slab than the unscattered ones \cite{Matt02}.

The lower panel in Figure~\ref{fig:jmm1} shows another model spectrum generated using {\sc mytorus} \cite{mytorus}. In this case, the obscuring column density was set to be $N_{H} = 2\times 10^{24}~{\rm cm}^{-2}$, making the emitting gas Compton-thick.  The first Compton shoulder is clearly evident from both the Fe~K$\alpha$ lines and Fe~K$\beta$ lines.  In this idealized example, the second K$\alpha$ Compton shoulder is also visible, extending down to 6.1~keV.  It is possible that the second shoulder may be detected in the best calorimeter spectra, but it may be more readily detected in an X-ray binary like GX 301-2 \cite{Watanabe03} than in an AGN.


\subsection{The nature and origin of ``narrow’’ Fe K emission lines in AGN}
\label{subsec:narrowfek}

``Does the narrow Fe~K line originate in the torus, or in the optical broad line region?”  

This simple question has been at the heart of many investigations using CCD and grating spectrometers over the last two decades.  It is built on solid expectations: above a certain threshold in the Eddington fraction, optical “broad line regions” and cold, obscuring torii appear to be ubiquitous in AGN.  Since it is the torus that determines whether or not the broad line region is visible in a given source, and since obscured and unobscured AGN  are roughly equal in number, it is logical to conclude that torii likely occupy approximately half of the sky as seen from the central engine.  A reasonable hypothesis, then, is that the torus overwhelmingly dominates the flux observed in narrow Fe K lines, and the dichotomy underlying this question is justified.

A few considerations argue otherwise and suggest that this is an ill-posed question.  First, {\it every} geometry that is at least partially composed of cold, dense gas will contribute Fe~K line flux when it is irradiated by hard X-rays.  Second, the line flux that is contributed by a given geometry depends on the hard X-ray flux received at that radius, not just its solid angle. Finally, but perhaps most importantly, the broad line region and torus may not be as physically distinct as some results would suggest.  Although some torii have been imaged in IR bands using interferometric techniques, and clearly span parsec scales, this does not convey the innermost extent of the torus.  Dust reverberation mapping in a growing number of quasars finds that the torus is only $\sim$5 times larger than the optical broad line region \cite{Minezaki19}.  If the presence of dust marks the innermost extent of a cold, dusty, molecular torus, then the torus is simply not much larger than the broad line region, and it may not be productive to treat them as entirely separate (see Figure~\ref{fig:agn}).

The \ch-High Energy Gratings (HEG) have a nominal resolution of 45~eV at 6.4~keV, in the first-order spectra (see Chap. 3).  While this is several times sharper than the resolution afforded by CCD spectrometers, such as the EPIC-pn aboard \xmm, the effective area of the HEG in the Fe~K band is $A_{eff} \simeq 30~{\rm cm}^{2}$, whereas that of the EPIC-pn is $A_{eff} \simeq 900~{\rm cm}^{2}$.  Observations with the \ch-HEG are therefore better suited to measurements of Fe~K line widths, and corresponding production radii and widths.  

An early, 83~ks \ch\ observation of NGC 5548 measured a line centroid energy of $E = 6.402^{+0.027}_{-0.025}$~keV, and a width of $FWHM = 4515^{+3525}_{-2645}~{\rm km}~{\rm s}^{-1}$ \cite{Yaqoob01}.  Even in a moderately deep grating spectrum, \ion{Fe}{xvii} (E = 6.43~keV) was not excluded.  The error bars on the line width are large in the fractional sense, but point to an origin in the optical BLR rather than in gas that is confined to parsec scales.  The uncertainties in these measurements partially reflect the limited effective area of the HEG.  

\ch\ made a much longer, 900~ks observation of NGC 3783.  The Fe K line centroid was measured to be $E = 6.3982~{\rm keV}\pm 3.3$~eV, and the line width was measured to be $FWHM = 1720\pm 360~{\rm km}~{\rm s}^{-1}$ \cite{Kaspi01}.  The line width is again consistent with the outer broad line region and/or the innermost extent of a small torus, rather than gas at the scale of a parsec.  This effort also detected the first Compton shoulder in the Fe~K line profile, indicating an origin in optically-thick material. 

Additional \ch-HEG spectra of Seyfert-1 made it possible to compare the width of Fe~K$\alpha$ emission lines to optical H$\beta$ lines from the BLR.  An early systematic comparison examined literature values in 14 sources, and found that (1) the average Fe~K$\alpha$ line width is a factor of $\sim2$ lower than the corresponding H$\beta$ line, and (2) that there is no correlation between the line widths \cite{Nandra06}.  The key conclusion of this analysis was that the narrow Fe~K$\alpha$ line originates in the torus.  Many intervening years and results make it possible to see this conclusion in context.  At the time, the torus was typically envisioned as a parsec-scale geometry; the FWHM differences do not necessitate that; rather, the contrast is broadly consistent with much smaller contrast indicated by dust reverberation mapping \cite{Minezaki19}. 

A more detailed examination of the growing number of sensitive \ch\ spectra of Seyfert~1 AGN, and a comparison to H$\beta$ lines in each source, was reported in 2010 \cite{Shu10}.  A total of 82 \ch\ observations from 36 sources were considered, explicitly allowing for variations in the line properties between observations.  In a subsample of 27 source, the mean Fe~K$\alpha$ line width is measured to be $\langle FWHM\rangle = 2060\pm 230~{\rm km}~{\rm s^{-1}}$, and no correlation is found between Fe~K$\alpha$ and H$\beta$ line widths.  The more detailed nature of this survey permitted a more nuanced and very important finding: {\it ``There is no universal location of the Fe~K$\alpha$ line-emitting region relative to the optical broad line region (BLR). In general, a given source may have contributions to the Fe~K$\alpha$ line flux from parsec-scale distances from the putative black hole, down to matter a factor of $\sim$2 closer to the black hole than the BLR”} \cite{Shu10}.

Andonie et al. \cite{Andonie22} have undertaken the most recent and expansive examination of narrow Fe~K$\alpha$ emission line regions.  Their analysis included 38 bright AGN in the {\it Neil Gehrels Swift Observatory} Burst Alert Telescope (BAT) Spectroscopic Survey.  Utilizing \ch\ images and spectra, \xmm\ spectra, and variability studies, independent estimates of the Fe~K$\alpha$ production radius ($R_{Fe~K} $) were obtained and compared to plausible estimates of the dust sublimation radius ($R_{subl.}$)  in each source.  In the cases where data permitted measurements of the line FWHM, the $R_{Fe~K} < R_{subl.}$ in 90\% of the sources (21/24 AGN).  Similarly, in the cases where significant line variability was detected, $R_{Fe~K} < R_{subl.}$ in 83\% of the sources.  Andonie et al. \cite{Andonie22} conclude that Fe~K$\alpha$ lines in unobscured AGN typically originate in the outer part of the BLR, or the outer disk, but carefully note that {\it ``the large diversity of continuum and narrow Fe~K$\alpha$ variability properties are not easily accommodated by a universal scenario.”}

\begin{figure*}
   \centering
   \advance\leftskip-0.5cm
   {\scalebox{0.38}{\includegraphics[trim= 0.cm 2cm 1cm 1cm, angle=0, clip=true]{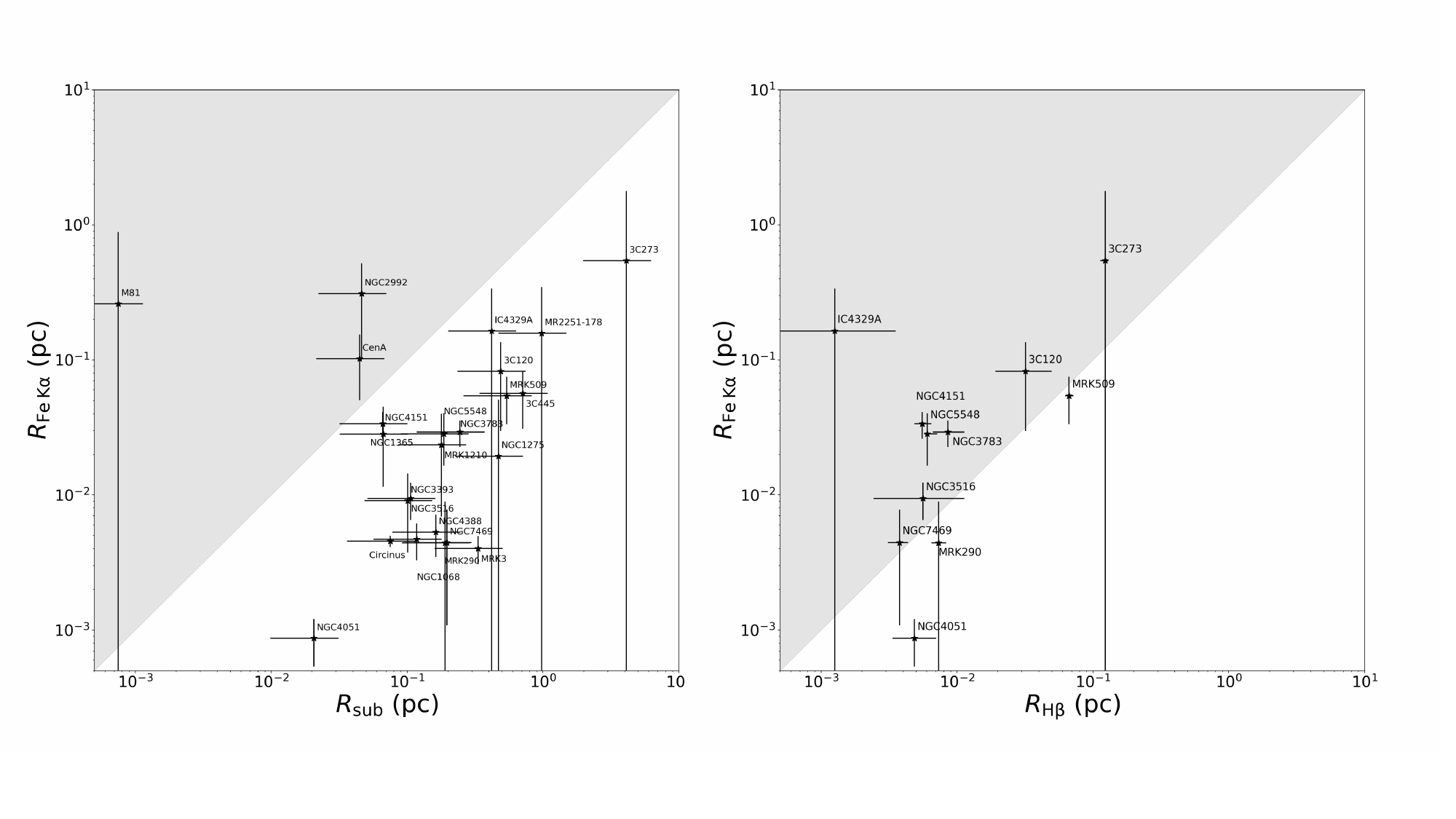}}}
\caption{These panels summarize some of the key results obtained in the survey of Seyfert Fe~K$\alpha$ lines undertaken by \cite{Andonie22}.  At left, the Fe~K$\alpha$ line production radius (measured via line widths) is compared to an estimate of the dust sublimation radius – a proxy for the innermost extent of the dusty torus – in each AGN.  The Fe~K$\alpha$ line production regions are found to occur at smaller radii, strongly suggesting that the bulk of the emission seen in unobscured AGN is produced within the BLR or in even more compact regions.  At right, the Fe~K$\alpha$ line production radius in each AGN is compared to the H$\beta$ production radius, the characteristic radius of the optical BLR.  The data do not appear to permit a systematic statement, but it is fair to conclude that the production radii are broadly comparable.  (This figure is reproduced from \cite{Andonie22}, with permission.) }
   \label{fig:jmm2}
\end{figure*}

%
The left hand panel in Figure~\ref{fig:jmm2} shows the Fe~K$\alpha$ emission radius versus an estimate of the dust sublimation radius.  The Fe~K$\alpha$ emission radius was calculated based on the velocity width of the line, and the dust sublimation radius was estimated using an expression derived by \cite{Nenkova08}, and assuming that graphite sublimates at $T = 1500$~K.  The figure illustrates that the Fe~K$\alpha$ line production radius is systematically smaller than the dust sublimation radius within the sample.  If we take the dust sublimation radius as indicative of the innermost extent of the torus, this finding is at least qualitatively consistent with the dust reverberation results of \cite{Minezaki19}.  The right hand panel in Figure~\ref{fig:jmm2} depicts the Fe~K$\alpha$ emission line radius versus the H$\beta$ production radius.  There is no clear trend within the data; only a few AGN clearly lie above or below the line that marks an equivalent production radius, and many Fe~K$\alpha$ line production radii carry relatively large uncertainties.

\begin{figure*}
   \centering
   \advance\leftskip-0cm
   {\scalebox{0.4}{\includegraphics[trim= 2cm 1cm 1cm 2cm, angle=-90, clip=true]{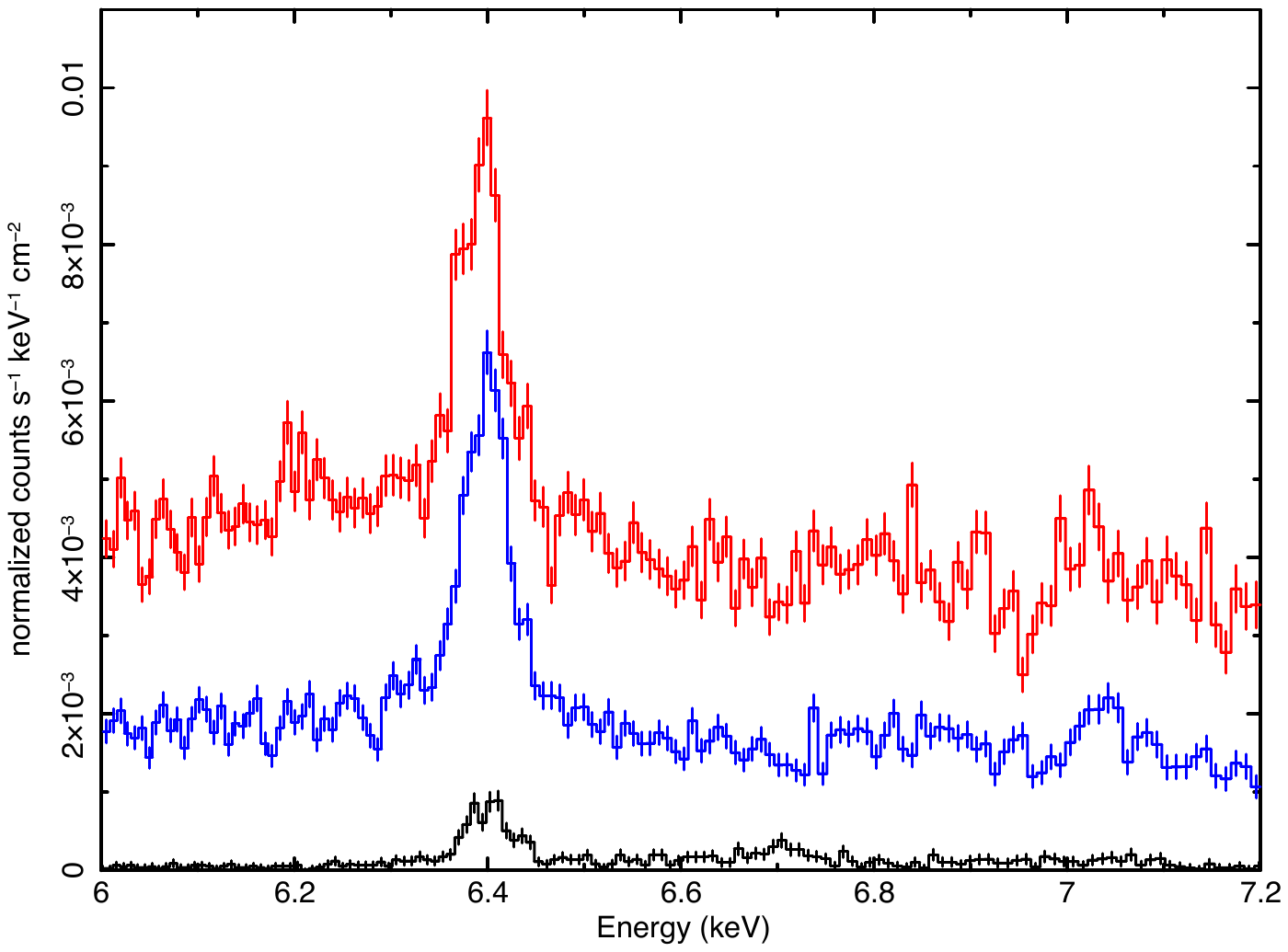}}}
   \caption{A comparison of spectra from the Compton-thick Seyfert~2 AGN NGC 1068 (black), and NGC 4151 in its high (red) and low (blue) flux states.  The spectra are the summed \ch-HEG spectra from each source, shifted to their host frame.  The flux of the NGC 4151 spectra have been shifted to the smaller distance to NGC 1068.  The inferred black hole masses and Eddington fractions in these AGN are broadly similar (see the text for details), enabling a consistent comparison.  If the covering factor of cold, dense gas were the only factor that determined line strength, the narrow Fe~K$\alpha$ line in NGC 1068 should be much stronger.  However, factors including the distance of the gas to the responding medium also matter, and this is likely the reason that the line is stronger in NGC 4151.  This figure illustrates that different viewing angles and obscuration properties can reveal different regions, not necessarily different aspects of the same emission region.
}
   \label{fig:jmm3}
\end{figure*}

Figure~\ref{fig:jmm3} shows a comparison of summed \ch-HEG spectra from the Seyfert~1 NGC 4151 in its high and low flux states, and the summed HEG spectrum of the Compton-thick Seyfert~2 NGC 1068.  The opposing first-order spectra were added, the spectra have been shifted in energy to their respective rest frames, and the flux of NGC 4151 has been adjusted to the distance of NGC 1068.  No spectral fits were made.  Recent estimates suggest that the mass of the black hole in NGC 1068 is $M = 1.5\times 10^{7}~M_{\odot}$ \cite{Morishima22}; optical reverberation mapping gives a formally equivalent black hole mass in NGC 4151: $M = 1.66^{+0.48}_{-0.34}\times 10^{7}~M_{\odot}$ \cite{Bentz22}.  For these mass estimates, NGC 1068 is likely accreting at an Eddington fraction at or below $\lambda \leq 0.2$ \cite{Bland97}, and NGC 4151 at a rate of $\lambda = 0.03-0.04$ \cite{Miller18}.  Given their similar masses, the fact that NGC 1068 is likely surrounded by gas with a higher filling factor and likely accreting at a higher rate than NGC 4151, it should show a stronger neutral Fe~K$\alpha$ emission line in this comparison.  However, the opposite is the case.  This can be explained if the gas that is emitting the Fe~K$\alpha$ line {\it within our line of sight} in NGC 4151 is closer to the central engine than the emitting gas {\it within our line of sight} is in NGC 1068.  Without any spectral modeling, this example illustrates that different views of the central engine actually reveal different regions, and that faulty conclusions may be drawn if a single emitting geometry is invoked.

\subsection{The approaching calorimeter era}
\label{subsec:}

A number of recent results highlight the potential of X-ray calorimeter spectroscopy to reveal the inner accretion flow onto massive black holes.  Here, we highlight four key results that likely provide an early glimpse of the science that will be enabled by the sharper resolutions and larger effective areas afforded by XRISM and Athena.

In a major departure from phenomenological Gaussian modeling, and simple comparisons of Fe~K$\alpha$ and H$\beta$ line widths, \cite{Costantini10} examined the spectrum of the Seyfert~1 Mrk 279 using the Local Optimally emitting Cloud model (LOC; \cite{Baldwin95}).  This model does not assume a geometry for the BLR; rather, it simply assumes that each line is the result of contributions from multiple regions that follow a power law distribution in gas density and radius.  Mrk 279 is particularly interesting among Seyferts, in that its ``narrow’’ Fe~K$\alpha$ line appears to be composed of an unresolved core and a $FWHM = 14,000~{\rm km}~{\rm s}^{-1}$ component.  

By first establishing the parameters
of a LOC model using UV and soft X-ray lines and then extending the model
to the Fe K band, \cite{costantini07, Costantini10} show
 that the BLR model can account for only 3-17\% of the Fe~K$\alpha$ broad line flux.
A subsequent study using the same technique, based on simultaneous optical, UV, and X-ray observations of Mrk~509, \cite{costantini16} established that the contribution of the BLR to the Fe~K$\alpha$ line could be up to 30\%. 
The clear implication is that part of the Fe~K$\alpha$ line
in these two objects originates at smaller radii than the optical, UV, and even soft X-ray BLR,
while the unresolved line core may be produced in the torus at larger radii.
If these cases are not unique, but rather just an AGN that offers a fortuitous view of its inner
disk, BLR, and torus, moderately deep observations of Seyfert-1 AGN with XRISM
may readily decompose seemingly monolithic lines into such components.

\begin{figure*}
   \centering
   \advance\leftskip-0cm
   {\scalebox{0.5}{\includegraphics[trim= 1cm 2cm 1cm 0cm, angle=0, clip=true]{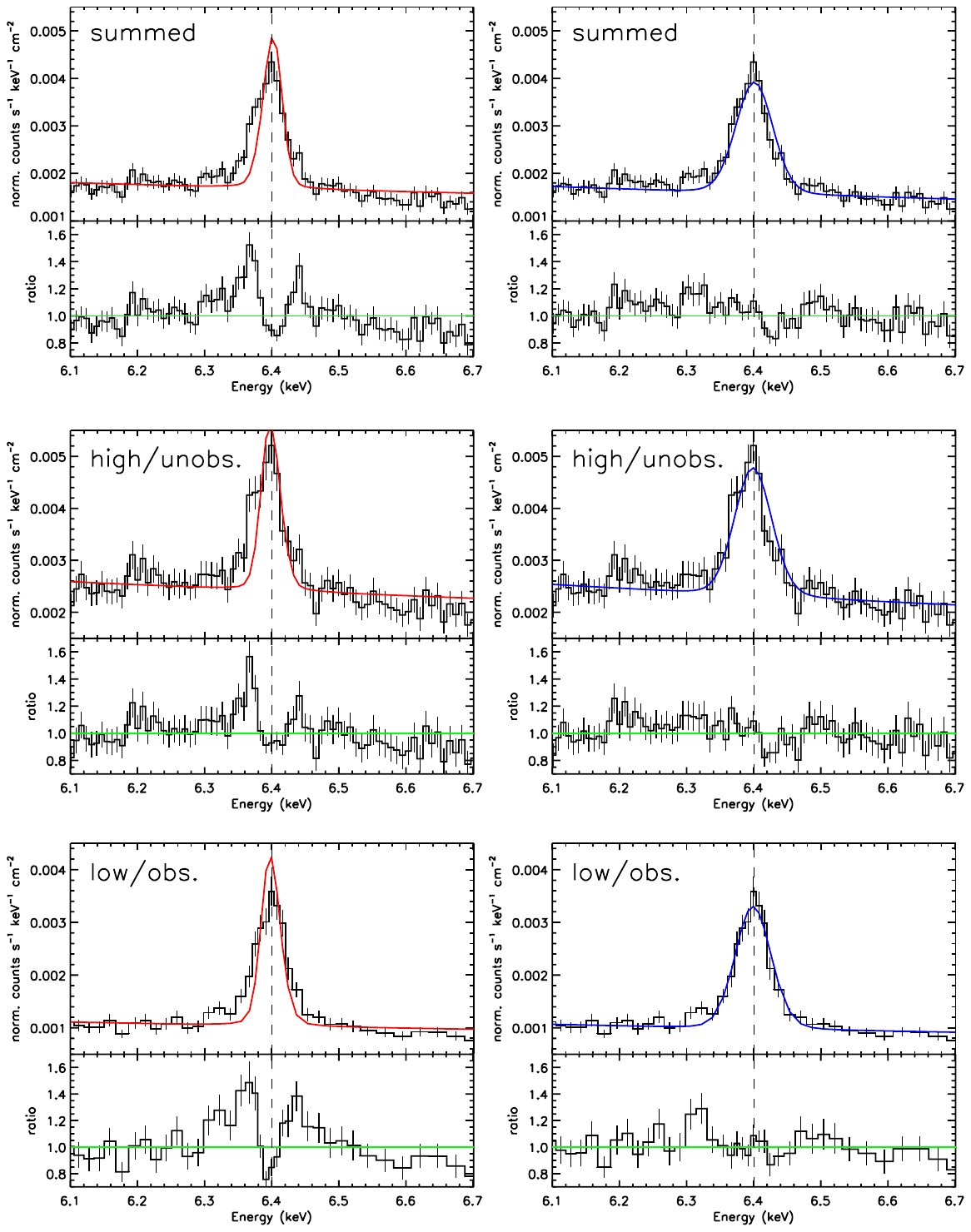}}}
   \caption{Fits to the ``narrow’’ Fe~K$\alpha$ line in summed \ch-HEG spectra of NGC 4151 \cite{Miller18}.  The spectra were shifted to the rest frame and binned for sensitivity and visual clarity.  The left hand panels depict the results of fits with Gaussian profiles that only have instrumental broadening; while the right hand panels depict fits with Gaussian profiles with a variable width.  The line is clearly asymmetrical, with excess flux to the red; a variety of models that include weak relativistic Doppler shifts and gravitational red-shifts suggest an inner production region of $R=500-1000~GM/c^{2}$, considerably smaller than the optical BLR.  (This figure is reproduced from \cite{Miller18}, with permission).
}
   \label{fig:jmm4}
\end{figure*}

NGC 4151 is the brightest Seyfert AGN in the 4-10~keV band, with the strongest narrow Fe~K$\alpha$ line observed in any Seyfert~1 (e.g.,  \cite{Shu10}).  As such, it can be expected to deliver the most sensitive spectra and important hints of the potential of calorimeter spectroscopy with XRISM and Athena.  These expectations led  \cite{Miller18} to examine \ch-HEG spectra of NGC 4151.  The total summed spectrum (631~ks of exposure), and spectra summed from low (337~ks) and high (294 ks) flux states were examined.  In each case, the line profile is found to be asymmetric and red-skewed, suggestive of weak relativistic Doppler shifts (see Figure~\ref{fig:jmm4}).  Fits with a number of independent models imply line production radii in the $R = 500-1000~GM/c^{2}$ range.  This again implies that the bulk of the ``narrow’’ Fe K line flux originates at smaller radii than the optical BLR.

\begin{figure*}
   \centering
   \advance\leftskip-0cm
   {\scalebox{0.6}{\includegraphics[trim= 1.cm 1cm 1.5cm 12cm, angle=0, clip=true]{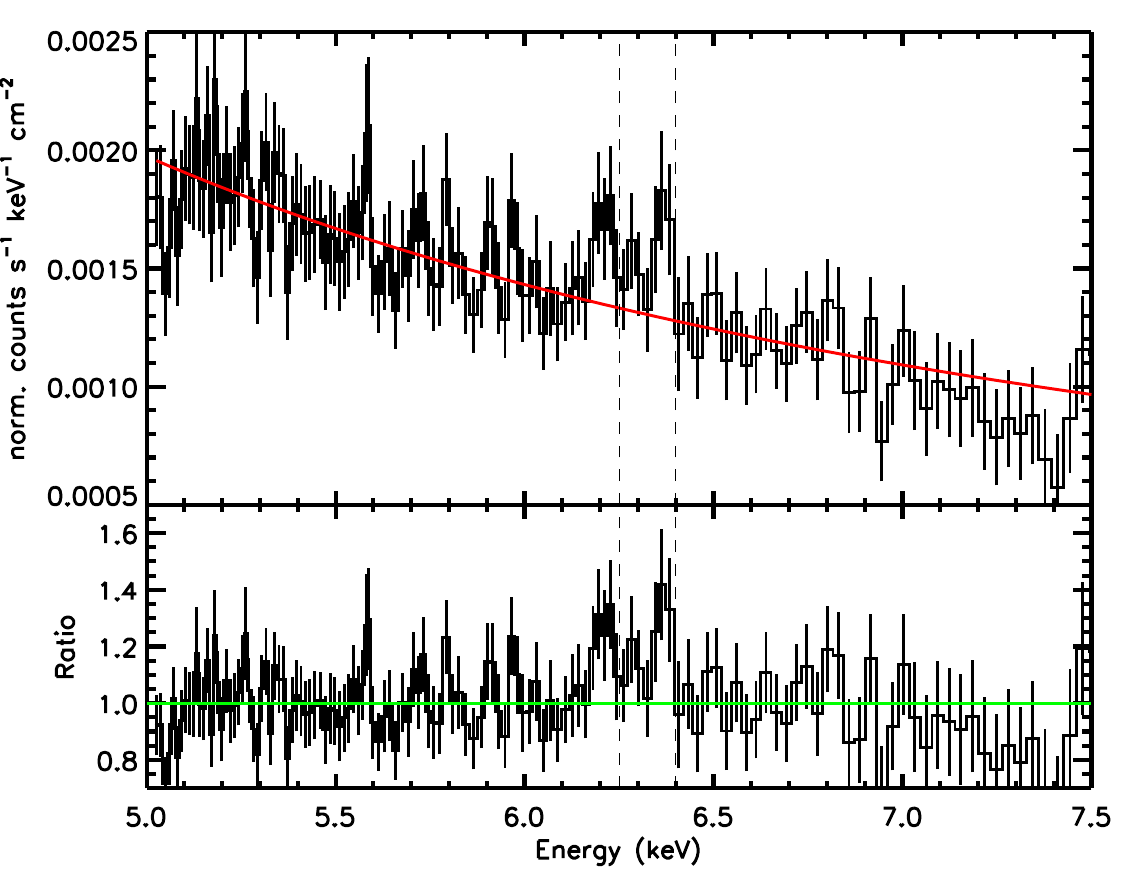}}}
   \caption{The difference spectrum obtained when the summed \ch-HEG exposure in the low flux state is subtracted from the summed exposure in the high flux state \cite{Miller18}.   The spectrum was shifted to its rest frame, binned, and fit with a simple power-law.  The lingering Fe~K$\alpha$ emission is red-shifted from the expected value of $E = 6.40$~keV, and equally strong emission is red-shifted from the Compton shoulder at $E = 6.25$~keV, though the narrow core and shoulder are not expected to be comparably strong.  The residual structure is suggestive of a warp or a ring in the inner disk extending between $R = 50-130~GM/c^{2}$, potentially enhanced with a higher mass accretion rate and/or stronger irradiation.  
(This figure is reproduced from \cite{Miller18}, with permission).
}
   \label{fig:jmm5}
\end{figure*}

NGC 4151 may also provide early hints that the region between the innermost disk and optical BLR is structured.  Miller et al. \cite{Miller18} find that the high-low flux difference spectrum reveals a line profile with two peaks, red-shifted from the expected narrow line core and Compton shoulder; fits to this profile require a narrow ring of emission between $R=70-120~GM/c^{2}$ (see Figure~\ref{fig:jmm5}).  Independently, within the spectra typified by a high continuum flux, the Fe~K$\alpha$ line flux appears to vary on time scales of $\Delta t \simeq 2\times 10^{4}~{\rm s}$, implying $R\simeq 50-130~GM/c^{2}$.  These findings are at least qualitatively consistent with a warp or ring-like structure, similar to the features seen in numerical simulation of accretion flows when the angular momenta of the black hole and accretion flow are misaligned (e.g., \cite{Nixon12}).

\begin{figure*}
   \centering
   \advance\leftskip-0cm
   {\scalebox{0.7}{\includegraphics[trim= 0.cm 0cm 0cm 0cm, angle=0, clip=true]{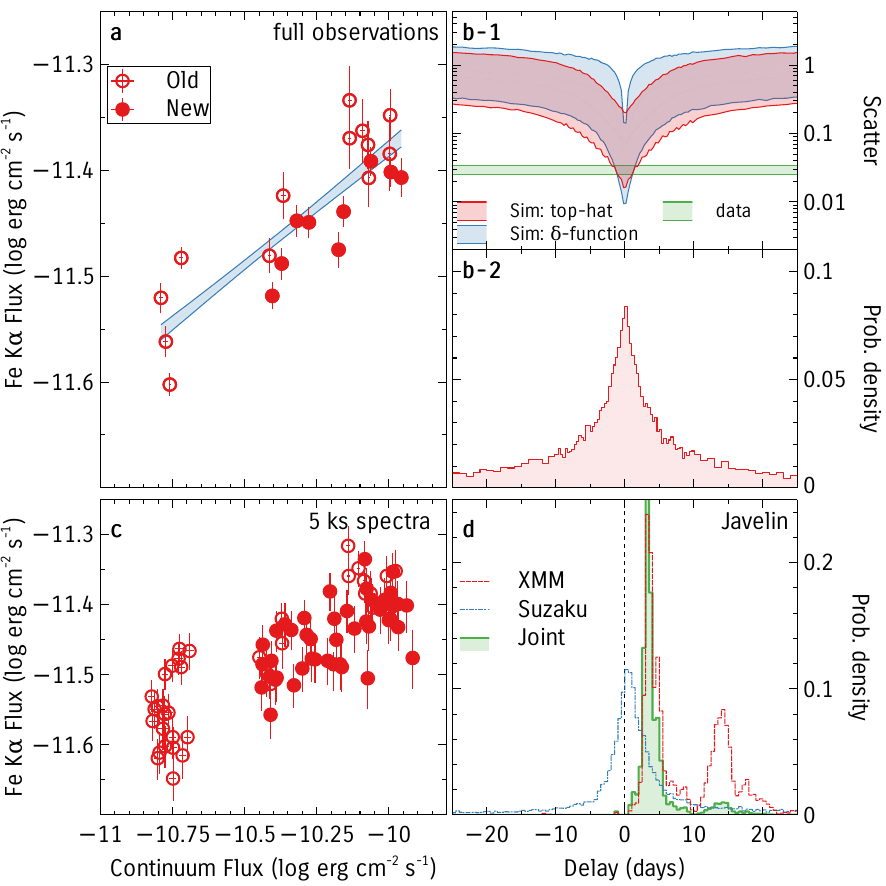}}}
   \caption{Evidence of a reverberation lag in the narrow Fe~K$\alpha$ line in NGC 4151 \cite{Zoghbi19}, based on numerous observations with \xmm\ and {\it Suzaku}.  Panel (a) shows a strong correlation between the Fe~K$\alpha$ flux and continuum flux measured across full exposures.  Panel (b1) shows the scatter in the line–continuum relation in different simulations that sample different delay times.  Panel (b2) shows the probability density of the different delay times.  Panel (c) shows the line–continuum relation, taking 5~ks segments instead of full observations.  Finally, panel (d) shows the probability density for different values of the lag using the 5~ks segments.  A lag of $\tau = 3.3^{+1.8}_{-0.7}$~days is measured, yielding a black hole mass of $M = 1.8^{+2.2}_{-1.1}\times 10^{7}~M_{\odot}$ assuming the line widths measured by \cite{Miller18}.  This is fully consistent with the most recent optical reverberation mass, $M_{BH} = 1.66^{+0.48}_{-0.34}\times 10^{7}~{\rm M}_{\odot}$ \cite{Bentz22}.
}
   \label{fig:jmm6}
\end{figure*}

Zoghbi et al. \cite{Zoghbi19} reported the first detection of reverberation in the narrow Fe~K$\alpha$ line in any AGN.  Again owing to its flux, this detection was made in NGC 4151, using data from \xmm\ and {\it Suzaku}.  The flux sensitivity of these data outweighed their modest resolution relative to the \ch-HEG.  Using the {\sc javelin} code that is widely implemented to measure lags from the optical BLR \cite{Zu13}, Zoghbi et al. \cite{Zoghbi19} measure a delay of $\tau = 3.3^{+1.8}_{-0.7}$~days (see Figure~\ref{fig:jmm6}).   For plausible black hole masses, this light travel time is broadly consistent with the radii measured via fits to the Fe~K$\alpha$ line shape observed with \ch\ \cite{Miller18}.

Using the standard equation to derive the black hole mass, $M_{BH} = f c \tau v^{2} / G$ (where $f$ is a geometric factor derived via comparisons to direct primary masses, and $v$ is the width of the line), \cite{Zoghbi19}  infer a mass of $M_{BH} = 1.8^{+2.2}_{-1.1}\times 10^{7}~M_{\odot}$, assuming a standard value for the geometric factor ($f = 4.13\pm 1.05$; \cite{Grier13}) and the line width measured using \ch.  This value agrees extremely well with the most recent optical reverberation mass, $M_{BH} = 1.66^{+0.48}_{-0.34}\times 10^{7}~{\rm M}_{\odot}$ \cite{Bentz22}.  Future studies may find that the geometric factor is not the same for the X-ray and optical gas, and/or that the responding geometries vary with time.  However, this initial detection and plausible mass estimate suggest a bright future for narrow Fe K$\alpha$ reverberation studies.

Another detection of reverberation in a narrow Fe~K$\alpha$ line was recently reported in archival {\it Suzaku} observations of the bright Seyfert~1 AGN NGC 3516 \cite{Noda22}.  Again using {\sc javelin} \cite{Zu13}, a lag of $\tau = 10.1^{+5.8}_{-5.6}$~days is measured.  Importantly, the lag is only found during a period when NGC 3516 was particularly faint in X-rays, and similar to a Seyfert~2 AGN.  In those cases, the BLR is typically blocked by the torus, so the fact that reverberation mapping was still possible using the Fe~K$\alpha$ line may signal that XRISM and Athena can measure lags and black hole masses in cases where optical attempts have not been successful.   

XRISM spectra will likely achieve a resolution of $5$~eV and an effective area close to $A_{eff} \simeq 300~{\rm cm}^{2}$ in the Fe K band.  Both represent order-of-magnitude improvements over the capabilities of the \ch-HEG.  Although this effective area does not exceed that of the EPIC-pn CCD camera aboard \xmm, its superior resolving power will make XRISM far more sensitive to lines.  It is therefore worth asking: in how many AGN will XRISM achieve similar and better results, assuming optimal conditions (e.g., a 10-year mission with consistent instrumental performance)?

XRISM will be revolutionary, but it is still a small telescope (in some sense, a pathfinder for Athena).  The extraordinary time required to achieve sensitive spectra of faint AGN would necessarily come at the expense of fully understanding the demographics of a brighter sample. For such reasons, it is pragmatic to set a flux limit.   In general, only Seyferts with a flux above $F \simeq 1\times 10^{-11}~{\rm erg}~{\rm cm}^{-2}~{\rm s}^{-1}$ in the \ch\ pass band have enabled significant line detections in exposures of $1-9 \times 10^{5}~{\rm s}$.   A broad view of surveys undertaken with {\it ROSAT}, \xmm, and {\it eROSITA} suggests that there are roughly 300~AGN with an X-ray flux of $F \geq 1\times 10^{-11}~{\rm erg}~{\rm cm}^{-2}~{\rm s}^{-1}$ in the 0.5-10~keV band \cite{Voges99, Saxton08, Brunner22}.  Observing each of these with XRISM for 100~ks would require 30~Ms of total exposure time, which is feasible over a 10-year mission.  Alternatively, a total program of 30~Ms would also make it possible to observe 100 bright AGN for 100~ks on three separate occasions, sampling a range of variations in intrinsic luminosity and transient obscuration.  Particularly if the putative sample of 100 AGN is selected from the 300 that exceed the flux threshold in a manner that samples key parameters (Eddington fraction, black hole mass, inclination, spectral type, etc.), the mission could create a legacy that benefits the entire field.

{\it Hitomi} spectroscopy of NGC~1275, the Fanaroff-Riley I (FRI) radio galaxy at the heart of the Perseus cluster, offers a glimpse of what XRISM is likely to achieve in very deep observations of the AGN that reshape clusters.  Those spectra reveal a very narrow line, with $FWHM = 500-1600~{\rm km}~{\rm s}^{-1}$ (90\% confidence) \cite{n1275}.  This places the line production region beyond the optical BLR, and likely associates the line with the cold molecular torus.  In AGN that provide fierce jet feedback, it is interesting to estimate the total mass reservoir that is available to eventually power the jet; the equivalent width of the line in NGC 1275 implies a gas mass of $M = 4\times 10^{7}~M_{\odot}$, sufficient to power the AGN for a Hubble time \cite{n1275}.  XRISM will make deep stares at a number of clusters, spanning a range of properties, making it possible to compare line production regions and the mass in gas reservoirs.

At the time this review is being written, the final configuration of Athena (or, NewAthena) is uncertain.  Whatever the details, the eventual combination of improved spectral resolution and larger collecting area is likely to make it at least 10 times as sensitive as XRISM (see, e.g., \cite{athena}).  Entirely new possibilities then open.  Excellent spectra could easily be obtained in 10-20 ks monitoring exposures in a large number of bright AGN, greatly expanding the number of AGN with measured time delays and reverberation masses.  It would be just as compelling to explore AGN evolution by obtaining excellent spectra from fainter sources at greater distances.   There are approximately 3000 AGN above a flux threshold of $F \geq 1\times 10^{-12}~{\rm erg}~{\rm cm}^{-2}~{\rm s}^{-1}$ in the Athena pass band \cite{Voges99, Saxton08, Brunner22}.  Selecting different subsets of this number could probe particular aspects of black hole growth and feedback.  All of these endeavors would comfortably fit within a 30~Ms envelope, which is again reasonable if the mission lifetime extends to 10 years.

\subsection{Relativistic Fe K emission lines}
\label{subsec:broadfek}

\begin{figure*}
   \centering
   \advance\leftskip-0cm
   {\scalebox{0.4}{\includegraphics[trim= 0cm 1cm 2cm 0cm, angle=90, clip=true]{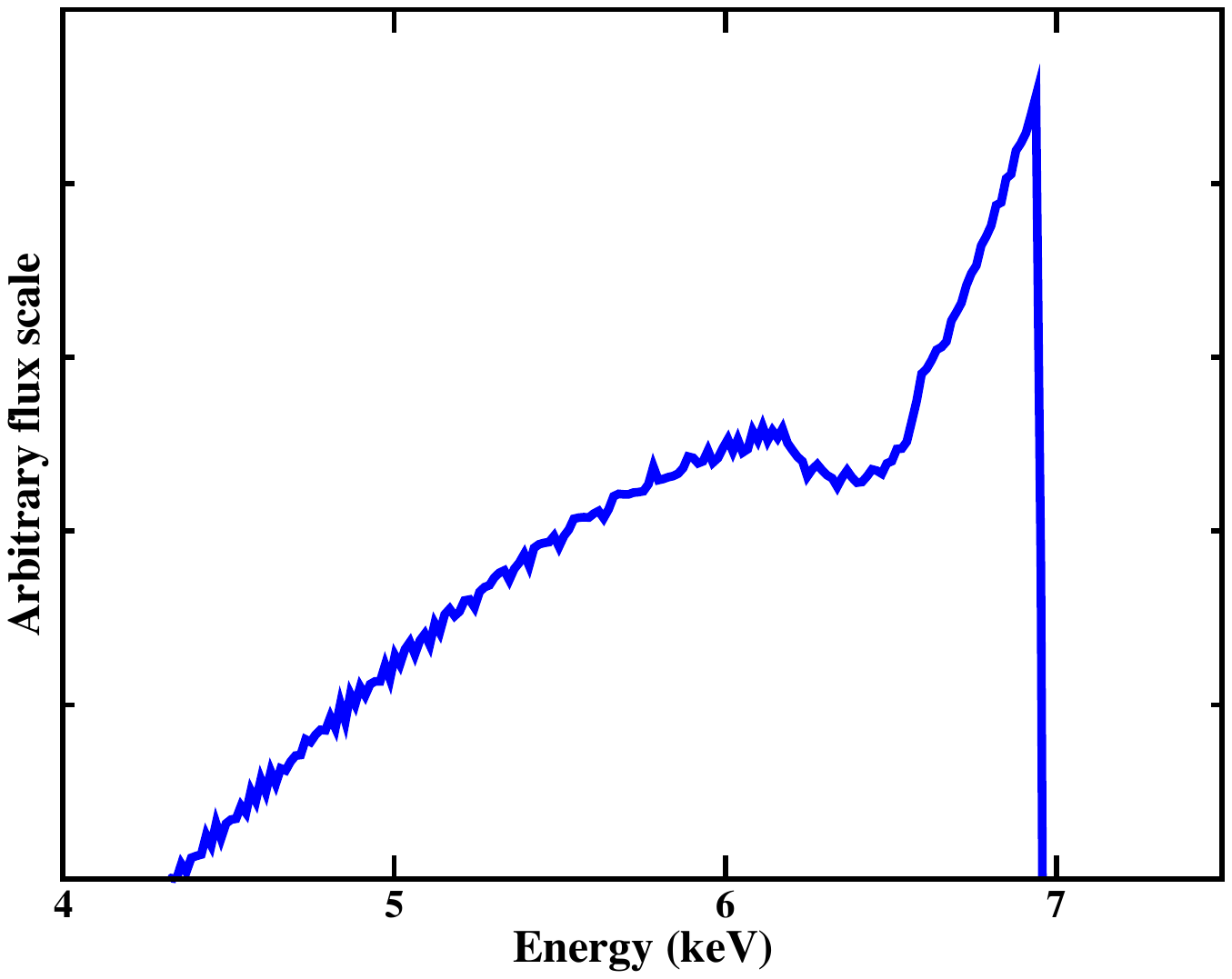}}}
   \caption{The predicted line profile \cite{Dauser10} from an intrinsically narrow Fe~K$\alpha$ emission line at $6.4$~keV that has been altered by Doppler broadening, special relativistic beaming, and general relativistic effects.}
   \label{fig:reline}
\end{figure*}

As described in Section~\ref{subsec:iron}, the reflection spectrum (e.g. \cite{GF91, RossFabian93}) can arise from the inner 10’s of gravitational radii, where the corona illuminates the inner accretion disc.  The spectrum resembles that of reflection in more distant, optically-thick material (e.g. the torus) as it produces strong Fe K emission, absorption edges, and a Compton hump.  In striking contrast to distant reflection, the material in the inner disc can be significantly ionised given its proximity to the corona (e.g. \cite{Ballantyne01, RossFabian05}), and substantially blurred from extreme orbital velocities and relativistic effects (e.g. \cite{Fabian89, Laor}).  In Figure~\ref{fig:reline}, the line profile of a relativistically broadened Fe~K$\alpha$ line is presented.  

In many ways, the study of ``blurred’ reflection (e.g. \cite{Miller07, Bambi21, Reynolds21}) is best suited for broadband spectroscopy and variability.  For example, even in the ``clean’’ Fe K region, the breadth of the Fe~K$\alpha$ can extend over several keV (Figure~\ref{fig:reline}).  The expanse of the reflection spectrum can overwhelm emission from other regions like distant reflection and even the primary continuum (e.g. \cite{Fabian09, Ponti10}).  This can be even more daunting when discerning relativistic features among the warm absorbers at low energy (e.g. \cite{steenbrugge09}). Great advances have been made utilizing \xmm\ and {\it NuSTAR} together to produce spectra between $0.3-79$~keV (e.g. \cite{Wilkins15, Jiang18, Walton20, Wilkins22}).  Relativistic, ionised reflection can also produce signatures in timing data generated from reverberation delays between the continuum and reflecting components (e.g. \cite{Zoghbi10, Zoghbi12}).   There is massive potential in discerning the geometry and environment in the inner few gravitational radii from the study of relativistic reflection (e.g. \cite{Alston20, Reynolds21,Wilkins22}).

\begin{figure*}
   \centering
   \advance\leftskip-0cm
   {\scalebox{0.4}{\includegraphics[trim= 0cm 1cm 2cm 0cm, angle=90, clip=true]{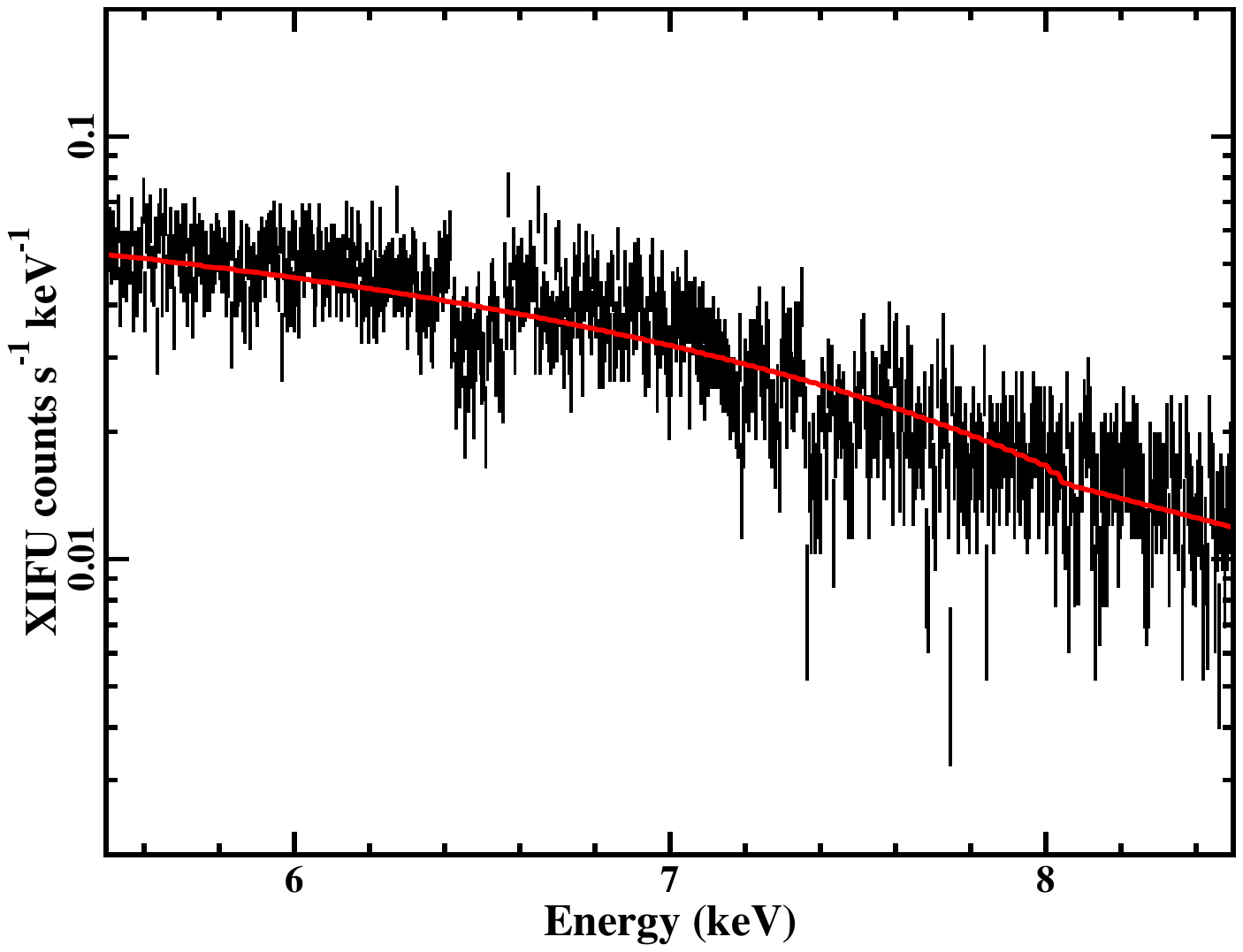}}}
   {\scalebox{0.4}{\includegraphics[trim= 1cm 1cm 2cm 2cm, angle=0,  clip=true]{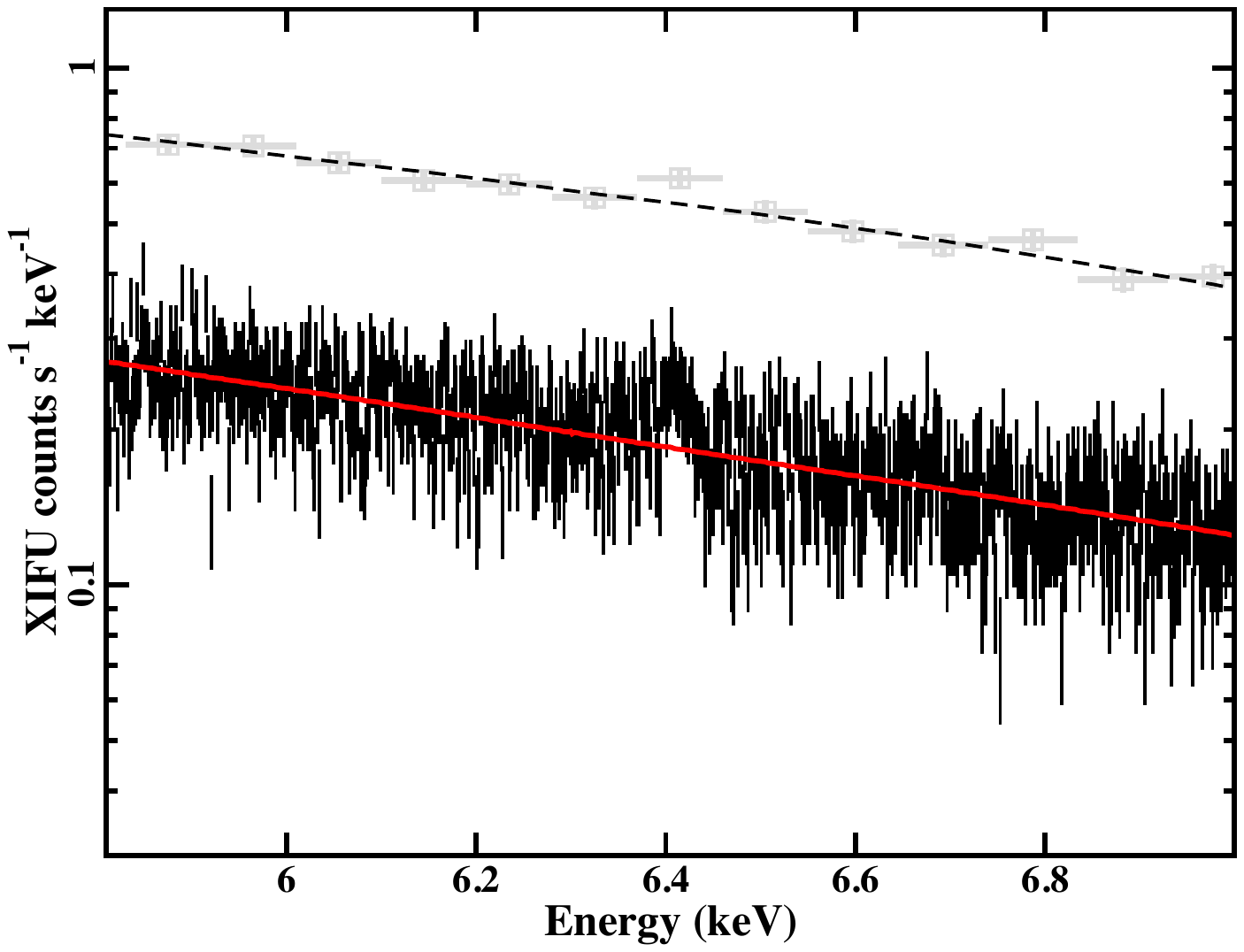}}}       
   \caption{Upper panel: A $100$~ks XRISM-Resolve simulation of a power law modified by ionised partial covering and fitted with blurred reflection.  Narrow absorption features from the absorbing material should be evident in XRISM data.  Lower panel: A $100$~ks Athena-XIFU simulation (black data) of a combined strong blurred reflector with a reflection fraction $\mathcal{R}=3$ and a weak distant reflector with $\mathcal{R}=0.3$, fitted with a single blurred reflector.  The weak distant emitter becomes evident in the Athena data.  A simulation for the EPIC-pn is shown in grey and is scaled for comparison.  The $2-10$~keV flux is $10^{-11} \rm~erg~cm^{-2}~s^{-1}$ in both panels and the data are optimally binned \cite{optbin}. }
   \label{fig:brsim}
\end{figure*}

A major challenge for realizing this potential is that alternate models like the two-coronae scenario (e.g. \cite{wc1,wc2,Ballantyne20}) and (ionised) partial covering (e.g. \cite{Holt80, Tanaka04}) can mimic the appearance of relativistic reflection.   High-resolution spectroscopy can provide some headway in breaking this degeneracy.  In Figure~\ref{fig:brsim}, an intrinsic power law spectrum modified by ionised partial covering is simulated for a $100$~ks observation with XRISM.  The data are then fitted with blurred reflection to show that XRISM can potential reveal narrow absorption features originating from the ionised partial covering material. 

Enhanced spectral resolution can also disentangle blurred reflection from other components.  In Figure~\ref{fig:brsim}, a weak distant reflector with a reflection fraction\footnote{The reflection fraction is the ratio of reflected flux to primary flux.} $\mathcal{R}=0.3$ is added to a strong blurred reflector with $\mathcal{R}=3$ and simulated for a $100$~ks Athena observation.  When fitted with blurred reflection alone, the weak narrow Fe~K$\alpha$ component from distant material is uncovered in the data.


\section{The nature of ultrafast outflows}
\label{sec:ufo}

For understanding ultrafast outflows (UFOs), the discovery space that will become accessible through high-resolution X-ray spectroscopy will be tremendous.  Various processes tied to the accretion mechanism can generate outflows from the disc (e.g. \cite{Murray95, Begelman83, BP82}).  If these outflows have sufficient energy to escape the inner kilo-parsec of the host galaxy, they can significantly impact star formation and abundances in the interstellar medium.  

\begin{figure*}
   \centering
   \advance\leftskip-0cm
   {\scalebox{0.4}{\includegraphics[trim= 1cm 1cm 2cm 2cm, angle=0, clip=true]{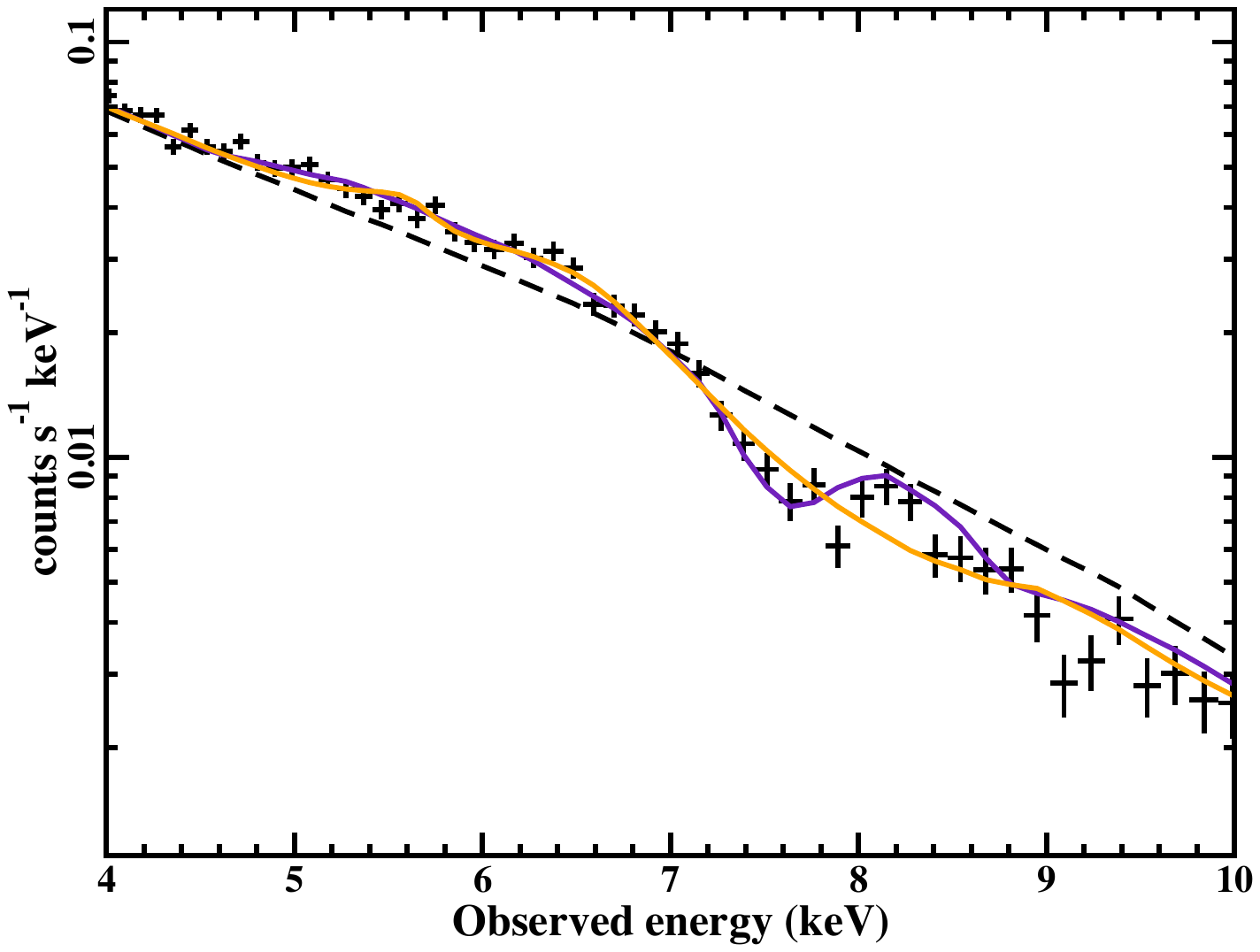}}}
   {\scalebox{0.4}{\includegraphics[trim= 1cm 1cm 2cm 2cm, angle=0,  clip=true]{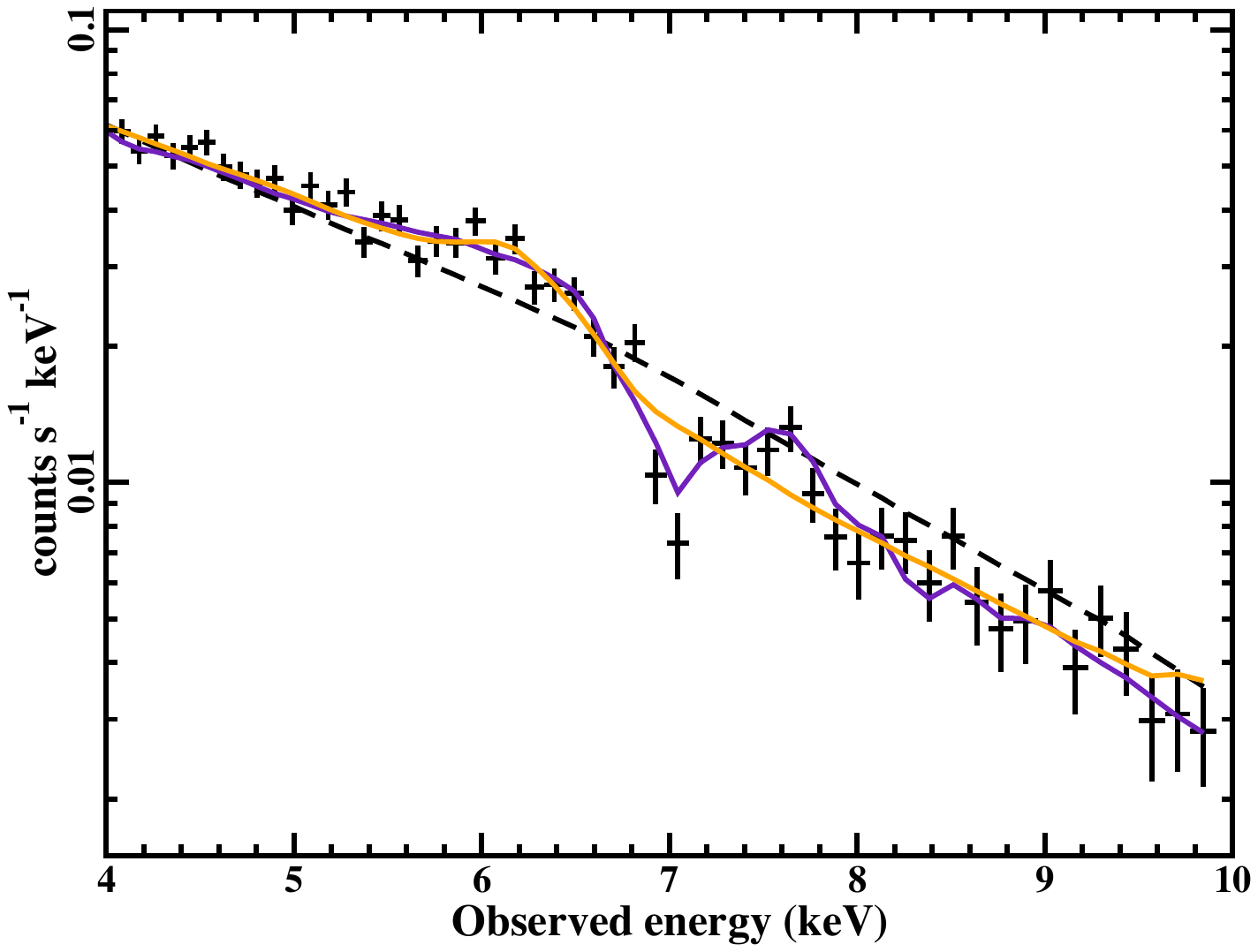}}}       
   \caption{EPIC-pn data from {\it XMM-Newton} observations of PDS~456 (upper panel; obsid:07210401) and PG~$1211+143$ (lower panel; obsid:01126101). In both cases, a single power law (black dashed curve) leaves absorption features above $7$~keV.  The MHD-driven wind ({\sc mhdwind}) \cite{Fukumura22} and line-driven wind ({\sc fast32rg}) \cite{Matzeu22}, represented by the orange and purple curves, respectively, reproduce the CCD data equally well.  Note, since {\sc mhdwind}  does not  currently account for scattered emission self-consistently, a Gaussian emission profile is included for more direct comparison to {\sc fast32rg}.}
   \label{fig:pdspg}
\end{figure*}

AGN feedback (e.g. \cite{Begelman04, King10, Fabian12}) will be important to galaxy evolution if the kinetic luminosity  (Eq.~\ref{eq:lkin}) of the wind deposits into the host galaxy approximately $0.5\%-5\%$ of the AGN bolometric luminosity \cite{SR98, DiMatteo05, SH05, HE10, KP15}.  Winds that originate at large distances from the black hole, for example from the torus or WAs, might be expected to have important effects on the host galaxy, however, these are rather slow-moving and may not carry sufficient kinetic luminosity or travel significant distances to influence the galaxy \cite{Crenshaw12, Fischer17}.   To this end, the highly blueshifted absorption features evident in some AGN X-ray spectra (Figure~\ref{fig:pdspg}), that are indicators of the so-called ultrafast outflows -- highly ionised material ejected from the black hole vicinity at substantial fractions of the speed of light -- might serve as the mechanism for delivering energy to the galaxy.  Since $L_{KE} \propto v_{r}^{3}$, these fast winds can potentially deposit the most amount of energy into the surroundings and alter galaxy evolution.

\subsection{UFO characteristics}
\label{subsec:ufochar}

{\em XMM-Newton}, {\em Chandra}, and {\em Suzaku} ushered in an era of high throughput spectroscopy in the Fe~K$\alpha$ band between $5-10$~keV.  Early observations revealed absorption-like features above $7$~keV (rest-frame) in the distant ($z=3.91$) lensed quasar APM~$08279+5255$ \cite{Chartas02}, the nearby ($z=0.184$) luminous quasar PDS~456 \cite{Reeves03}, and the narrow-line Seyfert~1 quasar, PG~$1211+143$  \cite{Pounds03}.  Attributed to K-shell absorption lines from \ion{Fe}{xxv} (He-like, $6.70$~keV) and \ion{Fe}{xxvi} (H-like, $6.97$~keV) implies relativistic velocities between $0.1-0.3$c.  

Sample studies \cite{Cappi06, Tombesi10a, Gofford13, Igo20} detect possible features with equivalent widths between $15-100$~eV in approximately $30\% - 40\%$ of sources. This population includes jetted \cite{Tombesi10b} and non-jetted AGN, as well as sources radiating at sub-Eddington and high-Eddington \cite{Pounds03, Hagino16, Parker17} values.  The high ionisation state of iron and depth of the absorption features implies high ionisation parameters of $\rm log( \xi / erg~cm~s^{-1}) \sim 3 - 6$ and column densities of $\rm log( N_H / cm^{-2}) \sim 22-24$ \cite{Tombesi11, Gofford13}.  The high detection rate indicates the covering factor ($\Omega$) of the wind is large.   Indeed, in PDS~456 an average solid angle of $\Omega \sim 3.2\pi$ is estimated \cite{Nardini15}.  In general, the application of P-Cygni profiles to simultaneously fit the absorption and emission from the wind imply it might be close to spherically symmetric \cite{Done07, Nardini15, Reeves19}.

UFO signatures have been reported in high-resolution grating data at lower energies between $0.3-2$~keV \cite{Longinotti15, Boissay19, Pounds03, Gupta13, Gupta15, Reeves20}.  The co-existence of slow-moving warm absorbers and ultrafast outflows \cite{Rogantini22, Pinto18, Parker17, Xu21} raise the question if these are different phases of the same stratified flow \cite{Serafinelli19}.  It is not yet clear if the AMD (Section~\ref{sec:wa}) can be described by an outflowing medium that is continuous, patchy, or in pressure equilibrium \cite{Behar09, Krongold03, Detmers11}.   

UFO features are significantly variable in equivalent width and velocities on all time scales down to hours \cite{Igo20, Parker17, Pinto18, Gallo19, Matzeu17, Reeves19}.  In some cases, these variations may arise from the wind responding to luminosity changes or they might be intrinsic to the launching mechanism (Section~\ref{subsec:wind}).  In these cases, variability studies will render a profound understanding of the wind and central engine.  Ascertaining the response time of the wind to changes in the ionising continuum reveals the wind density ($n$) since the recombination time is inversely proportional to $n$ (Eq.~\ref{eq:kk}).  The distance to the wind then follows since the ionisation parameter and luminosity are measured in the spectrum (i.e. $r = \frac{L_{ion}}{n\xi}$, from Eq.~\ref{eq:xi}).

It is still not completely possible to rule out that some wind features are consistent with random noise events.  In some cases, the wind features are based on the detection of a single absorption feature whose significance can depend on the continuum model and spectral binning. As illustrated in the Astro-H White paper on AGN winds \cite{Kaastra14}, the detection of two lines with a null hypothesis significance of $p_1$ and $p_2$ will elevate the significance of the wind to $p_1 \times p_2$.  With the potential to discern blended lines and distinguish weak features, high-resolution spectroscopy can better determine the occurrence rate of winds in AGN that is important for determining the wind geometry.  

\begin{figure*}
   \centering
   \advance\leftskip-0cm
   {\scalebox{0.4}{\includegraphics[trim= 1cm 1cm 2cm 2cm, angle=0, clip=true]{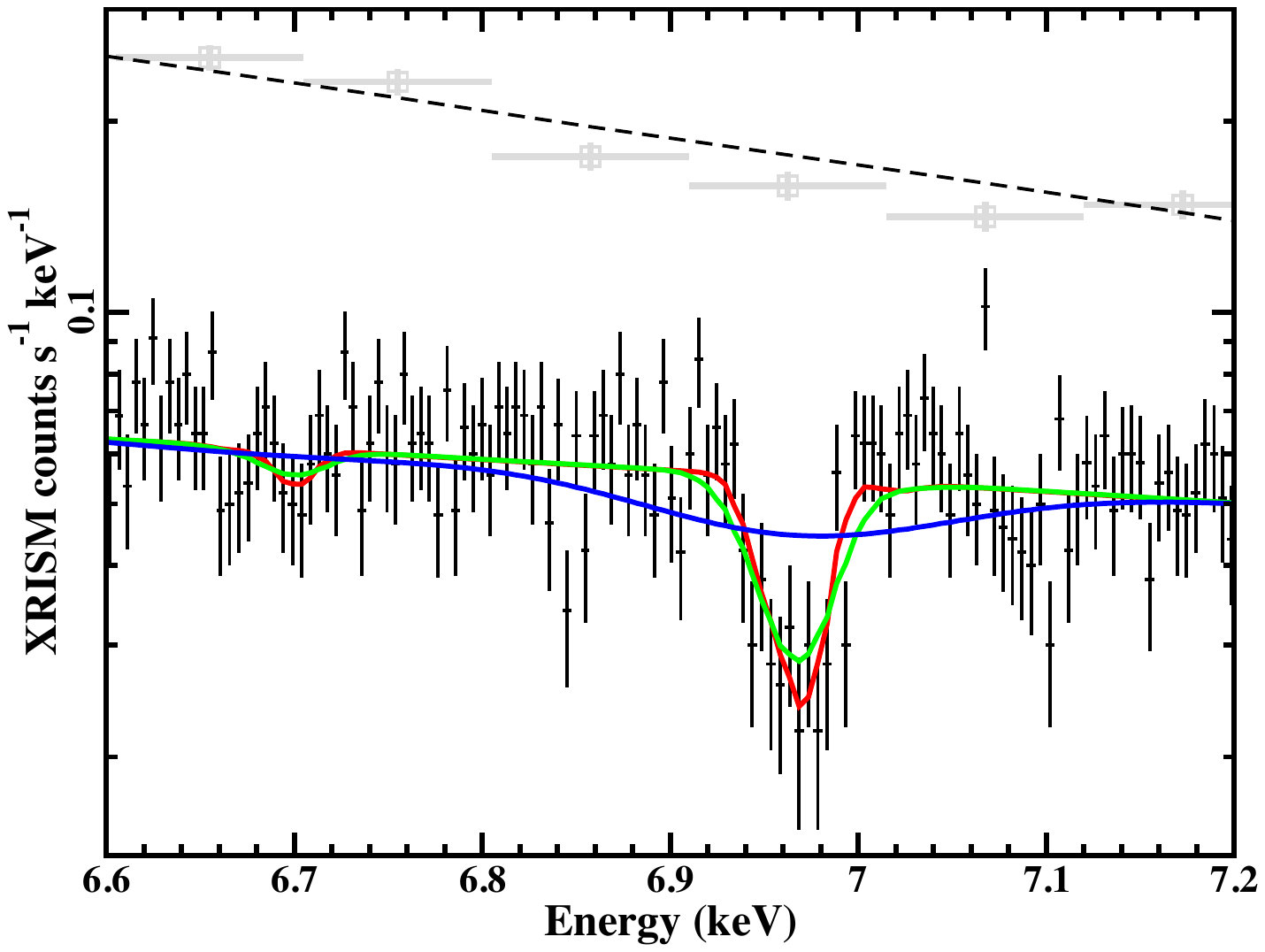}}}
   \caption{Simulated spectra of a photoionised plasma with $\rm log( N_H / cm^{-2})= 23$,  $\rm log( \xi / erg~cm~s^{-1})=4$ and $v_{turb} = 500$~km s$^{-1}$.  Both the EPIC-pn (grey) and XRISM-Resolve data (black) are based on $2-10$~keV brightness of $\sim10^{-11} \rm erg~cm^{-2}~s^{-1}$ and an exposure of $100$~ks.  The pn data are well fitted with a single power law.  The \ion{Fe}{xxvi} is significantly detected in the XRISM data and models with different turbulent velocities of $500$~km s$^{-1}$ (red), $1000$~km s$^{-1}$ (green), and $5000$~km s$^{-1}$ (blue) can be meaningfully examined. The pn spectrum is scaled for comparison. The data are optimally binned \cite{optbin}.}
   \label{fig:turb}
\end{figure*}

Figure~\ref{fig:turb} provides a demonstration of how line detection and line width will be significantly improved upon in the high-resolution era.  In the example, the H- and He-like iron lines have a turbulent velocity of $v_{turb} = 500$~km s$^{-1}$.  The line is undetected in the $100$~ks pn observation, but the \ion{Fe}{xxvi} is significantly detected in the $100$~ks XRISM-Resolve spectrum despite a smaller effective area.  Moreover, the Resolve data easily distinguishes from a high turbulent velocity of $5000$~km s$^{-1}$ and even from much more comparable velocities of $\sim1000$~km s$^{-1}$.

\subsection{The wind origin}
\label{subsec:wind}

The rapid variability, high degree of ionisation, and relativistic velocities all point to UFO features originating close to the black hole.  The inner accretion disc provides a natural environment for generating outflows in addition to the inward transport of matter.   The mechanisms by which winds are launched from an accretion disc are from (i) thermal (gas) pressure, (ii) radiation pressure, and (iii) magnetic fields.   Which mechanism  dominates depends on a number of factors including the degree of ionisation in the wind and the geometry of the system (see Figure~\ref{fig:agn}).

\subsubsection{Thermal driven winds}
\label{subsubsec:tdw}

The upper layer of the accretion disc will expand if it is heated to the Compton temperature by the X-rays coming from the inner region.  If the rate of expansion exceeds the escape velocity ($v_{esc} = \sqrt{2GM_{BH} / r }$) at a given radius ($r$), a thermal wind will be produced \cite{Begelman83}.  Thermal winds are commonly employed in stellar mass black holes \cite{Done18, Tomaru23}. Simulations show that thermal winds reach speeds of only $200-300$~km s$^{-1}$ because they are launched ballistically from the outer accretion disc (e.g. \cite{HP15, Higginbottom17}). Such winds are inadequate for explaining the UFO phenomenon in AGN.

\subsubsection{Radiative driven winds}
\label{subsubsec:rdw}

Radiation pressure from the accretion disc can launch a wind \cite{Murray95, Proga00, Proga04, KB01}.  This is best exemplified in broad absorption line (BAL) quasars that show broad UV absorption lines from, for example, \ion{C}{iv} and \ion{N}{v}, outflowing at velocities $\sim10,000 \rm km~s^{-1}$ (e.g. \cite{Turnshek84, Weymann91, Allen11}).  

If the gas is highly or completely ionised, Thomson and Compton scattering can be sufficient to power the wind.   This continuum-driving will be important in high luminosity systems radiating at values close to Eddington \cite{KP03}.  However, winds are commonly seen in sub-Eddington systems.  In these cases, line-driving will be more consequential.

In the line-driving scenario, the photoabsorption cross-section is larger than the electron scattering cross-section.  The opacity in the absorption lines serves as a force multiplier to enhance the effects of radiation pressure \cite{Castor75, Proga00, KB01, Dannen19}.  In this way, sub-Eddington sources can efficiently launch fast winds of weakly ionised material as is seen in BAL quasars.  However, the effects of the force multiplier are lost at $\rm log( \xi / erg~cm~s^{-1})>3$ \cite{Dannen19} even though ionisation parameters of $\rm log( \xi / erg~cm~s^{-1})>4$ are required to describe the spectral features seen in X-ray winds.  The ionisation of the gas likely occurs after the wind is launched, once it reaches a sufficient height above the disc (above the shielding failed wind; see Figure~\ref{fig:agn}) to be exposed to the central X-rays \cite{Proga04, Higginbottom14, Hagino15, Nomura17, Nomura20, Mizumoto21}.

\begin{figure*}
   \centering
   \advance\leftskip-0.1cm
   {\scalebox{0.37}{\includegraphics[trim= 0.5cm 4cm 2.cm 3cm, angle=0, clip=true]{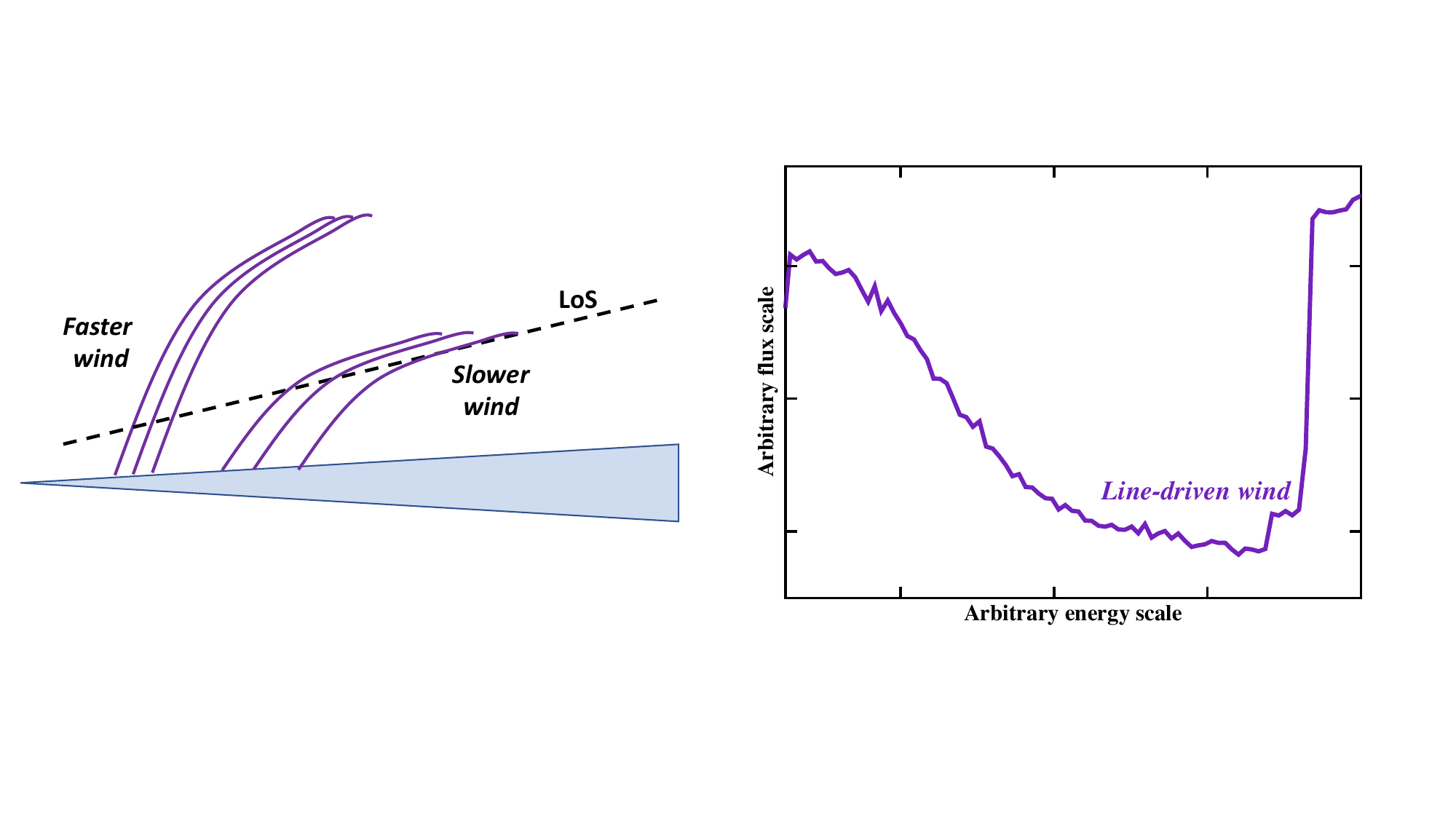}}}
   \caption{Left: The faster streamlines originating closer to the X-ray source will be more vertical compared to those of the slower wind at larger distances.  Right:  The scenario results in a line profile with a sharp blue edge and an extend red wing, which is a signature of the line-driven wind models.}
   \label{fig:rdline}
\end{figure*}

The observer looking down the line-driven wind funnel will see a large range of velocities.  The slow wind at large distance is generally moving along the line-of-sight, but the fast winds  will move on streamlines that are increasingly slanted closer to the origin (Figure~\ref{fig:rdline}).  This produces an absorption line profile with a sharp blue edge and extended red tail \cite{Knigge95, LK02, Sim08, Hagino15, Matzeu22}. Such a profile is depicted in Figure~\ref{fig:rdline} using the {\sc fast32rg} disc wind model \cite{Matzeu22, Sim08, Sim10} and can be compared to the MHD-driven wind profiles in Section~\ref{subsubsec:mhd}.
%

\subsubsection{Magnetic driven winds}
\label{subsubsec:mhd}

An outflow driven by magnetohydrodynamics (MHD) in the accretion disc is widely expected \cite{BP82, CL94, Fukumura10a, Fukumura10b, Kazanas12}.  In many sub-Eddington sources that possess UFOs, the force multiplier is small (i.e. the material is too highly ionised) to radiatively accelerate the wind \cite{Kraemer18}.  MHD winds can naturally explain the high velocity of highly ionised material without the need for line-driving.   

The wind will be launched from a continuous region over a large area of the accretion disc.  The wind density will fall with increasing distance.  The ionised gas will be accelerated in the poloidal magnetic field by magnetic-centrifugal and magnetic pressure forces  \cite{BP82, CL94, Fukumura10a, Fukumura22, Kazanas12}.  The faster gas will originate closer to the black hole.  Consequently, the fastest gas is also the most ionised gas. The line profile will depend on the velocity gradient along the line-of-sight and this can yield a characteristic line profile for MHD-driven winds when combined with the photoionisation balance \cite{Fukumura22}.  

\begin{figure*}
   \centering
   \advance\leftskip-0.1cm
   {\scalebox{0.37}{\includegraphics[trim= 0.5cm 3.5cm 1.5cm 2.5cm, angle=0, clip=true]{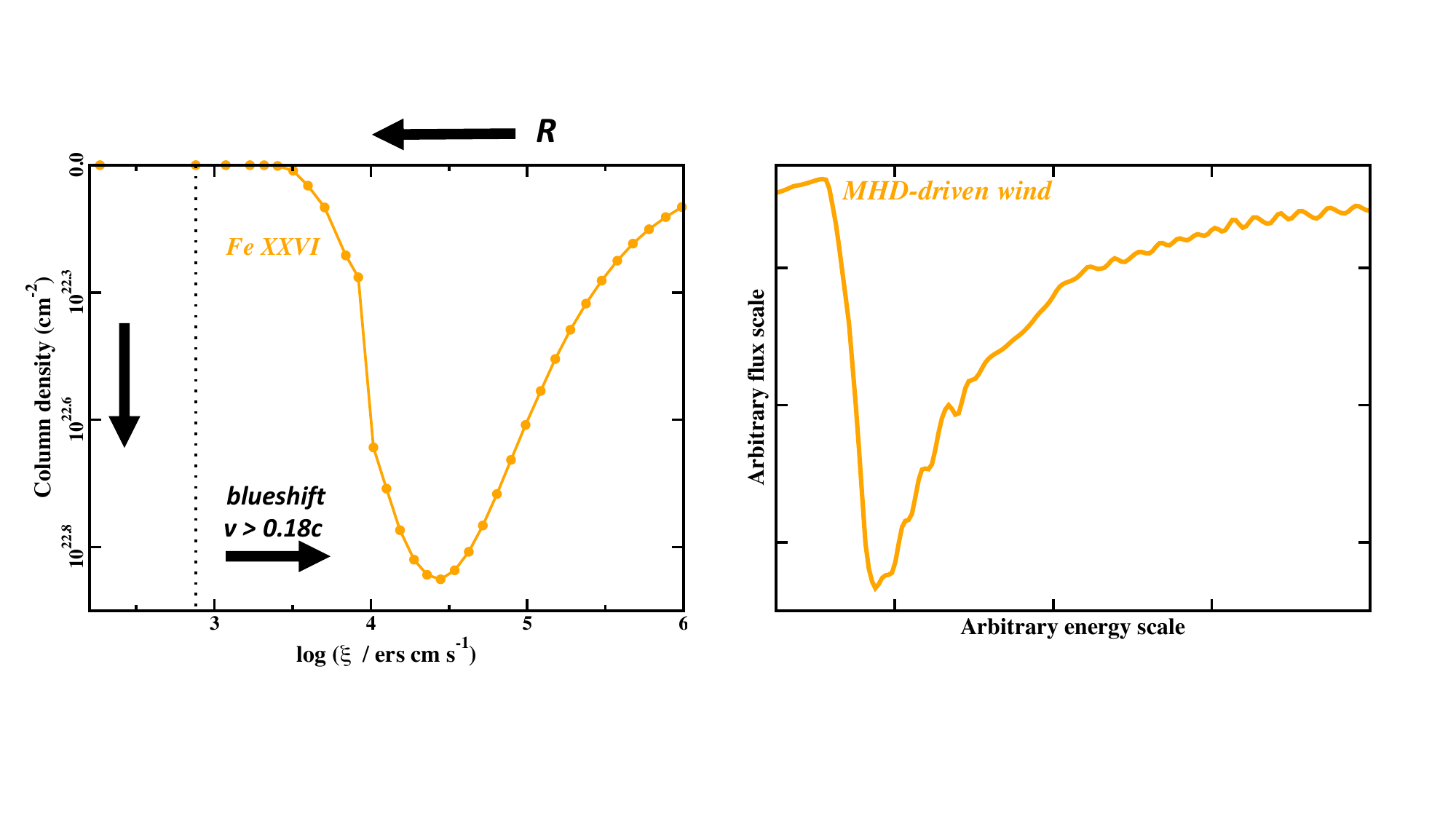}}}
   \caption{Left: The calculated column density (inverted on the vertical axis) as a function of ionisation parameter in the \ion{Fe}{xxvi} line for an adopted AMD (data provide courtesy of K. Fukumura). The ionisation parameter and velocity decrease with increasing distance ($R$) from the primary X-rays source.  Right:  The sharp red edge and extended blue wing in the $\rm N_{H}-\xi$ relation is replicated in the absorption line profile produced by an MHD-driven wind.}
   \label{fig:mhdline}
\end{figure*}

For a given ion, the column density will depend on the density gradient.  At the peak column density the optical depth in the line is maximum.  Moving toward a lower ionisation parameter (i.e. increasing distance from the ionising source) the column density drops rapidly producing a sharp red edge in the line.  The column density drops off more gradually as the ionisation parameter increases (i.e. decreasing distance) such that a blue wing forms in the line.  An example of the opacity in the \ion{Fe}{xxvi} line for an adopted AMD is shown in the right panel of Figure~\ref{fig:mhdline} (data kindly provided by K. Fukumura). Using the {\sc mhdwind} disc wind model \cite{Fukumura22}, it is seen that the MHD-driven wind will produce a line profile that replicates the same behaviour (Figure~\ref{fig:mhdline}, left panel).  This can be compared to the red asymmetry that is generated in a line-driven wind (see Figure~\ref{fig:rdline}). 

\subsubsection{Distinguishing the winds with high resolution spectroscopy}
\label{subsubsec:hipds}

Despite the striking differences evident in the line profiles (Figures~\ref{fig:turb}, ~\ref{fig:rdline}, and ~\ref{fig:mhdline}), the CCD resolution afforded by current observatories is insufficient to distinguish wind models.  As shown in Figure~\ref{fig:pdspg}, the MHD- and radiative-driven winds describe {\it XMM-Newton} data of PDS~456 and PG~1211+143 relatively well.  

\begin{figure*}
   \centering
   \advance\leftskip-0.cm
   {\scalebox{0.4}{\includegraphics[trim= 1cm 1cm 2cm 2cm, angle=0, clip=true]{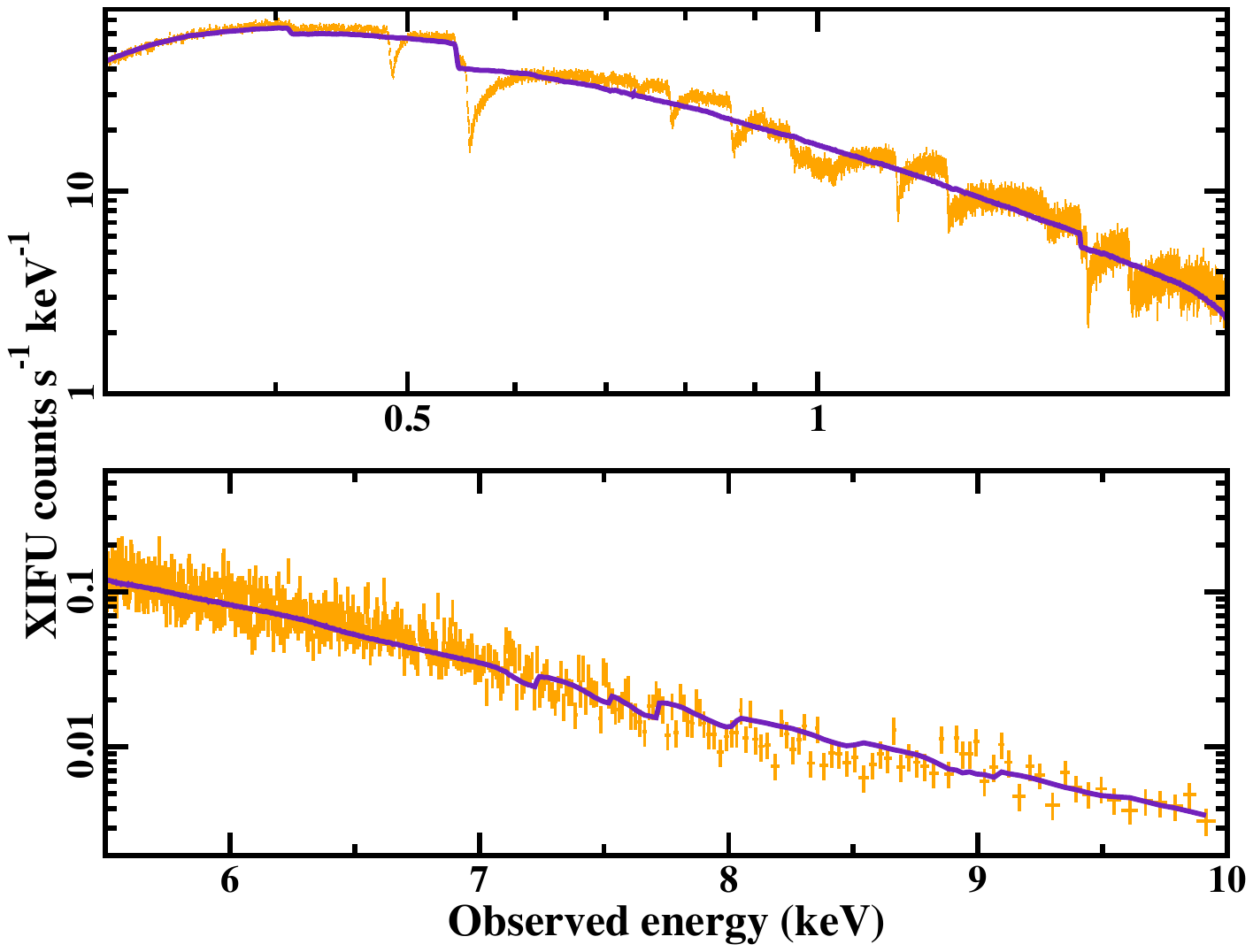}}}
   \caption{Simulated Athena-XIFU spectrum of PDS~456 ($100\rm~ks$) assuming the best-fit MHD-wind ({\sc mhdwind}) shown in Figure~\ref{fig:pdspg}.  The data are then fitted (purple curve) with the line-driven wind model ({\sc fast32rg}). The line profiles can be clearly distinguished with Athena revealing the nature of the launching mechanism.}
   \label{fig:athenapds}
\end{figure*}

The power of the high spectral resolution and large effective area that will be delivered by  Athena-XIFU is on display in Figure~\ref{fig:athenapds}.  Here, the best-fit MHD-driven wind model ({\sc mhdwind}) used to describe the EPIC-pn data (Figure~\ref{fig:pdspg}) is simulated for a $100\rm~ks$ observation with the Athena-XIFU.  The XIFU data cannot be well fitted with the line-driven wind model ({\sc fast32rg}) because of the different line profiles (Figure~\ref{fig:athenapds}).

\subsection{A windless alternative?}
\label{subsec:noufo}

Less fashionable, but equally important, is to consider alternative interpretations for the UFO features seen in X-ray spectra. In 2011, Gallo \& Fabian \cite{GF11} proposed that the blueshifted absorption features might not be due to outflows, but simply arising from low density, ionised gas on the surface of the orbiting inner disc.  In this scenario, the absorption feature is effectively an inverted disc line \cite{Fabian89, Laor} that is modified by orbital motions and relativity, and is imprinted on the reflection spectrum.  Orbital velocities within $10~R_{G}$ are more than sufficient to reproduce the blueshifts seen in UFOs features \cite{GF11}.

\begin{figure*}
   \centering
   \advance\leftskip-0.cm
   {\scalebox{0.4}{\includegraphics[trim= 1.5cm 2.7cm 2.cm 1.35cm, angle=+90,  clip=true]{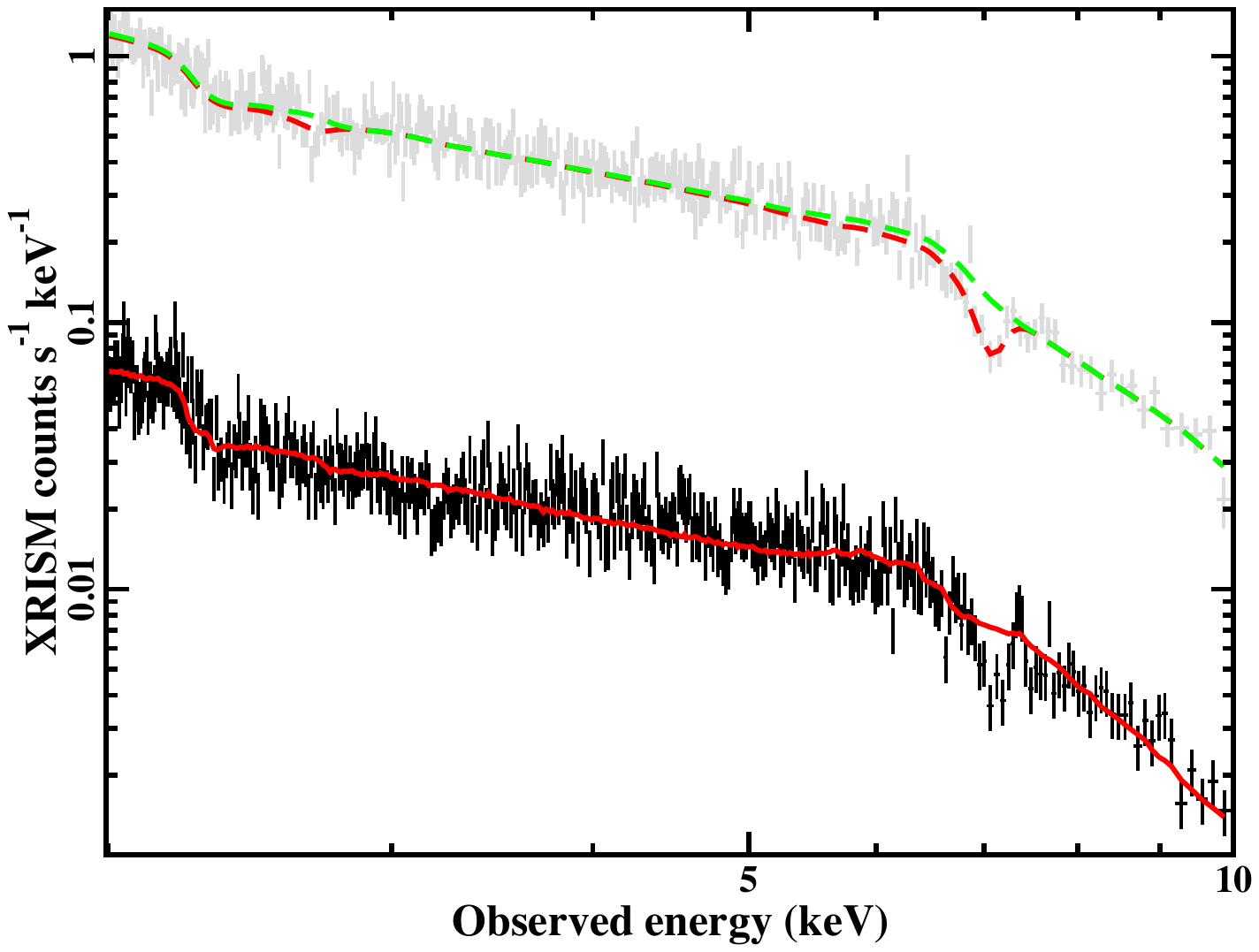}}}
   \caption{The disc-line absorption model fitted to the $2-10~\rm keV$ {\it XMM-Newton} pn data (grey) of PG~$1211+143$.  The green dashed line is a blurred ionised reflection fit to the spectrum.  The red dashed curve shows the fit to the data when including an inverted disc line originating beyond $6~R_G$.  The figure is reproduced with permission from \cite{GF13}.  The model is simulated for a $100\rm~ks$ XRISM-Resolve observation (black) and fitted with a line-driven wind model with no blurred reflection (solid red curve).  The wind model fits the redward emission well with a P-Cygni profile, but is unable to simultaneously describe the absorption.  The alternative disc-line absorption model should be testable with XRISM.}
   \label{fig:refabs}
\end{figure*}

The model was shown to fit the spectra of PG~1211+143 (Figure~\ref{fig:refabs}) \cite{GF13} and IRAS~$13224-3809$ \cite{Fabian20}.  For IRAS~$13224-3809$, the long, near continuous observations permitted variability studies of the UFO features \cite{Parker17, Jiang18, Pinto18}.  The manner in which the features were stronger in the low-flux state and disappeared when the AGN was bright was attributed to overionising the gas in the high-flux state.  In the disc-absorption scenario, the variability was consistent with the increased reflection fraction at low flux levels \cite{Fabian20}.  

This alternative model can be tested with high-resolution spectroscopy.  A $100\rm~ks$  XRISM-Resolve simulation of PG~$1211+143$ is shown in Figure~\ref{fig:refabs}.  When the simulated data are fitted with a  line-driven wind model with no blurred reflection, the redward excess and blueshifted absorption cannot be simultaneously reproduced.  Such a scenario cannot replace the wind scenario in all cases but might be viable for AGN exhibiting strong blurred reflection like some narrow-line Seyfert~1 galaxies \cite{Gallo18}.  This model can be scrutinized with XRISM and Athena.

\subsection{Progress and caveats}
\label{subsec:noufo}

High-resolution X-rays spectroscopy will carry the massive potential to enhance our understanding of ultrafast outflows in AGN.  With XRISM, Arcus, and Athena we will truly begin probing the origins, physics, and mechanisms at work in winds; and finally, test alternative models.  Not only will this improve our understanding of UFOs themselves, but better our knowledge of AGN physics on the black hole scales and AGN feedback on galactic scales. 

High-resolution spectroscopy will expose our misunderstandings.  The current wind models are incredibly sophisticated, but are incomplete as they are based on simplifying assumptions, limited understanding, and finite computational power.  Line profiles will depend on a number of factors, for example, the slope of the AMD. Even with high resolution, in some situations the different launching mechanisms might produce similar profiles \cite{Fukumura22}.  The reader is advised to understand the assumptions and limitations of each model before employing them in research.  

The physical situation will undoubtedly be much more complex.  There is no reason that hybrid winds, which are combinations of thermal, magnetic and radiative, cannot co-exist in a given AGN \cite{Everett05}.  Likewise, the blurred reflection will also complicate modelling \cite{Parker22}, and both the intrinsic AGN and wind can be simultaneously varying on different time scales. 

\section{Conclusion}
\label{sec:con}

The AGN community has benefited immensely from \xmm, \ch, and other great X-ray observatories over the past 20 years.  Many of these instruments will be producing excellent science for years to come. It is because of the tremendous success we have had with these instruments that the community can move forward with an eye on high resolution spectroscopy.

The launch of XRISM is imminent.  This will be followed by the launch
of an Athena-like mission (NewAthena) in the early or mid-2030s, covering the same pass band with
even sharper resolution and higher sensitivity.  In between, mission
concepts such as Arcus may provide particularly high resolution grating
spectroscopy below 1 keV, potentially simultaneously with UV
spectroscopy, providing pristine line profiles to perform different density diagnostic tests, the detection of metastable levels, and time-resolved spectroscopy.

Assuming that the calorimeter aboard XRISM will operate
for ten years, what are some optimistic goals for our understanding of
AGN?  And what are some equivalent goals for the more advanced
missions that will follow XRISM?

The sensitivity of XRISM should clearly determine the demographics of
both slow X-ray winds, and UFOs in bright, local Seyferts.  We should
learn if some sources display particularly line-rich spectra because
they afford a fortuitous viewing angle, because their Eddington
fraction is just right, or because their wind components have
fortuitous ionisation levels.  It is particularly important to
understand the demographics of wind feedback as a function of
Eddington fraction and XRISM will begin examining this for local Seyferts.
If XRISM finds that UFOs provide
insufficient feedback in the Seyfert phase, this does not preclude a
larger role for winds at higher Eddington fractions.  Athena will make
it possible to extend detailed wind studies to quasars at modest
red-shifts, and to local sources with lower Eddington fractions, and
thus answer this key question.

In a subset of the brightest Seyferts, XRISM should be able to extend
the range in ionisation over which AMDs are constructed by two orders
of magnitude, offering an improved understanding of wind driving
mechanisms.  In the same subset, ionisation time scales should reveal
gas densities, and therefore the absorption radius in the given wind
zone, providing an independent angle on wind-driving mechanisms.

Since new reverberation studies suggest that the innermost wall of the
``torus'' is only a factor of a few more distant from the central
engine than the optical BLR, few composite narrow Fe~K$\alpha$
emission line profiles might be expected in XRISM spectra.  However,
observations of bright Seyferts should clearly reveal the narrow
Fe~K$\alpha$ line production region and its relationship to the BLR
and torus.  In a subset of the brightest Seyferts, reverberation
mapping may offer an independent view on this problem, as well as
independent constraints on black hole masses.  If the disk is warped
between the ISCO and the BLR owing to a misalignment of the disk with
the black hole spin vector, resulting in ``extra'' reflecting area per
unit radius, XRISM may also be able to reveal hints of this structure.
In the longer term, the added sensitivity afforded by Athena should be
able to test related theories in a much larger set of systems.

The epoch of high-resolution X-ray spectroscopy is upon us.  For AGN, this is an exciting era that is full of massive potential for uncovering the ins and outs of black hole accretion.


\begin{acknowledgement}
The authors would like to thank Margaret Buhariwalla, Susmita Chackravorty, Keigo Fukumura, Adam Gonzalez, Jelle Kaastra, Tim Kallman, Gabriele Matzeu, Missagh Mehdipour, Daniel Proga, John Raymond, Daniele Rogantini for discussion, data, code, and help with figures.
\end{acknowledgement}







\begin{thebibliography}{99}

\bibitem{Allen11}
J. T. Allen, P. C. Hewett, N. Maddox, G. T. Richards, \& V. Belokurov, MNRAS \textbf{410}, 860-884 (2011)
 doi:10.1111/j.1365-2966.2010.17489.x [arXiv:1007.3991]
 \bibitem{Alston20}
W. N. Alston, A. C. Fabian, E. Kara, et al., NatAs \textbf{4}, 597-602 (2020)
 doi:10.1038/s41550-019-1002-x [arXiv:2001.06454]
\bibitem{Andonie22}
C. Andonie, F. E. Bauer, R. Carraro, et al., A\&A \textbf{664}, A46 (2022)
 doi:10.1051/0004-6361/202142473 [arXiv:2204.09469]
\bibitem{Antonucci93}
R. Antonucci, ARA\&A \textbf{31}, 473-521 (1993)
 doi:10.1146/annurev.aa.31.090193.002353 
 \bibitem{arav15}
N. Arav, C. Chamberlain, G. A. Kriss, et al., A\&A \textbf{577}, A37 (2015)
 doi:10.1051/0004-6361/20142530210.48550/arXiv.1411.2157 [arXiv:1411.2157]
\bibitem{arav08}
N. Arav, M. Moe, E. Costantini, et al., ApJ \textbf{681}, 954-964 (2008)
 doi:10.1086/588651 [arXiv:0807.0228]
 \bibitem{BH91}
S. A. Balbus \& J. F. Hawley, ApJ \textbf{376}, 214 (1991)
 doi:10.1086/170270 
\bibitem{Baldwin95}
J. Baldwin, G. Ferland, K. Korista, \& D. Verner, ApJL \textbf{455}, L119 (1995)
 doi:10.1086/309827 [arXiv:astro-ph/9510080]
 \bibitem{Ballantyne20}
D. R. Ballantyne, MNRAS \textbf{491}, 3553-3561 (2020)
 doi:10.1093/mnras/stz3294 [arXiv:1911.10029]
 \bibitem{Ballantyne01}
D. R. Ballantyne, R. R. Ross, \& A. C. Fabian, MNRAS \textbf{327}, 10-22 (2001)
 doi:10.1046/j.1365-8711.2001.04432.x [arXiv:astro-ph/0102040]
  \bibitem{Bambi21}
C. Bambi, L. W. Brenneman, T. Dauser, et al., SSRv \textbf{217}, 65 (2021)
 doi:10.1007/s11214-021-00841-8 [arXiv:2011.04792]
 \bibitem{Bambynek72}
W. Bambynek, B. Crasemann, R. W. Fink, et al., Reviews of Modern Physics, \textbf{44}, 716-813 (1972)
\bibitem{athena}
D. Barret, T. Lam Trong, J.-W. den Herder, et al., SPIE \textbf{10699}, 106991G (2018)
 doi:10.1117/12.2312409 [arXiv:1807.06092]
 \bibitem{Bautista03}
M. A. Bautista, C. Mendoza, T. R. Kallman, \& P. Palmeri, A\&A \textbf{403}, 339-355 (2003)
 doi:10.1051/0004-6361:20030367 [arXiv:astro-ph/0207323]
\bibitem{Bautista04}
M. A. Bautista, C. Mendoza, T. R. Kallman, \& P. Palmeri, A\&A \textbf{418}, 1171-1178 (2004)
 doi:10.1051/0004-6361:20034198 
  \bibitem{Begelman04}
M. C. Begelman, cbhg.symp \textbf 374 (2004)
 doi: [arXiv:astro-ph/0303040]
 \bibitem{Begelman83}
M. C. Begelman, C. F. McKee, \& G. A. Shields, ApJ \textbf{271}, 70-88 (1983)
 doi:10.1086/161178 
 \bibitem{Behar09}
E. Behar, ApJ \textbf{703}, 1346-1351 (2009)
 doi:10.1088/0004-637X/703/2/1346 [arXiv:0908.0539]
 \bibitem{behar01}
E. Behar, M. Sako, \& S. M. Kahn, ApJ \textbf{563}, 497-504 (2001)
 doi:10.1086/323966 [arXiv:astro-ph/0109314]
\bibitem{Bentz22}
M. C. Bentz, P. R. Williams, \& T. Treu, ApJ \textbf{934}, 168 (2022)
 doi:10.3847/1538-4357/ac7c0a [arXiv:2206.03513]
\bibitem{bianchi09}
S. Bianchi, M. Guainazzi, G. Matt, N. Fonseca Bonilla, \& G. Ponti, A\&A \textbf{495}, 421-430 (2009)
 doi:10.1051/0004-6361:200810620 [arXiv:0811.1126]
  \bibitem{BP82}
R. D. Blandford \& D. G. Payne, MNRAS \textbf{199}, 883-903 (1982)
 doi:10.1093/mnras/199.4.883 
\bibitem{Bland97}
J. Bland-Hawthorn, S. L. Lumsden, G. M. Voit, G. N. Cecil, \& J. C. Weisheit, Ap\&SS \textbf{248}, 177-190 (1997)
 doi:10.1023/A:1000525513140 [arXiv:astro-ph/9706211]
 \bibitem{blustin05}
A. J. Blustin, M. J. Page, S. V. Fuerst, G. Branduardi-Raymont, \& C. E. Ashton, A\&A \textbf{431}, 111-125 (2005)
 doi:10.1051/0004-6361:20041775 [arXiv:astro-ph/0411297]
 \bibitem{Boissay19}
R. Boissay-Malaquin, A. Danehkar, H. L. Marshall, \& M. A. Nowak, ApJ \textbf{873}, 29 (2019)
 doi:10.3847/1538-4357/ab0082 [arXiv:1901.06641]
 \bibitem{boller96}
T. Boller, W. N. Brandt, \& H. Fink, A\&A \textbf{305}, 53 (1996)
 doi:10.48550/arXiv.astro-ph/9504093 [arXiv:astro-ph/9504093]
\bibitem{brandt97}
W. N. Brandt, S. Mathur, \& M. Elvis, MNRAS \textbf{285}, L25-L30 (1997)
 doi:10.1093/mnras/285.3.L2510.48550/arXiv.astro-ph/9703100 [arXiv:astro-ph/9703100]
 \bibitem{Brunner22}
H. Brunner, T. Liu, G. Lamer, et al., A\&A \textbf{661}, A1 (2022)
 doi:10.1051/0004-6361/202141266 [arXiv:2106.14517]
  \bibitem{Cappi06}
M. Cappi, AN \textbf{327}, 1012 (2006)
 doi:10.1002/asna.200610639 [arXiv:astro-ph/0610117] 
\bibitem{Castor75}
J. I. Castor, D. C. Abbott, \& R. I. Klein, ApJ \textbf{195}, 157-174 (1975)
 doi:10.1086/153315 
 \bibitem{susmita09}
S. Chakravorty, A. K. Kembhavi, M. Elvis, \& G. Ferland, MNRAS \textbf{393}, 83-98 (2009)
 doi:10.1111/j.1365-2966.2008.14249.x10.48550/arXiv.0811.2404 [arXiv:0811.2404]
\bibitem{susmita12}
S. Chakravorty, R. Misra, M. Elvis, A. K. Kembhavi, \& G. Ferland, MNRAS \textbf{422}, 637-651 (2012)
 doi:10.1111/j.1365-2966.2012.20641.x10.48550/arXiv.1201.5435 [arXiv:1201.5435]
\bibitem{Chartas02}
G. Chartas, W. N. Brandt, S. C. Gallagher, \& G. P. Garmire, ApJ \textbf{579}, 169-175 (2002)
 doi:10.1086/342744 [arXiv:astro-ph/0207196]
 \bibitem{CL94}
J. Contopoulos \& R. V. E. Lovelace, ApJ \textbf{429}, 139 (1994)
 doi:10.1086/174307 
 \bibitem{costantini10}
E. Costantini, SSRv \textbf{157}, 265-277 (2010)
 doi:10.1007/s11214-010-9706-310.48550/arXiv.1011.2384 [arXiv:1011.2384]
 \bibitem{costantini07}
E. Costantini, J. S. Kaastra, N. Arav, et al., A\&A \textbf{461}, 121-134 (2007)
 doi:10.1051/0004-6361:20065390 [arXiv:astro-ph/0609385]
 \bibitem{Costantini10}
E. Costantini, J. S. Kaastra, K. Korista, et al., A\&A \textbf{512}, A25 (2010)
 doi:10.1051/0004-6361/200912555 [arXiv:1001.2712]
 \bibitem{costantini16}
E. Costantini, G. Kriss, J. S. Kaastra, et al., A\&A \textbf{595}, A106 (2016)
doi:10.1051/0004-6361/201527956 [arXiv:1606.06579]
 \bibitem{costantini00}
E. Costantini, F. Nicastro, A. Fruscione, et al., ApJ \textbf{544}, 283-292 (2000)
 doi:10.1086/31720010.48550/arXiv.astro-ph/0007158 [arXiv:astro-ph/0007158]
 \bibitem{Crenshaw12}
D. M. Crenshaw \& S. B. Kraemer, ApJ \textbf{753}, 75 (2012)
 doi:10.1088/0004-637X/753/1/75 [arXiv:1204.6694]
 \bibitem{crenshaw03}
D. M. Crenshaw, S. B. Kraemer, J. R. Gabel, et al., ApJ \textbf{594}, 116-127 (2003)
 doi:10.1086/37679210.48550/arXiv.astro-ph/0305154 [arXiv:astro-ph/0305154]
 \bibitem{wc2}
B. Czerny, M. Nikołajuk, A. Różańska, et al., A\&A \textbf{412}, 317-329 (2003)
 doi:10.1051/0004-6361:20031441 [arXiv:astro-ph/0309242]
 \bibitem{Dannen19}
R. C. Dannen, D. Proga, T. R. Kallman, \& T. Waters, ApJ \textbf{882}, 99 (2019)
 doi:10.3847/1538-4357/ab340b [arXiv:1812.01773]
 \bibitem{Dauser10}
T. Dauser, J. Wilms, C. S. Reynolds, \& L. W. Brenneman, MNRAS \textbf{409}, 1534-1540 (2010)
 doi:10.1111/j.1365-2966.2010.17393.x [arXiv:1007.4937]
 \bibitem{dehghanian19}
M. Dehghanian, G. J. Ferland, B. M. Peterson, et al., ApJL \textbf{882}, L30 (2019)
 doi:10.3847/2041-8213/ab3d4110.48550/arXiv.1908.07686 [arXiv:1908.07686]
 \bibitem{Detmers11}
R. G. Detmers, J. S. Kaastra, K. C. Steenbrugge, et al., A\&A \textbf{534}, A38 (2011)
 doi:10.1051/0004-6361/201116899 [arXiv:1107.0658]
 \bibitem{digesu16}
L. Di Gesu \& E. Costantini, A\&A \textbf{594}, A88 (2016)
 doi:10.1051/0004-6361/20162867010.48550/arXiv.1607.01943 [arXiv:1607.01943]
\bibitem{digesu13}
L. Di Gesu, E. Costantini, N. Arav, et al., A\&A \textbf{556}, A94 (2013)
 doi:10.1051/0004-6361/201321416 [arXiv:1305.6232]
 \bibitem{digesu15}
L. Di Gesu, E. Costantini, J. Ebrero, et al., A\&A \textbf{579}, A42 (2015)
 doi:10.1051/0004-6361/201525934 [arXiv:1505.02562]
 \bibitem{digesu14a}
L. Di Gesu, E. Costantini, E. Piconcelli, et al., A\&A \textbf{563}, A95 (2014)
 doi:10.1051/0004-6361/201322916 [arXiv:1401.5614]
  \bibitem{DiMatteo05}
T. Di Matteo, V. Springel, \& L. Hernquist, Natur \textbf{433}, 604-607 (2005)
 doi:10.1038/nature03335 [arXiv:astro-ph/0502199]
 \bibitem{Done07}
C. Done, M. A. Sobolewska, M. Gierliński, \& N. J. Schurch, MNRAS \textbf{374}, L15-L19 (2007)
 doi:10.1111/j.1745-3933.2006.00255.x [arXiv:astro-ph/0610078]
 \bibitem{Done18}
C. Done, R. Tomaru, \& T. Takahashi, MNRAS \textbf{473}, 838-848 (2018)
 doi:10.1093/mnras/stx2400 [arXiv:1612.09377]
\bibitem{ebrero21}
J. Ebrero, V. Domček, G. A. Kriss, \& J. S. Kaastra, A\&A \textbf{653}, A125 (2021)
 doi:10.1051/0004-6361/202040045 [arXiv:2107.05676]
\bibitem{ebrero16}
J. Ebrero, G. A. Kriss, J. S. Kaastra, \& J. C. Ely, A\&A \textbf{586}, A72 (2016)
 doi:10.1051/0004-6361/20152749510.48550/arXiv.1511.07169 [arXiv:1511.07169]
\bibitem{edmonds11}
D. Edmonds, B. Borguet, N. Arav, et al., ApJ \textbf{739}, 7 (2011)
 doi:10.1088/0004-637X/739/1/7 
\bibitem{Everett05}
J. E. Everett, ApJ \textbf{631}, 689-706 (2005)
 doi:10.1086/432678 [arXiv:astro-ph/0506321]
  \bibitem{Fabian12}
A. C. Fabian, ARA\&A \textbf{50}, 455-489 (2012)
 doi:10.1146/annurev-astro-081811-125521 [arXiv:1204.4114]
 \bibitem{Fabian89}
A. C. Fabian, M. J. Rees, L. Stella, \& N. E. White, MNRAS \textbf{238}, 729-736 (1989)
 doi:10.1093/mnras/238.3.729 
 \bibitem{Fabian20}
A. C. Fabian, C. S. Reynolds, J. Jiang, et al., MNRAS \textbf{493}, 2518-2522 (2020)
 doi:10.1093/mnras/staa482 [arXiv:2002.06388]
  \bibitem{Fabian09}
A. C. Fabian, A. Zoghbi, R. R. Ross, et al., Natur \textbf{459}, 540-542 (2009)
 doi:10.1038/nature08007
 \bibitem{FM00}
L. Ferrarese \& D. Merritt, ApJL \textbf{539}, L9-L12 (2000)
 doi:10.1086/312838 [arXiv:astro-ph/0006053]
 \bibitem{Fischer17}
T. C. Fischer, C. Machuca, M. R. Diniz, et al., ApJ \textbf{834}, 30 (2017)
 doi:10.3847/1538-4357/834/1/30 [arXiv:1609.08927]
  \bibitem{Fukumura22}
K. Fukumura, M. Dadina, G. Matzeu, et al., ApJ \textbf{940}, 6 (2022)
 doi:10.3847/1538-4357/ac9388 [arXiv:2205.08894]
\bibitem{Fukumura10b}
K. Fukumura, D. Kazanas, I. Contopoulos, \& E. Behar, ApJ \textbf{715}, 636-650 (2010)
 doi:10.1088/0004-637X/715/1/636 [arXiv:0910.3001]
\bibitem{Fukumura10a}
K. Fukumura, D. Kazanas, I. Contopoulos, \& E. Behar, ApJL \textbf{723}, L228-L232 (2010)
 doi:10.1088/2041-8205/723/2/L228 [arXiv:1009.5644]
\bibitem{Fukumura15}
K. Fukumura, F. Tombesi, D. Kazanas, et al., ApJ \textbf{805}, 17 (2015)
 doi:10.1088/0004-637X/805/1/17 [arXiv:1503.04074]
 \bibitem{gabel03}
J. R. Gabel, D. M. Crenshaw, S. B. Kraemer, et al., ApJ \textbf{583}, 178-191 (2003)
 doi:10.1086/34509610.48550/arXiv.astro-ph/0209484 [arXiv:astro-ph/0209484]
\bibitem{gabel05}
J. R. Gabel, S. B. Kraemer, D. M. Crenshaw, et al., ApJ \textbf{631}, 741-761 (2005)
 doi:10.1086/43268210.48550/arXiv.astro-ph/0506323 [arXiv:astro-ph/0506323]
 \bibitem{gallo06}
L. C. Gallo, MNRAS \textbf{368}, 479-486 (2006)
 doi:10.1111/j.1365-2966.2006.10137.x10.48550/arXiv.astro-ph/0602145 [arXiv:astro-ph/0602145]
 \bibitem{Gallo18}
L. Gallo, rnls.conf \textbf 34 (2018)
 doi:10.22323/1.328.0034 [arXiv:1807.09838]
 \bibitem{GF11}
L. C. Gallo \& A. C. Fabian, MNRAS \textbf{418}, L59-L63 (2011)
 doi:10.1111/j.1745-3933.2011.01143.x [arXiv:1108.5060]
\bibitem{GF13}
L. C. Gallo \& A. C. Fabian, MNRAS \textbf{434}, L66-L69 (2013)
 doi:10.1093/mnrasl/slt080 [arXiv:1306.3404]
\bibitem{Gallo19}
L. C. Gallo, A. G. Gonzalez, S. G. H. Waddell, et al., MNRAS \textbf{484}, 4287-4297 (2019)
 doi:10.1093/mnras/stz274 [arXiv:1901.07899]
 \bibitem{Gebhardt00}
K. Gebhardt, R. Bender, G. Bower, et al., ApJL \textbf{539}, L13-L16 (2000)
 doi:10.1086/312840 [arXiv:astro-ph/0006289]
 \bibitem{GF91}
I. M. George \& A. C. Fabian, MNRAS \textbf{249}, 352 (1991)
 doi:10.1093/mnras/249.2.352 
 \bibitem{george98}
I. M. George, T. J. Turner, H. Netzer, et al., ApJS \textbf{114}, 73-120 (1998)
 doi:10.1086/31306710.48550/arXiv.astro-ph/9708046 [arXiv:astro-ph/9708046]
 \bibitem{GP19}
M. Giustini \& D. Proga, A\&A \textbf{630}, A94 (2019)
 doi:10.1051/0004-6361/201833810 [arXiv:1904.07341]
 \bibitem{GP21}
M. Giustini \& D. Proga, IAUS \textbf{356}, 82-86 (2021)
 doi:10.1017/S1743921320002628 [arXiv:2002.07564]
\bibitem{Gofford13}
J. Gofford, J. N. Reeves, F. Tombesi, et al., MNRAS \textbf{430}, 60-80 (2013)
 doi:10.1093/mnras/sts481 [arXiv:1211.5810]
 \bibitem{Grier13}
C. J. Grier, P. Martini, L. C. Watson, et al., ApJ \textbf{773}, 90 (2013)
 doi:10.1088/0004-637X/773/2/90 [arXiv:1305.2447]
\bibitem{gu06}
M. F. Gu, T. Holczer, E. Behar, \& S. M. Kahn, ApJ \textbf{641}, 1227-1232 (2006)
 doi:10.1086/500640 [arXiv:astro-ph/0512410]
 \bibitem{gu01}
M. F. Gu, S. M. Kahn, D. W. Savin, et al., ApJ \textbf{563}, 462-471 (2001)
 doi:10.1086/323683 
\bibitem{Gupta13}
A. Gupta, S. Mathur, Y. Krongold, \& F. Nicastro, ApJ \textbf{772}, 66 (2013)
 doi:10.1088/0004-637X/772/1/66 [arXiv:1301.6139]
\bibitem{Gupta15}
A. Gupta, S. Mathur, \& Y. Krongold, ApJ \textbf{798}, 4 (2015)
 doi:10.1088/0004-637X/798/1/4 [arXiv:1406.5968]
\bibitem{gupta18}
M. Gupta, M. Sikora, K. Rusinek, \& G. M. Madejski, MNRAS \textbf{480}, 2861-2871 (2018)
 doi:10.1093/mnras/sty204310.48550/arXiv.1808.07170 [arXiv:1808.07170]

 

\bibitem{Hagino15}
K. Hagino, H. Odaka, C. Done, et al., MNRAS \textbf{446}, 663-676 (2015)
 doi:10.1093/mnras/stu2095 [arXiv:1410.1640]
\bibitem{Hagino16}
K. Hagino, H. Odaka, C. Done, et al., MNRAS \textbf{461}, 3954-3963 (2016)
 doi:10.1093/mnras/stw1579 [arXiv:1509.05645]
 \bibitem{halpern84}
J. P. Halpern, ApJ \textbf{281}, 90-94 (1984)
 doi:10.1086/162077 
\bibitem{HP15}
N. Higginbottom \& D. Proga, ApJ \textbf{807}, 107 (2015)
 doi:10.1088/0004-637X/807/1/107 [arXiv:1504.03328]
 \bibitem{Higginbottom14}
N. Higginbottom, D. Proga, C. Knigge, et al., ApJ \textbf{789}, 19 (2014)
 doi:10.1088/0004-637X/789/1/19 [arXiv:1402.1849]
\bibitem{Higginbottom17}
N. Higginbottom, D. Proga, C. Knigge, \& K. S. Long, ApJ \textbf{836}, 42 (2017)
 doi:10.3847/1538-4357/836/1/42 [arXiv:1612.08996]
 \bibitem{perseus}
Hitomi Collaboration, F. Aharonian, H. Akamatsu, et al., Natur \textbf{535}, 117-121 (2016)
 doi:10.1038/nature18627 [arXiv:1607.04487]
 \bibitem{n1275}
Hitomi Collaboration, F. Aharonian, H. Akamatsu, et al., PASJ \textbf{70}, 13 (2018)
 doi:10.1093/pasj/psx147 [arXiv:1711.06289]
 \bibitem{holczer07}
T. Holczer, E. Behar, \& S. Kaspi, ApJ \textbf{663}, 799-807 (2007)
 doi:10.1086/51841610.48550/arXiv.astro-ph/0703351 [arXiv:astro-ph/0703351]
 \bibitem{Holt80}
S. S. Holt, R. F. Mushotzky, R. H. Becker, et al., ApJL \textbf{241}, L13-L17 (1980)
 doi:10.1086/183350
 \bibitem{HE10}
P. F. Hopkins \& M. Elvis, MNRAS \textbf{401}, 7-14 (2010)
 doi:10.1111/j.1365-2966.2009.15643.x [arXiv:0904.0649]
 \bibitem{Igo20}
Z. Igo, M. L. Parker, G. A. Matzeu, et al., MNRAS \textbf{493}, 1088-1108 (2020)
 doi:10.1093/mnras/staa265 [arXiv:2001.08208]
 \bibitem{Jiang18}
J. Jiang, M. L. Parker, A. C. Fabian, et al., MNRAS \textbf{477}, 3711-3726 (2018)
 doi:10.1093/mnras/sty836 [arXiv:1804.00349]
 \bibitem{jiang19}
J. Jiang, D. J. Walton, A. C. Fabian, \& M. L. Parker, MNRAS \textbf{483}, 2958-2967 (2019)
 doi:10.1093/mnras/sty3228 [arXiv:1811.10932]
 \bibitem{juranova22}
A. Juráňová, E. Costantini, \& P. Uttley, MNRAS \textbf{510}, 4225-4235 (2022)
 doi:10.1093/mnras/stab3731 [arXiv:2201.02640]
 \bibitem{optbin}
J. S. Kaastra \& J. A. M. Bleeker, A\&A \textbf{587}, A151 (2016)
 doi:10.1051/0004-6361/201527395 [arXiv:1601.05309]
 \bibitem{kaastra12}
J. S. Kaastra, R. G. Detmers, M. Mehdipour, et al., A\&A \textbf{539}, A117 (2012)
 doi:10.1051/0004-6361/201118161 [arXiv:1201.1855]
 \bibitem{kaastra14}
J. S. Kaastra, G. A. Kriss, M. Cappi, et al., Sci \textbf{345}, 64-68 (2014)
 doi:10.1126/science.125378710.48550/arXiv.1406.5007 [arXiv:1406.5007]
 \bibitem{kaastra18}
J. S. Kaastra, M. Mehdipour, E. Behar, et al., A\&A \textbf{619}, A112 (2018)
 doi:10.1051/0004-6361/201832629 [arXiv:1805.03538]
 \bibitem{kaastra00}
J. S. Kaastra, R. Mewe, D. A. Liedahl, S. Komossa, \& A. C. Brinkman, A\&A \textbf{354}, L83-L86 (2000)
 doi: [arXiv:astro-ph/0002345]
\bibitem{kaastra96}
J. S. Kaastra, R. Mewe, \& H. Nieuwenhuijzen, uxsa.conf \textbf 411-414 (1996)
 doi: 
  \bibitem{kaastra04}
J. S. Kaastra, A. J. J. Raassen, R. Mewe, et al., A\&A \textbf{428}, 57-66 (2004)
 doi:10.1051/0004-6361:2004143410.48550/arXiv.astro-ph/0406199 [arXiv:astro-ph/0406199]
 \bibitem{Kaastra14}
J. S. Kaastra, Y. Terashima, T. Kallman, et al., arXiv \textbf arXiv:1412.1171 (2014)
 doi: [arXiv:1412.1171]
 \bibitem{KB01}
T. Kallman \& M. Bautista, ApJS \textbf{133}, 221-253 (2001)
 doi:10.1086/319184 
 \bibitem{Kammoun20}
E. S. Kammoun, J. M. Miller, M. Koss, et al., ApJ \textbf{901}, 161 (2020)
 doi:10.3847/1538-4357/abb29f [arXiv:2007.02616]
 \bibitem{kara21}
E. Kara, M. Mehdipour, G. A. Kriss, et al., ApJ \textbf{922}, 151 (2021)
 doi:10.3847/1538-4357/ac2159 [arXiv:2105.05840]
 \bibitem{kaspi00}
S. Kaspi, W. N. Brandt, H. Netzer, et al., ApJL \textbf{535}, L17-L20 (2000)
 doi:10.1086/312697
\bibitem{Kaspi01}
S. Kaspi, W. N. Brandt, H. Netzer, et al., ApJ \textbf{554}, 216-232 (2001)
 doi:10.1086/321333 [arXiv:astro-ph/0101540]
 \bibitem{Kazanas12}
D. Kazanas, K. Fukumura, E. Behar, I. Contopoulos, \& C. Shrader, AstRv \textbf{7}, 92-123 (2012)
 doi:10.1080/21672857.2012.11519707 [arXiv:1206.5022]
\bibitem{keshet22}
N. Keshet \& E. Behar, ApJ \textbf{934}, 124 (2022)
 doi:10.3847/1538-4357/ac7c6b [arXiv:2206.15074]
 \bibitem{King10}
A. R. King, MNRAS \textbf{402}, 1516-1522 (2010)
 doi:10.1111/j.1365-2966.2009.16013.x [arXiv:0911.1639]
 \bibitem{KP03}
A. R. King \& K. A. Pounds, MNRAS \textbf{345}, 657-659 (2003)
 doi:10.1046/j.1365-8711.2003.06980.x [arXiv:astro-ph/0305541]
 \bibitem{KP15}
A. King \& K. Pounds, ARA\&A \textbf{53}, 115-154 (2015)
 doi:10.1146/annurev-astro-082214-122316 [arXiv:1503.05206]
\bibitem{Knigge95}
C. Knigge, J. A. Woods, \& J. E. Drew, MNRAS \textbf{273}, 225-248 (1995)
 doi:10.1093/mnras/273.2.225 
 \bibitem{komossa00}
S. Komossa, NewAR \textbf{44}, 483-485 (2000)
 doi:10.1016/S1387-6473(00)00084-110.48550/arXiv.astro-ph/9810105 [arXiv:astro-ph/9810105]
\bibitem{komossa01}
S. Komossa \& S. Mathur, A\&A \textbf{374}, 914-918 (2001)
 doi:10.1051/0004-6361:2001081910.48550/arXiv.astro-ph/0106242 [arXiv:astro-ph/0106242]
 \bibitem{korista08}
K. T. Korista, M. A. Bautista, N. Arav, et al., ApJ \textbf{688}, 108-115 (2008)
 doi:10.1086/59214010.48550/arXiv.0807.0230 [arXiv:0807.0230]
 \bibitem{Koyama96}
K. Koyama, Y. Maeda, T. Sonobe, et al., PASJ \textbf{48}, 249-255 (1996)
 doi:10.1093/pasj/48.2.249 
  \bibitem{Kraemer18}
S. B. Kraemer, F. Tombesi, \& M. C. Bottorff, ApJ \textbf{852}, 35 (2018)
 doi:10.3847/1538-4357/aa9ce0 [arXiv:1711.07965]
  \bibitem{NIST}
 A. Kramida, Y. Ralchenko, J. Reader, NIST AST Team, (2022) NIST Atomic Spectra Database Version 5.10, available online
\bibitem{KK87}
J. H. Krolik \& T. R. Kallman, ApJL \textbf{320}, L5 (1987)
 doi:10.1086/184966 
 \bibitem{krolik_kriss95}
J. H. Krolik \& G. A. Kriss, ApJ \textbf{447}, 512 (1995)
 doi:10.1086/17589610.48550/arXiv.astro-ph/9501089 [arXiv:astro-ph/9501089]
 \bibitem{krolik81}
J. H. Krolik, C. F. McKee, \& C. B. Tarter, ApJ \textbf{249}, 422-442 (1981)
 doi:10.1086/159303 
 \bibitem{Krongold03}
Y. Krongold, F. Nicastro, N. S. Brickhouse, et al., ApJ \textbf{597}, 832-850 (2003)
 doi:10.1086/378639 [arXiv:astro-ph/0306460]
 \bibitem{krongold05}
Y. Krongold, F. Nicastro, M. Elvis, et al., ApJ \textbf{620}, 165-182 (2005)
 doi:10.1086/425293 [arXiv:astro-ph/0409490]
 \bibitem{krongold07}
 Y. Krongold, F. Nicastro, M. Elvis, et al. ApJ \textbf{659} 1022-1039 (2007)
 doi:10.1086/512476
 \bibitem{Laha16}
S. Laha, M. Guainazzi, S. Chakravorty, G. C. Dewangan, \& A. K. Kembhavi, MNRAS \textbf{457}, 3896-3911 (2016)
 doi:10.1093/mnras/stw211 [arXiv:1601.06369]
 \bibitem{Laha14}
S. Laha, M. Guainazzi, G. C. Dewangan, S. Chakravorty, \& A. K. Kembhavi, MNRAS \textbf{441}, 2613-2643 (2014)
 doi:10.1093/mnras/stu669 [arXiv:1404.0899]
 \bibitem{Laor}
A. Laor, ApJ \textbf{376}, 90 (1991)
 doi:10.1086/170257 
 \bibitem{LK02}
K. S. Long \& C. Knigge, ApJ \textbf{579}, 725-740 (2002)
 doi:10.1086/342879 [arXiv:astro-ph/0208011]
\bibitem{Longinotti15}
A. L. Longinotti, Y. Krongold, M. Guainazzi, et al., ApJL \textbf{813}, L39 (2015)
 doi:10.1088/2041-8205/813/2/L39 [arXiv:1511.01165]
 \bibitem{longinotti13}
A. L. Longinotti, Y. Krongold, G. A. Kriss, et al., ApJ \textbf{766}, 104 (2013)
 doi:10.1088/0004-637X/766/2/10410.48550/arXiv.1301.5463 [arXiv:1301.5463]
 \bibitem{wc1}
P. Magdziarz, O. M. Blaes, A. A. Zdziarski, W. N. Johnson, \& D. A. Smith, MNRAS \textbf{301}, 179-192 (1998)
 doi:10.1046/j.1365-8711.1998.02015.x 
 \bibitem{markowitz14}
A. G. Markowitz, M. Krumpe, \& R. Nikutta, MNRAS \textbf{439}, 1403-1458 (2014)
 doi:10.1093/mnras/stt249210.48550/arXiv.1402.2779 [arXiv:1402.2779]
 \bibitem{Matt02}
G. Matt, MNRAS \textbf{337}, 147-150 (2002)
 doi:10.1046/j.1365-8711.2002.05890.x [arXiv:astro-ph/0207615]
 \bibitem{Matzeu22}
G. A. Matzeu, M. Lieu, M. T. Costa, et al., MNRAS \textbf{515}, 6172-6190 (2022)
 doi:10.1093/mnras/stac2155 [arXiv:2207.13731]
\bibitem{Matzeu17}
G. A. Matzeu, J. N. Reeves, V. Braito, et al., MNRAS \textbf{472}, L15-L19 (2017)
 doi:10.1093/mnrasl/slx129 [arXiv:1708.03546]
 \bibitem{mao17}
J. Mao, J. S. Kaastra, M. Mehdipour, et al., A\&A \textbf{607}, A100 (2017)
 doi:10.1051/0004-6361/20173137810.48550/arXiv.1707.09552 [arXiv:1707.09552]
 \bibitem{mao22}
J. Mao, J. S. Kaastra, M. Mehdipour, et al., A\&A \textbf{665}, A72 (2022)
 doi:10.1051/0004-6361/20214263710.48550/arXiv.2204.06813 [arXiv:2204.06813]
 \bibitem{mehdipour19}
M. Mehdipour \& E. Costantini, A\&A \textbf{625}, A25 (2019)
 doi:10.1051/0004-6361/20193520510.48550/arXiv.1903.11605 [arXiv:1903.11605]
\bibitem{mehdipour17}
M. Mehdipour, G. A. Kriss, E. Costantini, et al., ApJL \textbf{934}, L24 (2022)
 doi:10.3847/2041-8213/ac822f [arXiv:2207.09464]
\bibitem{mehdipour22}
M. Mehdipour, G. A. Kriss, E. Costantini, et al., ApJL \textbf{934}, L24 (2022)
 doi:10.3847/2041-8213/ac822f10.48550/arXiv.2207.09464 [arXiv:2207.09464]
\bibitem{Mendoza04}
C. Mendoza, T. R. Kallman, M. A. Bautista, \& P. Palmeri, A\&A \textbf{414}, 377-388 (2004)
 doi:10.1051/0004-6361:20031621 [arXiv:astro-ph/0306320]
 \bibitem{Miller07}
J. M. Miller, ARA\&A \textbf{45}, 441-479 (2007)
 doi:10.1146/annurev.astro.45.051806.110555 [arXiv:0705.0540]
\bibitem{Miller18}
J. M. Miller, E. Cackett, A. Zoghbi, et al., ApJ \textbf{865}, 97 (2018)
 doi:10.3847/1538-4357/aadbaa [arXiv:1808.07435]
 \bibitem{Mizumoto21}
M. Mizumoto, M. Nomura, C. Done, K. Ohsuga, \& H. Odaka, MNRAS \textbf{503}, 1442-1458 (2021)
 doi:10.1093/mnras/staa3282 [arXiv:2003.01137]
\bibitem{Minezaki19}
T. Minezaki, Y. Yoshii, Y. Kobayashi, et al., ApJ \textbf{886}, 150 (2019)
 doi:10.3847/1538-4357/ab4f7b [arXiv:1910.08722]
\bibitem{Morishima22}
Y. Morishima, H. Sudou, A. Yamauchi, Y. Taniguchi, \& N. Nakai, PASJ..tmp \textbf (2022)
 doi:10.1093/pasj/psac092 [arXiv:2211.02280]
 \bibitem{mytorus}
K. D. Murphy \& T. Yaqoob, MNRAS \textbf{397}, 1549-1562 (2009)
 doi:10.1111/j.1365-2966.2009.15025.x [arXiv:0905.3188]
 \bibitem{Murray95}
N. Murray, J. Chiang, S. A. Grossman, \& G. M. Voit, ApJ \textbf{451}, 498 (1995)
 doi:10.1086/176238 
\bibitem{Nandra06}
K. Nandra, MNRAS \textbf{368}, L62-L66 (2006)
 doi:10.1111/j.1745-3933.2006.00158.x [arXiv:astro-ph/0602081]
\bibitem{Nardini15}
E. Nardini, J. N. Reeves, J. Gofford, et al., Sci \textbf{347}, 860-863 (2015)
 doi:10.1126/science.1259202 [arXiv:1502.06636]
\bibitem{Nenkova08}
M. Nenkova, M. M. Sirocky, R. Nikutta, Ž. Ivezić, \& M. Elitzur, ApJ \textbf{685}, 160-180 (2008)
 doi:10.1086/590483 [arXiv:0806.0512]
 \bibitem{nicastro99a}
F. Nicastro, F. Fiore, \& G. Matt, ApJ \textbf{517}, 108-122 (1999)
 doi:10.1086/30718710.48550/arXiv.astro-ph/9812392 [arXiv:astro-ph/9812392]
 \bibitem{nicastro99b}
F. Nicastro, F. Fiore, G. C. Perola, \& M. Elvis, ApJ \textbf{512}, 184-196 (1999)
 doi:10.1086/30673610.48550/arXiv.astro-ph/9808316 [arXiv:astro-ph/9808316]
 \bibitem{nicastro00}
F. Nicastro, L. Piro, A. De Rosa, et al., ApJ \textbf{536}, 718-728 (2000)
 doi:10.1086/30895010.48550/arXiv.astro-ph/0001201 [arXiv:astro-ph/0001201]
\bibitem{Nixon12}
C. Nixon, A. King, D. Price, \& J. Frank, ApJL \textbf{757}, L24 (2012)
 doi:10.1088/2041-8205/757/2/L24 [arXiv:1209.1393]
\bibitem{Noda22}
H. Noda, T. Mineta, T. Minezaki, et al., arXiv \textbf arXiv:2212.02731 (2022)
 doi: [arXiv:2212.02731]
 \bibitem{Nomura17}
M. Nomura \& K. Ohsuga, MNRAS \textbf{465}, 2873-2879 (2017)
 doi:10.1093/mnras/stw2877 [arXiv:1610.08511]
 \bibitem{Nomura20}
M. Nomura, K. Ohsuga, \& C. Done, MNRAS \textbf{494}, 3616-3626 (2020)
 doi:10.1093/mnras/staa948 [arXiv:1811.01966]
\bibitem{Palmeri03a}
P. Palmeri, C. Mendoza, T. R. Kallman, \& M. A. Bautista, A\&A \textbf{403}, 1175-1184 (2003)
 doi:10.1051/0004-6361:20030405 
\bibitem{Palmeri03b}
P. Palmeri, C. Mendoza, T. R. Kallman, M. A. Bautista, \& M. Meléndez, A\&A \textbf{410}, 359-364 (2003)
 doi:10.1051/0004-6361:20031262 [arXiv:astro-ph/0306321]
 \bibitem{Parker17}
M. L. Parker, C. Pinto, A. C. Fabian, et al., Natur \textbf{543}, 83-86 (2017)
 doi:10.1038/nature21385 [arXiv:1703.00071]
 \bibitem{Parker22}
M. L. Parker, G. A. Matzeu, J. H. Matthews, et al., MNRAS \textbf{513}, 551-572 (2022)
 doi:10.1093/mnras/stac877 [arXiv:2203.14789]
 \bibitem{Pinto18}
C. Pinto, W. Alston, M. L. Parker, et al., MNRAS \textbf{476}, 1021-1035 (2018)
 doi:10.1093/mnras/sty231 [arXiv:1708.09422]
 \bibitem{Ponti10}
G. Ponti, L. C. Gallo, A. C. Fabian, et al., MNRAS \textbf{406}, 2591-2604 (2010)
 doi:10.1111/j.1365-2966.2010.16852.x [arXiv:0911.1003]
 \bibitem{Ponti15}
G. Ponti, M. R. Morris, R. Terrier, et al., MNRAS \textbf{453}, 172-213 (2015)
 doi:10.1093/mnras/stv1331 [arXiv:1508.04445]
 \bibitem{Pounds03}
K. A. Pounds, J. N. Reeves, A. R. King, et al., MNRAS \textbf{345}, 705-713 (2003)
 doi:10.1046/j.1365-8711.2003.07006.x [arXiv:astro-ph/0303603]
 \bibitem{Proga00}
D. Proga, J. M. Stone, \& T. R. Kallman, ApJ \textbf{543}, 686-696 (2000)
 doi:10.1086/317154 [arXiv:astro-ph/0005315]
 \bibitem{Proga04}
D. Proga \& T. R. Kallman, ApJ \textbf{616}, 688-695 (2004)
 doi:10.1086/425117 [arXiv:astro-ph/0408293]
 \bibitem{Reeves19}
J. N. Reeves \& V. Braito, ApJ \textbf{884}, 80 (2019)
 doi:10.3847/1538-4357/ab41f9 [arXiv:1909.05039]
 \bibitem{Reeves20}
J. N. Reeves, V. Braito, G. Chartas, et al., ApJ \textbf{895}, 37 (2020)
 doi:10.3847/1538-4357/ab8cc4 [arXiv:2004.12439]
 \bibitem{Reeves03}
J. N. Reeves, P. T. O'Brien, \& M. J. Ward, ApJL \textbf{593}, L65-L68 (2003)
 doi:10.1086/378218 [arXiv:astro-ph/0307127]
 \bibitem{reeves09}
J. N. Reeves, R. M. Sambruna, V. Braito, \& M. Eracleous, ApJL \textbf{702}, L187-L190 (2009)
 doi:10.1088/0004-637X/702/2/L18710.48550/arXiv.0908.1715 [arXiv:0908.1715]
 \bibitem{reynolds97}
C. S. Reynolds, MNRAS \textbf{286}, 513-537 (1997)
 doi:10.1093/mnras/286.3.51310.48550/arXiv.astro-ph/9610127 [arXiv:astro-ph/9610127]
 \bibitem{Reynolds21}
C. S. Reynolds, ARA\&A \textbf{59}, 117-154 (2021)
 doi:10.1146/annurev-astro-112420-035022 [arXiv:2011.08948]
\bibitem{risaliti11}
G. Risaliti, E. Nardini, M. Salvati, et al., MNRAS \textbf{410}, 1027-1035 (2011)
 doi:10.1111/j.1365-2966.2010.17503.x10.48550/arXiv.1008.5067 [arXiv:1008.5067]
 \bibitem{Rogantini22}
D. Rogantini, E. Costantini, L. C. Gallo, et al., MNRAS \textbf{516}, 5171-5186 (2022)
 doi:10.1093/mnras/stac2552 [arXiv:2209.02747]
 \bibitem{rogantini22}
D. Rogantini, M. Mehdipour, J. Kaastra, et al., ApJ \textbf{940}, 122 (2022)
 doi:10.3847/1538-4357/ac9c01 
  \bibitem{RossFabian93}
R. R. Ross \& A. C. Fabian, MNRAS \textbf{261}, 74-82 (1993)
 doi:10.1093/mnras/261.1.74 
 \bibitem{RossFabian05}
R. R. Ross \& A. C. Fabian, MNRAS \textbf{358}, 211-216 (2005)
 doi:10.1111/j.1365-2966.2005.08797.x [arXiv:astro-ph/0501116]
\bibitem{rozanska08}
A. Różańska, I. Kowalska, \& A. C. Gonçalves, A\&A \textbf{487}, 895-900 (2008)
 doi:10.1051/0004-6361:200809549 
 


\bibitem{sako01}
M. Sako, S. M. Kahn, E. Behar, et al., A\&A \textbf{365}, L168-L173 (2001)
 doi:10.1051/0004-6361:20000081 [arXiv:astro-ph/0010660]
 \bibitem{Saxton08}
R. D. Saxton, A. M. Read, P. Esquej, et al., A\&A \textbf{480}, 611-622 (2008)
 doi:10.1051/0004-6361:20079193 [arXiv:0801.3732]
 \bibitem{Serafinelli19}
R. Serafinelli, F. Tombesi, F. Vagnetti, et al., A\&A \textbf{627}, A121 (2019)
 doi:10.1051/0004-6361/201935275 [arXiv:1906.02765]
 \bibitem{SS73}
N. I. Shakura \& R. A. Sunyaev, A\&A \textbf{24}, 337-355 (1973)
 doi: 
\bibitem{Shu10}
X. W. Shu, T. Yaqoob, \& J. X. Wang, ApJS \textbf{187}, 581-606 (2010)
 doi:10.1088/0067-0049/187/2/581 [arXiv:1003.1790]
 \bibitem{SR98}
J. Silk \& M. J. Rees, A\&A \textbf{331}, L1-L4 (1998)
 doi: [arXiv:astro-ph/9801013]
  \bibitem{silva16}
C. V. Silva, P. Uttley, \& E. Costantini, A\&A \textbf{596}, A79 (2016)
 doi:10.1051/0004-6361/20162855510.48550/arXiv.1607.01065 [arXiv:1607.01065]
\bibitem{Sim08}
S. A. Sim, K. S. Long, L. Miller, \& T. J. Turner, MNRAS \textbf{388}, 611-624 (2008)
 doi:10.1111/j.1365-2966.2008.13466.x [arXiv:0805.2251]
\bibitem{Sim10}
S. A. Sim, L. Miller, K. S. Long, T. J. Turner, \& J. N. Reeves, MNRAS \textbf{404}, 1369-1384 (2010)
 doi:10.1111/j.1365-2966.2010.16396.x [arXiv:1002.0544]
\bibitem{smith16}
R. K. Smith, M. H. Abraham, R. Allured, et al., SPIE \textbf{9905}, 99054M (2016)
 doi:10.1117/12.2231778 
 \bibitem{SH05}
V. Springel \& L. Hernquist, ApJL \textbf{622}, L9-L12 (2005)
 doi:10.1086/429486 [arXiv:astro-ph/0411379]
 \bibitem{steenbrugge09}
K. C. Steenbrugge, M. Fenovčík, J. S. Kaastra, E. Costantini, \& F. Verbunt, A\&A \textbf{496}, 107-119 (2009)
 doi:10.1051/0004-6361/200810416 [arXiv:0901.3108]
\bibitem{steenbrugge05}
K. C. Steenbrugge, J. S. Kaastra, D. M. Crenshaw, et al., A\&A \textbf{434}, 569-584 (2005)
 doi:10.1051/0004-6361:20047138 [arXiv:astro-ph/0501122]
\bibitem{hitomi}
T. Takahashi, M. Kokubun, K. Mitsuda, et al., JATIS \textbf{4}, 021402 (2018)
 doi:10.1117/1.JATIS.4.2.021402 
 \bibitem{Tanaka04}
Y. Tanaka, T. Boller, L. Gallo, R. Keil, \& Y. Ueda, PASJ \textbf{56}, L9-L13 (2004)
 doi:10.1093/pasj/56.3.L9 [arXiv:astro-ph/0405158]
 \bibitem{tarter69}
C. B. Tarter, W. H. Tucker, \& E. E. Salpeter, ApJ \textbf{156}, 943 (1969)
 doi:10.1086/150026 
  \bibitem{xrism}
M. Tashiro, H. Maejima, K. Toda, et al., SPIE \textbf{11444}, 1144422 (2020)
 doi:10.1117/12.2565812 
 \bibitem{titarchuk94}
L. Titarchuk, ApJ \textbf{434}, 570 (1994)
 doi:10.1086/174760 
 \bibitem{Tomaru23}
R. Tomaru, C. Done, \& J. Mao, MNRAS \textbf{518}, 1789-1801 (2023)
 doi:10.1093/mnras/stac3210 [arXiv:2204.08802]
 \bibitem{Tombesi10a}
F. Tombesi, M. Cappi, J. N. Reeves, et al., A\&A \textbf{521}, A57 (2010)
 doi:10.1051/0004-6361/200913440 [arXiv:1006.2858]
 \bibitem{Tombesi11}
F. Tombesi, M. Cappi, J. N. Reeves, et al., ApJ \textbf{742}, 44 (2011)
 doi:10.1088/0004-637X/742/1/44 [arXiv:1109.2882]
 \bibitem{Tombesi10b}
F. Tombesi, R. M. Sambruna, J. N. Reeves, et al., ApJ \textbf{719}, 700-715 (2010)
 doi:10.1088/0004-637X/719/1/700 [arXiv:1006.3536]
 \bibitem{torresi12}
E. Torresi, P. Grandi, E. Costantini, \& G. G. C. Palumbo, MNRAS \textbf{419}, 321-329 (2012)
 doi:10.1111/j.1365-2966.2011.19694.x10.48550/arXiv.1108.4890 [arXiv:1108.4890]
\bibitem{turner11}
T. J. Turner, L. Miller, S. B. Kraemer, \& J. N. Reeves, ApJ \textbf{733}, 48 (2011)
 doi:10.1088/0004-637X/733/1/48 [arXiv:1103.3709]
 \bibitem{Turnshek84}
D. A. Turnshek, ApJ \textbf{280}, 51-65 (1984)
 doi:10.1086/161967 
\bibitem{Urry95}
C. M. Urry \& P. Padovani, PASP \textbf{107}, 803 (1995)
 doi:10.1086/133630 [arXiv:astro-ph/9506063]
 \bibitem{Voges99}
W. Voges, B. Aschenbach, T. Boller, et al., A\&A \textbf{349}, 389-405 (1999)
 doi: [arXiv:astro-ph/9909315]
 \bibitem{Walton20}
D. J. Walton, W. N. Alston, P. Kosec, et al., MNRAS \textbf{499}, 1480-1498 (2020)
 doi:10.1093/mnras/staa2961 [arXiv:2009.10734]
\bibitem{Watanabe03}
S. Watanabe, M. Sako, M. Ishida, et al., ApJL \textbf{597}, L37-L40 (2003)
 doi:10.1086/379735 [arXiv:astro-ph/0309344]
 \bibitem{Wilkins22}
D. R. Wilkins, L. C. Gallo, E. Costantini, W. N. Brandt, \& R. D. Blandford, MNRAS \textbf{512}, 761-775 (2022)
 doi:10.1093/mnras/stac416 [arXiv:2202.06958]
\bibitem{Wilkins15}
D. R. Wilkins, L. C. Gallo, D. Grupe, et al., MNRAS \textbf{454}, 4440-4451 (2015)
 doi:10.1093/mnras/stv2130 [arXiv:1510.07656]
 \bibitem{Weymann91}
R. J. Weymann, S. L. Morris, C. B. Foltz, \& P. C. Hewett, ApJ \textbf{373}, 23 (1991)
 doi:10.1086/170020 
 \bibitem{Xu21}
Y. Xu, C. Pinto, S. Bianchi, et al., MNRAS \textbf{508}, 6049-6067 (2021)
 doi:10.1093/mnras/stab2984 [arXiv:2110.06633]
 \bibitem{Yaqoob01}
T. Yaqoob, I. M. George, K. Nandra, et al., ApJ \textbf{546}, 759-768 (2001)
 doi:10.1086/318315 [arXiv:astro-ph/0008471]
 \bibitem{Yaqoob12}
T. Yaqoob, MNRAS \textbf{423}, 3360-3396 (2012)
 doi:10.1111/j.1365-2966.2012.21129.x [arXiv:1204.4196]
 \bibitem{Zoghbi12}
A. Zoghbi, A. C. Fabian, C. S. Reynolds, \& E. M. Cackett, MNRAS \textbf{422}, 129-134 (2012)
 doi:10.1111/j.1365-2966.2012.20587.x [arXiv:1112.1717]
\bibitem{Zoghbi10}
A. Zoghbi, A. C. Fabian, P. Uttley, et al., MNRAS \textbf{401}, 2419-2432 (2010)
 doi:10.1111/j.1365-2966.2009.15816.x [arXiv:0910.0367]
 \bibitem{Zoghbi19}
A. Zoghbi, J. M. Miller, \& E. Cackett, ApJ \textbf{884}, 26 (2019)
 doi:10.3847/1538-4357/ab3e31 [arXiv:1908.09862]
 \bibitem{Zu13}
Y. Zu, C. S. Kochanek, S. Kozłowski, \& A. Udalski, ApJ \textbf{765}, 106 (2013)
 doi:10.1088/0004-637X/765/2/106 [arXiv:1202.3783]
 
 
 
 
 

  
  
 
 
 
 


\end{thebibliography}
\end{document}